\documentclass[%
 aip,
 amsmath,amssymb,
floatfix,%
reprint,%
]{revtex4-1}

\usepackage{graphicx}
\usepackage{dcolumn}
\usepackage{bm}
\usepackage{xcolor}
\usepackage[normalem]{ulem}

\usepackage[utf8]{inputenc}
\usepackage[T1]{fontenc}
\usepackage{mathptmx}
\usepackage{etoolbox}
\usepackage[shortlabels]{enumitem}

\newcommand{\fref}[1]{Fig.~\ref{#1}}
\newcommand{\tref}[1]{Table~\ref{#1}}
\newcommand{\eref}[1]{Eq.~\ref{#1}}
\newcommand{\cref}[1]{Chapter~\ref{#1}}
\newcommand{\sref}[1]{Sec.~\ref{#1}}
\newcommand{\aref}[1]{Appendix~\ref{#1}}

\newcommand{\newtxt}[1]{#1}

\makeatletter
\def\@email#1#2{%
 \endgroup
 \patchcmd{\titleblock@produce}
  {\frontmatter@RRAPformat}
  {\frontmatter@RRAPformat{\produce@RRAP{*#1\href{mailto:#2}{#2}}}\frontmatter@RRAPformat}
  {}{}
}%
\makeatother
\begin{document}

\preprint{AIP/123-QED}

\title[Dilute polymer solutions under shear flow]{Dilute polymer solutions under shear flow: comprehensive qualitative analysis using a bead-spring chain model with a FENE-Fraenkel spring}
\author{I. Pincus}
 \affiliation{Department of Chemical Engineering, Monash University, 
Melbourne, VIC~3800, Australia}
 
\author{A. Rodger}%
\affiliation{ 
Molecular Sciences, Macquarie University, NSW, 2109, Australia
}%

\author{J. Ravi Prakash}
 \email{ravi.jagadeeshan@monash.edu}
 \homepage{https://users.monash.edu.au/~rprakash/}
\affiliation{%
Department of Chemical Engineering, Monash University, 
Melbourne, VIC~3800, Australia
}%

\date{\today}

\begin{abstract}
Although the non-equilibrium behaviour of polymer solutions is generally well understood, particularly in extensional flow, there remain several unanswered questions for dilute solutions in simple shear flow, and full quantitative agreement with experiments has not been achieved.
For example, experimental viscosity data exhibit qualitative differences in shear-thinning exponents, shear rate for onset of shear-thinning and high-shear Newtonian plateaus depending on polymer semiflexibility, contour length and solvent quality.
While polymer models are able to incorporate all of these effects through various spring force laws, bending potentials, excluded volume (EV) potentials, and hydrodynamic interaction (HI), the inclusion of each piece of physics has not been systematically matched to experimentally observed behaviour.
Furthermore, attempts to develop multiscale models \newtxt{(in the sense of representing an arbitrarily small or large polymer chains)} which can make quantitative predictions are hindered by the lack of ability to fully match the results of bead-rod models, often used to represent a polymer chain at the Kuhn step level, with bead-spring models, which take into account the entropic elasticity.
In light of these difficulties, this work aims to develop a general model based on the so-called FENE-Fraenkel spring, originally formulated by Larson and coworkers [J. Chem. Phys. \textbf{124} (2006), 10.1063/1.2161210], which can span the range from rigid rod to traditional entropic spring, as well as include a bending potential, EV and HI.
As we show, this model can reproduce, and smoothly move between, a wide range of previously observed polymer solution rheology in shear flow.
\end{abstract}

\maketitle

\section{Introduction}
\label{Introduction} 

Decades of research have resulted in a mature understanding of the behaviour of dilute polymer solutions in flow, to the point where direct quantitative comparisons with experimental data in extensional flow are possible \cite{Prabhakar2004SFG, Saadat2015SFG, Sunthar2005parameterfree}.
However, there remain several unresolved qualitative questions regarding behaviour in shear flow, and complete quantitative analysis of experimental results continues to be challenging \cite{Larson2005review, schroeder2005dynamics, Prakash2019review}.
\newtxt{Experimentally, in shear flow, one observes a decrease in polymer viscosity and first normal stress difference with shear rate \cite{Larson2005review} (shear thinning), as well as a `flattening' of the end-to-end distribution function and changes in gyration tensor components due to stretching and tumbling of the polymer chain \cite{Schroeder2018review, Schroeder2005_characteristic}.
It is generally accepted that a combination of chain-solvent friction, Brownian motion, finite chain extensibility, semiflexibility, hydrodynamic interactions (HI) and excluded volume (EV) effects can account for these observations \cite{Prakash2019review, Larson2005review, pan2018shear} (as well as internal viscosity, self-entanglement and charge effects for certain polymers \cite{Larson2005review, Kailasham2018, Kailasham2021RouseModel, kailasham2021important, kailasham2022shear}).}

\begin{figure}[t]
    \centering
    \includegraphics[width=8.5cm,height=!]{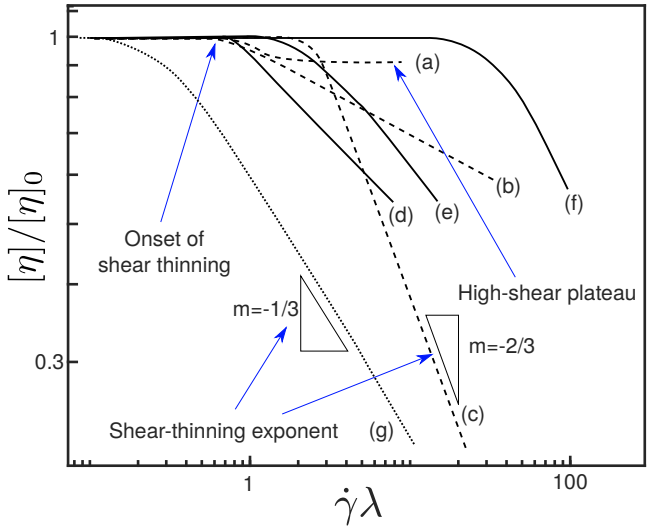}
    \caption{Schematic of observed experimental shear-thinning behaviour in polymer solutions. Here $[\eta]/[\eta]_0$ is the intrinsic viscosity at some shear rate divided by the intrinsic viscosity at zero shear. The shear rate is  $\dot{\gamma}$, normalised by the polymer relaxation time $\lambda$. The value of $\lambda$ is determined from the zero-shear viscosity via $\lambda = [\eta]_0 \eta_s M/ N_A k_\mathrm{B} T$, where $\eta_s$ is the solvent viscosity, $M$ is the polymer molecular weight, $k_\mathrm{B}$ is Boltzmann's constant and $T$ is the temperature. Data is traced from plots of several authors \cite{noda1968rate, Hua2006, pan2018shear, yang1958non}. Values should not be considered exact, only qualitatively correct. Dashed lines (curves a, b and c) are polystyrene of $13.6 \times 10^6$ g/mol \cite{noda1968rate} in a theta solvent (a) and good solvent (b), as as well as $2 \times 10^6$ g/mol \cite{Hua2006} in a theta solvent (c). Solid lines (curves d, e and f) are DNA at various solvent qualities \cite{pan2018shear} with lengths $25$ kbp (d), $48.5$ kbp (e), and $165.6$ kbp (f). Dotted line (curve g) is PBLG (a rigid molecule) of $M=2.08 \times 10^5$ g/mol in m-cresol solvent \cite{yang1958non}. Note the differences in onset of shear thinning, shear thinning exponent, and high-shear plateau as polymer length, flexibility and solvent quality are changed.}
    \label{fig:Experimental schematic}
\end{figure}

\begin{figure}[t]
    \centering
    \includegraphics[width=8.5cm,height=!]{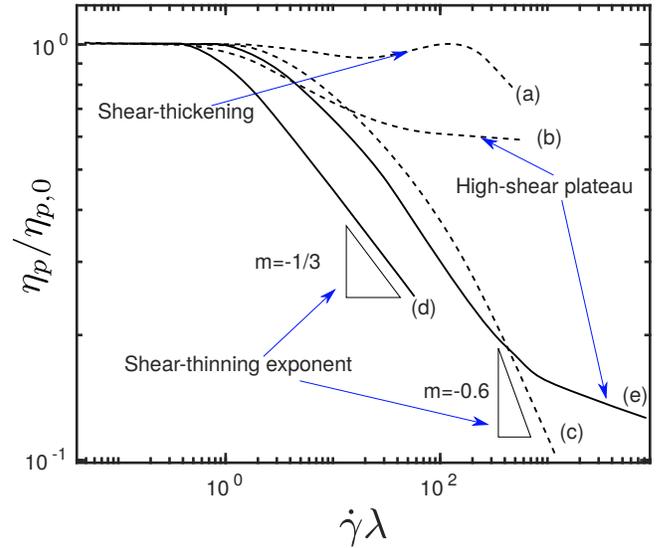}
    \caption{Schematic of theoretical and simulation findings, traced from previous works and not exact. Here $\eta_p/\eta_{p,0}$ is the polymer contribution to the solution viscosity divided by that at zero shear rate. The shear rate is $\dot{\gamma}$, while $\lambda$ is some measure of longest relaxation time of the chain, generally the decay time of the end-to-end vector autocorrelation function. Dotted lines (curves a, b and c) are (a) `spring-like' results for FENE chains with HI \cite{kishbaugh1990discussion}, (b) Hookean chains with EV \cite{ahn1993bead} and (c) Marko-Siggia force-law \cite{marko1995statistical} chains with EV and HI \cite{schroeder2005dynamics}. Solid lines (curves d and e) are `rod-like' results, namely (d) stiff Fraenkel springs with a strong bending potential \cite{Ryder2006}, and (e) a bead-rod chain with HI but no EV \cite{Petera1999}.}
    \label{fig:Simulations schematic}
\end{figure}

\newtxt{However, current models are somewhat narrow in their inclusion of the key physics.
There are two broad classes - bead-rod models, which represent the physical polymer chain at the level of a Kuhn step, and incorporate HI and EV to match the local chain friction and chain self-exclusioin \cite{Petera1999, Liu2004, hur2000brownian}, as well as bead-spring models, which further coarse-grain many Kuhn steps into a single extensible segment, and incorporate HI and EV in universal terms through the radius of gyration swelling and chain relaxation time \cite{Prabhakar2004SFG, hur2000brownian, Hsieh2004, schroeder2005dynamics, Sunthar2005parameterfree}.
On one hand, both models have been generally successful at reproducing the aforementioned experimental behaviour, but on the other hand, they disagree on specific details in quite important ways.}
One key difficulty is in correctly describing the change in viscosity as shear rate is increased, where experiments and simulations give confusingly varied results \cite{pan2018shear, Prakash2019review}.
For example, changes in polymer molecular weight, backbone semiflexibility and solvent-polymer interactions contribute to differences in shear-thinning exponents, shear rates for onset of shear-thinning, and appearance of a high-shear plateau, as depicted in \fref{fig:Experimental schematic}.
Additionally, \fref{fig:Simulations schematic} gives several results derived from simulated and theoretical models, with considerable differences in behaviour depending on the type of bead-bead connection (rod or spring), inclusion of hydrodynamic interactions (HI) or excluded volume effects (EV), as well as use of a bending potential.
\newtxt{It is also difficult to correlate these effects with other polymer properties - if we see a particular change in the viscosity scaling, what should we expect this to say about the tumbling frequency or gyration tensor?}
Clearly, it would be useful to have a single model which can span the entire range of previously-modelled behaviour, in order to systematically investigate the effects of each piece of added physics.

\newtxt{In light of these difficulties, the aim of this paper is to introduce a model based upon the so-called FENE-Fraenkel spring force law (originally used by Larson and coworkers \cite{Hsieh2006}) which can behave as both a bead-rod and bead-spring model \cite{pincus2020viscometric}, while also including HI, EV and semiflexibility.
While we will present some very preliminary comparisons with experimental results for a short, rigid polymer chain, our primary purpose is to examine the qualitative features of this model, particularly in the crossover between bead-spring and bead-rod behaviour.
In doing so, we hope to elucidate the effects of EV, HI and semiflexibility in these two limits, as well as carefully describe and consolidate the wide range of measured properties currently reported in the literature.
}

\begin{figure*}[!ht]
  \centerline{
  \includegraphics[width=17cm,height=!]{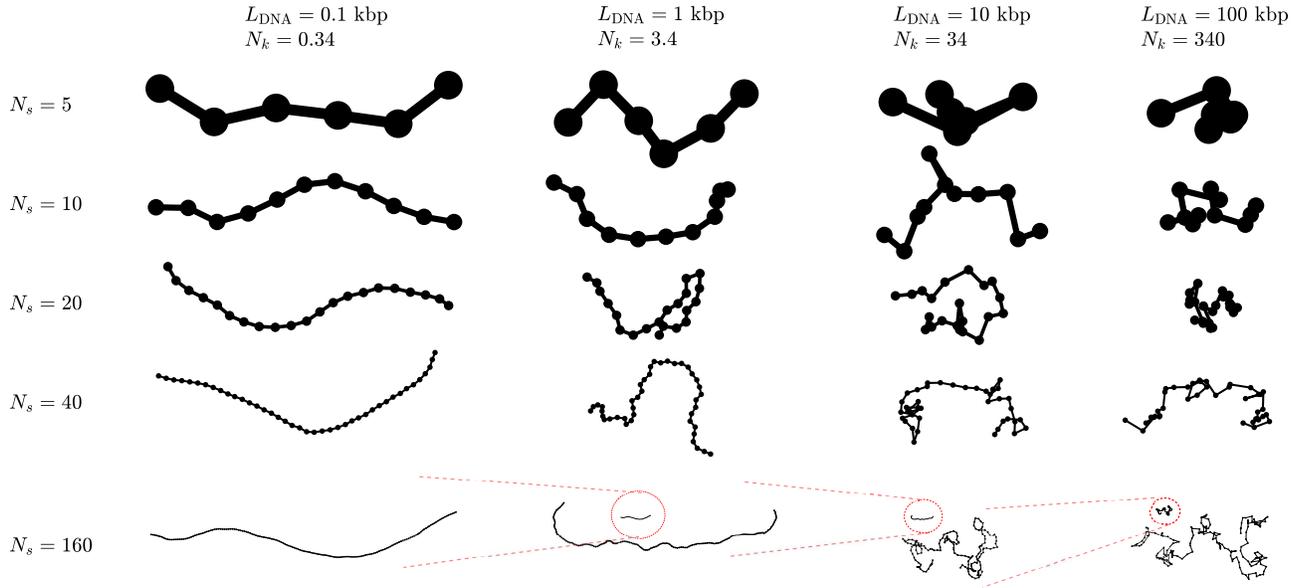}
  }
  \caption[Example equilibrium configurations for several DNA fragment lengths and levels of model discretisation]{\newtxt{Example equilibrium configurations for several DNA fragment lengths and levels of model discretisation generated using our modelling procedure. $L_\mathrm{DNA}$ is the length of each DNA fragment in base pair units, $N_k \equiv L/2l_p$ is the number of Kuhn steps in the DNA fragment, and $N_s$ is the number of springs in the model chain. Overall 3D structures have been projected onto the principal gyration axes for ease of comparison. Columns correspond to different lengths of DNA, while rows are the number of springs used to represent the DNA fragment. Note that although solid lines are used to connect beads in this schematic, in reality each connection is a spring of variable length. Final row of $N_s = 160$ shows comparisons of coil sizes between columns, such that each circled configuration represents the previous configuration at the same scale as the next.}}
  \label{fig: example eq configs FF}
\end{figure*}

\newtxt{Before continuing with prior literature on this subject, we believe it insightful to `start at the end' so to speak, and show how our model relates to a real polymer chain, to provide context to qualitative findings both in the current work and in prior work.
Imagine that one wishes to simulate a semiflexible polymer chain with a wide range of molecular weights, of which DNA is perhaps the ideal example.
The equilibrium properties of DNA in a $\theta$-solvent can be characterised by two properties, the persistence length $l_p$, related to the semiflexibility and backbone stiffness, and the contour length $L$, proportional to the molecular weight.
If one wishes to represent a DNA fragment with arbitrary $l_p$ and $L$ using a coarse-grained bead-spring-chain polymer model with fixed bead number $N$ (ideally in such a way that results are in fact independent of $N$ for sufficiently large $N$), one must set the bending potential to capture the semiflexibility (quantified by $l_p$), and then set the spring potential to capture both the contour length $L$ as well as the DNA fragment gyration radius (which is itself a function of $L$ and $l_p$).
One therefore requires not only a bending potential but also a spring force law which can represent a DNA sub-segment of arbitrary size, which may be of only several base pairs length, or up to tens of thousands of base pairs length.
The short segment of several to hundreds of base pairs has been modelled by a rigid rod \cite{hur2000brownian}, while a longer segment of tens of thousands of base pairs has been modelled by a WLC or similar potential \cite{schroeder2005dynamics, Sunthar2005parameterfree}.}

\newtxt{Our model aims to move smoothly between these limits, as shown in \fref{fig: example eq configs FF}.
For very short chains with high bead numbers, a very stiff, inextensible spring and strong bending potential is required, while for long chains with low bead numbers, the bending potential vanishes and the spring essentially has a FENE or WLC form.
At fixed bead number, decreasing the length of the underlying polymer chain is equivalent to moving from a `spring-like' to a `rod-like' force law, an area which is as-yet largely unexplored in the literature.}

\newtxt{Having introduced this model with relation to a real DNA chain, we wish to strongly emphasise that our primary purpose is not to quantitatively reproduce the behaviour of a real polymer chain (but note that this has been performed for Linear Dichroism measurements \cite{Kubista1993} in separate work pending publication \cite{Pincus2022}).
Instead, we explore the properties of our model in the crossover between bead-spring and bead-rod regimes, and in fact will show that the specific form of the spring potential is less important in qualitative terms than whether it behaves `like a spring' or `like a rod'. 
To do so, we first summarise prior literature on shear flow behaviour of dilute polymer solutions.
}

While chains with Hookean springs and preaveraged hydrodynamic interactions (the Zimm model) are able to accurately describe the linear viscoelastic properties of many polymer solutions, their infinite stretchability leads to inaccurate predictions in flow \cite{bird1987dynamics}.
The finite extensibility of a chain is generally included in one of two ways, either using rigid rods to represent each Kuhn step of the backbone individually, or a finitely extensible entropic spring which approximates the force-extension behaviour of a large segment of the underlying molecule \cite{underhill2004coarse}.
Bead-rod dumbbells are known to have a shear-thinning exponent of $(-1/3)$ (\fref{fig:Simulations schematic}, curve d) and approximate the behaviour of highly rigid molecules \cite{Stewart:1972gt, pincus2020viscometric}, while FENE-spring dumbbells show a $(-2/3)$ exponent and are able to qualitatively predict the shear-thinning behaviour of some flexible polymer solutions \cite{bird1987dynamics, mochimaru1981fast, Fan:1985jk}.
Additionally, the Weissenberg number for onset of shear-thinning increases with increasing extensibility (quantified by the FENE $b$-parameter) \cite{ahn1993bead}, a behaviour which is also found as the molecular weight of experimental systems is increased \cite{noda1968rate, pan2018shear} (\fref{fig:Experimental schematic} curves d, e and f).

For chains, a well-known and somewhat counter-intuitive result is that a bead-rod chain with connections modelled as hard constraints gives a different equilibrium configurational distribution from an infinitely-stiff Fraenkel chain \cite{bird1987dynamics, van1981statistical, VanKampen1984}, although the difference seems unimportant in practice \cite{Hsieh2006}.
In the following discussion, we will refer to both simply as `bead-rod' chains, except where explicitly distinguished.
While finitely extensible bead-spring chains show a $(-1/2)$ to $(-0.6)$ shear-thinning exponent in the viscosity (without excluded volume effects) \cite{Doyle1997, lyulin1999brownian, Petera1999, Aust1999, Hsieh2004, schroeder2005dynamics, Colby2007}, bead-rod chains display some unexpected behaviours at high shear rates.
Most notably, there is an apparent second Newtonian plateau in the viscosity at high shear rates, which appears to be exacerbated by the inclusion of HI \cite{Petera1999, Doyle1997}.
This behaviour is generally not seen in experimental studies, although there are some hints of it in the polystyrene data of Hua et al. \cite{Hua2006}, and of Noda for much longer polymer chains \cite{noda1968rate}.

This viscosity plateau is somewhat correlated with a decrease in polymer extension in the flow direction at high shear rates \cite{Petera1999, lyulin1999brownian, Moghani2017}.
In general, all models show a compression (measured in terms of the components of the gyration tensor) in the gradient and neutral direction, which is also seen in single-molecule imaging of DNA \cite{schroeder2005dynamics}.
However, the compression in flow direction is unexpected, and was explored in detail first by Netz and Sendner \cite{Sendner2009} and then by Larson and coworkers \cite{Dalal2012regimes, Dalal2012tumbling, Dalal2014}.
The conclusion is that HI increases this effect, while an appropriately chosen EV potential largely eliminates it, as does fine-graining beyond the Kuhn-step level in the form of a stiff bending potential between segments \cite{Dalal2012tumbling, Dalal2014}. 
Since these extremely high shear rates are largely out of the reach of experiments, it's unclear the extent to which this effect is a real physical result and not an artefact of the coarse-graining.
These authors did not study the effects of this behaviour on the viscosity or normal stress, however they did examine the end-on-end tumbling behaviour of chains, deriving power-law expressions for the tumbling time as a function of shear rate based on the segmental diffusion and convection \cite{Dalal2012tumbling}.
Some studies seem to suggest a $(-3/4)$ power law slope in tumbling period with shear rate \cite{Dalal2014} (or $-1.1$ on inclusion of HI but with no EV), while others show a $(-2/3)$ slope for experimental and simulation data \cite{Schroeder2005_characteristic}.

We note that it is possible to include a parallel dashpot in a bead-spring model, which models the internal viscosity (IV) of a polymer chain \cite{bird1987dynamics, kailasham2021important, Kailasham2018, Kailasham2021RouseModel, kailasham2022shear, manke1987stress, manke1992stress}.
Such models display a $-1/3$ slope in the viscosity for bead-spring-dashpot dumbbells \cite{Kailasham2018}, a high-shear plateau in viscosity which depends on the IV parameter \cite{Kailasham2021RouseModel}, and instantaneous stress jumps \cite{manke1992stress, Kailasham2018}, all of which are also characteristic features of bead-rod models.
However, including them in a BD simulation is extremely complicated without preaveraging \cite{kailasham2021important, Kailasham2021RouseModel, kailasham2022shear}, and measuring the strength of IV experimentally so as to determine dashpot viscosity is not common or straightforward \cite{kailasham2020wet}.
While this is a promising avenue of investigation, in this paper we will restrict our focus to bead-spring models without IV.

Briefly returning to the entropic bead-spring models, there is again some uncertainty in the power-law exponent, with bead FENE-spring chains showing a $(-2/3)$ or $-0.6$ power-law slope \cite{lyulin1999brownian,Aust1999}, while some results suggest a Cohen-Pade or Marko-Siggia wormlike-chain (MS-WLC) force law could give a $(-1/2)$ power-law slope \cite{Hsieh2004, schroeder2005dynamics}.
However, it is important to consider whether the slope is truly `terminal', as these models tend to have a large crossover region between the low-shear Newtonian plateau and the high-shear behaviour.
For example, Schroeder et al. \cite{schroeder2005dynamics} report a $(-1/2)$ power-law exponent in the viscosity when simulations are compared with available experimental data at the relevant shear rates, but note a $-0.61$ exponent at higher shear rates for the MS-WLC spring force law when carrying out BD simulations.

While the chain connectivity is crucially important to the shear-flow behaviour, there has also been a large body of work incorporating the effects of hydrodynamic interactions (HI) and excluded volume (EV).
It's well known that a bead-Hookean-spring chain with preaveraged HI (Zimm model) does not lead to shear-thinning.
However, when consistently-averaged, treated using a Gaussian or similar approximation, or with full fluctuating effects \cite{prabhakar2006, Ottinger1989, zylka1991, kishbaugh1990discussion}, a chain of Hookean springs shows slight shear-thinning and then shear-thickening.
The intuitive explanation of this behaviour is that the Zimm zero-shear viscosity is lower than the Rouse zero-shear viscosity, but shear flow pulls the beads apart and lessens the effects of HI.
Therefore, the chain \newtxt{shear-}thins slightly due to the `backflow' from HI, then \newtxt{shear-}thickens to reach the Rouse viscosity at high shear rates.
This shear-thickening is also seen for sufficiently extensible non-Hookean springs at sufficiently high bead numbers, before the onset of further shear-thinning due to finite extensibility \cite{Hsieh2004, kishbaugh1990discussion}.
\newtxt{This behaviour has not been uncontroversially established for experimental measurements of dilute polymer solutions, since the shear-thickening seen in the measurements of Layec et al. \cite{layec1976shear} is expected to disappear in the infinitely-dilute limit.
The thickening for semi-dilute solutions is then thought to be related to entanglements rather than HI \cite{kishbaugh1990discussion}, but this question unfortunately seems not to have been re-visited in detail.}

As has been mentioned, HI also causes a compression of bead-rod models at high shear rates, as well as a second Newtonian plateau in the viscosity \cite{Liu1989,Petera1999, Sendner2009, Dalal2014}.
Additionally, as predicted by the Zimm model, zero-shear viscosity is reduced when HI is included, despite no change in the equilibrium structure.
Note that the intermediate-shear-rate viscosity thickening of bead-spring models due to HI has not been observed in bead-rod models, however it is possible that longer chains are required.
For example, there are slight hints of the effect in the bead-rod simulations of Khomami and Moghani \cite{Moghani2017} who used 350 beads, although their findings are not definitive.

The effects of excluded volume are generally characterised via the chain swelling at equilibrium, which is related to the solvent quality parameter $z$ \cite{schafer2012excluded, Prakash2019review}.
Rigorous theoretical developments treating EV using a delta function potential and renormalisation group approaches \cite{schafer2012excluded} find that the EV contribution to the shear-thinning exponent should be $(-1/4)$ for sufficiently long Rouse chains \cite{ottinger1989renormalization}.
In simulations, EV potentials can generally be grouped into soft-core repulsive (such as the  Gaussian potential \cite{Prakash2002, Kumar2003}), hard-core repulsive (such as the Weeks-Chandler-Andersen (WCA) potential \cite{Petera1999}), or hard-core potentials with repulsive and attractive components (such as the Lennard-Jones (LJ) \cite{Dalal2014} or Soddemann-D\"{u}nweg-Kremer (SDK) \cite{santra2019universality, Soddemann2001} potentials).
A theta-solvent is one either without any EV interactions, or with a potential equal parts repulsive and attractive such that there is no swelling at equilibrium - importantly, Dalal et al. \cite{Dalal2014} showed that these are emphatically NOT equivalent away from equilibrium for bead-rod chains.
In fact, several authors have shown that besides causing chain-swelling at equilibrium, a hard-core EV potential also suppresses the high-shear decrease in chain stretch seen in bead-rod chains, as well as the high-shear plateau in viscosity \cite{Dalal2012regimes, Dalal2014, Petera1999, Moghani2017}.
This effect occurs even using a theta-solvent LJ potential, constructed as to cause no swelling at equilibrium \cite{Dalal2014}.
For Rouse chains and FENE chains, BD simulations show the expected $(-1/4)$ decrease in viscosity with shear rate for sufficiently strong EV in the long-chain limit \cite{Kumar2004, Prabhakar2004Separation}.

The inclusion of semiflexibility, generally modelled through a potential energy cost for backbone bending or twisting \cite{Yamakawa2016}, also has somewhat uncertain effects upon the shear-flow behaviour.
Generally, a strong bending potential is associated with a $(-1/3)$ power law slope in the viscosity with shear rate for BD simulations of bead-rod models \cite{lyulin1999brownian, Ryder2006}.
This is expected, as a bead-rod chain with a very strong bending potential is essentially a rigid multibead-rod, which has a $(-1/3)$ power-law shear-thinning exponent.
However, the mean-field model of Winkler \cite{Winkler2006, Winkler2010} suggests a $(-2/3)$ slope irrespective of bending stiffness, which may be due to the backbone extensibility inherent to the model.
Additionally, as has already been mentioned, the use of a bending potential to increase the level of fine-graining in a bead-rod model beyond the Kuhn length, in order to accurately model the true polymer persistence length, reduces the compression of the bead-rod chain at high shear rates \cite{Dalal2012tumbling, Dalal2014}.

To conclude our discussion of prior results, we briefly touch upon a few additional measures of chain behaviour in shear flow which have been explored in the literature.
The first and second normal stress coefficients $\Psi_1$ and $\Psi_2$, are the two experimentally observable material properties besides viscosity.
The first normal stress coefficient $\Psi_1$ is generally found to show a $(-4/3)$ power-law scaling with shear rate irrespective of the model \cite{Stewart:1972gt, Fan:1985jk, Liu1989, Doyle1997, lyulin1999brownian, Petera1999, Hsieh2004, schroeder2005dynamics}, however some long bead-rod-chains with HI and EV seem to display a $-1.1$ power law slope \cite{Liu2004, Moghani2017}.
The second normal stress difference is difficult to get accurate statistics on, both in experiments \cite{keentok1980measurement} and simulations, but is thought to have a positive value for bead-rod chains \cite{Liu2004} and a negative value for bead-spring chains \cite{Hsieh2004}.
One can also measure optical properties, such as the birefringence \cite{Doyle1997}, extinction or orientation angles \cite{Doyle1997, Aust1999, pincus2020viscometric, Kumar2004, schroeder2005dynamics, Winkler2010}, or linear dichroism \cite{Kubista1993, Liu2004}.
Generally, simulations give similar results to experimental single-molecule imaging, which shows an extension in the flow direction and contraction in the gradient direction.
Finally, we have the power spectral density, which essentially allows one to examine frequency components belonging to different time scales of polymer motion \cite{schroeder2005dynamics}. 
This can be matched with experimental data \cite{Schroeder2005_characteristic}, and was analysed extensively by Hur et al. \cite{hur2000brownian}, but will not be calculated directly here.

In light of the wide variety of expected behaviour based on the physics included in a given polymer model, it can be difficult to predict what effect a given component will have on the qualitative shear-flow behaviour.
This ambiguity may be resolved by using a singular model which can span the entire range of previously-identified behaviour, as well as move smoothly between each limit, allowing one to successively add each piece of physics in turn to investigate the effects.
In \sref{Methods}, we will describe such a model based on the so-called FENE-Fraenkel spring, along with a bending potential, EV, and full hydrodynamic interactions.
Additionally, we will give a brief overview of the Brownian dynamics (BD) simulation algorithm, as well as expressions for our measured rheological, conformation and optical properties.
We will then present results in \sref{Results}, first showing that our FENE-Fraenkel-spring chain can reproduce both FENE-spring behaviour, as well as show exact agreement with the bead-rod simulations of Petera and Muthukumar \cite{Petera1999}.
The behaviour of material properties, gyration tensor components and tumbling frequencies are then carefully investigated in the crossover between bead-rod and bead-spring behaviour, and as a function of bending stiffness, EV and HI.
Finally, to conclude in \sref{Conclusions}, we will qualitatively compare our simulations with the previous experimental, theoretical and numerical results in \fref{fig:Experimental schematic} and \fref{fig:Simulations schematic}.
We show that the whole range of behaviour can be qualitatively reproduced by our model, although further work is needed for exact quantitative predictions.
\newtxt{We do also present some very brief semi-quantitative comparisons with experimental data for PBLG \cite{yang1958non}, but do not extend these findings to a wider class of polymers.}

\section{Methods}
\label{Methods}

\begin{figure}[t]
    \centering
    \includegraphics[width=8.5cm,height=!]{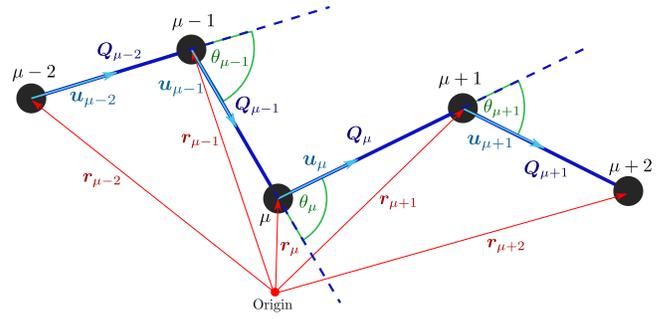}    \caption{Diagram of bead, segment and segment angle labelling scheme. Bead $\mu$ is at position $\bm{r}_\mu$ relative to the center of mass, with some angle $\theta_\mu$. The segment from $\mu$ to $\mu + 1$ has unit vector $\bm{u}_\mu$, with length $Q_\mu$. Note that for N beads, the beads are numbered from $\mu=1,2,3,\ldots,N$, the segments from $\mu=1,2,3,\ldots,N-1$, and the angles from $\mu=2,3,4,\ldots,N-1$.}
    \label{fig:angles schematic}
\end{figure}

Our current model is a bead-spring chain of $N$ beads and $N_\mathrm{s} = N-1$ segments with bead $\mu$ at position $\bm{r}_\mu$ relative to the chain center of mass, bead-bead vectors $\bm{Q}_\mu = \bm{r}_\mu-\bm{r}_{\mu-1}$ and segment angles $\theta_\mu$.
This is displayed schematically in \fref{fig:angles schematic}, which also gives the numbering scheme for beads, segments and angles.
We impose some connector force law which acts along the segments, as well as bending forces between segments, EV forces between every set of spatially nearby beads, and HI perturbations to represent the effects of the implicit solvent, all of which will be described in detail shortly.
The solvent is represented implicitly such that beads have solvent friction $\zeta = 6 \pi \eta_s a$, where $\eta_s$ is the solvent viscosity and $a$ is the effective bead radius.
Flow is imposed through the tensor $\bm{\kappa}$, where the velocity field of the Newtonian solvent is $\bm{v} = \bm{\kappa} \cdot \bm{r}$, assuming the background flow $\bm{v}_0 = 0$.
For the case of shear flow considered here, the only non-zero component of $\bm{\kappa}$ is $\kappa_{x,y} = \dot{\gamma}$, the shear rate. 

\subsection{FENE-Fraenkel force law}

The connector forces act along the segments, and we will predominately use the FENE-Fraenkel form.
This force law was introduced by Hsieh et al. in a 2006 paper with the purpose of reproducing a bead-rod chain while avoiding the complications of BD simulations with constraints \cite{Hsieh2006}.
However, this force law also has the useful property that it can simultaneously represent other commonly-used force laws, such as the FENE, Hookean and Fraenkel springs.
In a previous paper, we have discussed the properties of a FENE-Fraenkel dumbbell in detail, including how to correctly choose a timestep during simulations, how various rheological properties scale with shear rate, and how one can smoothly move between bead-rod and bead-spring behaviour \cite{pincus2020viscometric}. 
Here, we show that it is possible to use a bead-FF-spring-chain to recover the full range of bead-spring-chain and bead-rod-chain behaviour, including all the qualitative features of \fref{fig:Experimental schematic} and \fref{fig:Simulations schematic}.

\begin{figure}[t]
    \centering
    \includegraphics[width=8.5cm,height=!]{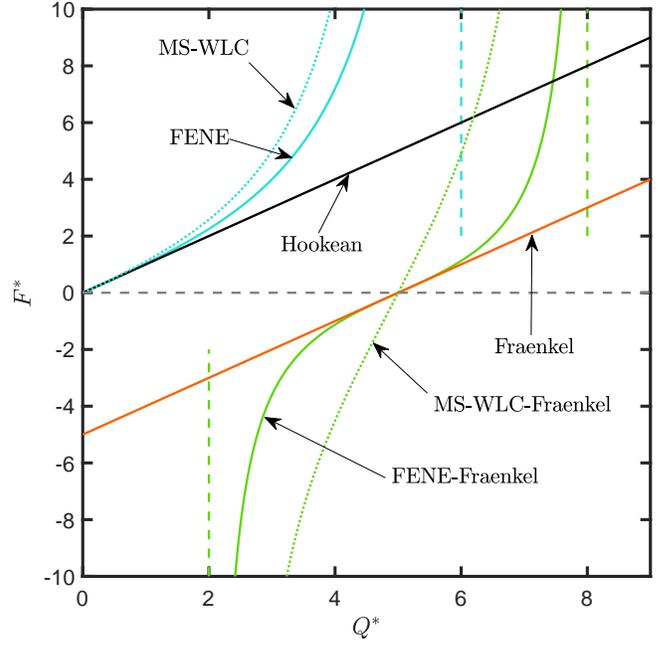}
    \caption{FENE-Fraenkel (full lines, \eref{eq:FF_force_eqn}) and MS-WLC-Fraenkel (dotted lines, \eref{MS-MOD non dim eqn}) force limits. Vertical lines show maximum and minimum extensibilities. `Hookean' spring has $\delta Q^* = 10000$, $\sigma^* = 0$. Fraenkel spring has $\delta Q^* = 10000$, $\sigma^* = 5$. FENE and MS-WLC springs have $\delta Q^* = 6$, $\sigma^* = 0$. FENE-Fraenkel spring has $\delta Q^* = 3$, $\sigma^* = 5$, while MS-WLC-Fraenkel spring has $\delta Q^* = 8$, $\sigma^* = 5$ (which has equivalent minimum and maximum extensibility compared to the shown FENE-Fraenkel spring).}
    \label{fig:FF force limits}
\end{figure}

Written in dimensional form, the FENE-Fraenkel spring force law is given by:
\begin{equation}
    \bm{F}^{(c)} = \frac{H(Q-\sigma)}{1-(Q-\sigma)^2/(\delta Q)^2} \frac{\bm{Q}}{Q}
\label{eq:FF_force_eqn dimensional}
\end{equation}
Here $\bm{F}^{(c)}$ is the force vector between the beads with bead-bead vector $\bm{Q}$ and length $Q$, $\sigma$ is the natural length of the spring ($Q=\sigma$ in the absence of any additional forces), $\delta Q$ is the maximum extensibility around $\sigma$, and $H$ is the effective elastic modulus of the spring (with units of force per length). 
Furthermore, we generally rescale lengths to a non-dimensional form using the spring stiffness, namely $l_H = \sqrt{k_\mathrm{B} T/H}$, where $k_\mathrm{B}$ is Boltzmann's constant and $T$ is the solution temperature.
Since energies are also scaled by $k_\mathrm{B} T$, forces are additionally non-dimensionalised by $\sqrt{k_\mathrm{B} T H}$.
In this form, the spring force law reads:
\begin{equation}
    \bm{F}^{(c)*} = \frac{(Q^*-\sigma^*)}{1-(Q^*-\sigma^*)^2/(\delta Q^*)^2} \frac{\bm{Q}^*}{Q^*}
\label{eq:FF_force_eqn}
\end{equation}
with non-dimensional qualtities denoted by an asterisk (e.g. $Q^* = Q/l_H$).
As can be seen in \fref{fig:FF force limits}, setting $\sigma^* = 0$ recovers the FENE and Hookean (in the limit $\delta Q^* \rightarrow \infty$) force laws, while $\delta Q^* \rightarrow \infty$ for finite $\sigma^*$ gives the Fraenkel force law.
Note that in the $\sigma = 0$ (FENE) case, the parameter $\delta Q$ is equivalent to the more common label $Q_0$, and the non-dimensional $\delta Q^*$ is equivalent to the square root of the FENE $b$-parameter, $\sqrt{b} = Q_0/l_H$.

We also briefly investigate the properties of a new spring force law, which we have called the `MS-WLC-Fraenkel' (Marko-Siggia Wormlike-Chain Fraenkel) spring, which has force-extension behaviour given by:
\begin{multline}
\label{Modified MS in terms of H}
    \bm{F}^{(c)}= H \bm{Q} \frac{2}{3} \frac{\delta Q}{Q} \Bigg\{ \frac{(1-\alpha)^{-2}-1}{4}+\alpha \\ 
    - \frac{\sigma}{\delta Q} \left[\frac{(1+\alpha)^{-2}-1}{4}-\alpha \right] \Bigg\}
\end{multline}
where $\alpha$ is a non-dimensional quantity given by:
\begin{equation}
     \alpha = \frac{Q-\sigma}{\delta Q-\sigma} 
\end{equation}
Therefore, scaling lengths by $l_H$ as for the FENE-Fraenkel spring, this can be written as:
\begin{multline}
\label{MS-MOD non dim eqn}
    \bm{F}^{(c)*}= \bm{Q}^*  \frac{2}{3} \frac{\delta Q^*}{Q^*} \Bigg\{ \frac{(1-\alpha)^{-2}-1}{4}+\alpha \\ 
    - \frac{\sigma^*}{\delta Q^*} \left[\frac{(1+\alpha)^{-2}-1}{4}-\alpha \right] \Bigg\}
\end{multline}
with $\alpha$ defined as before via $\sigma^*$ and $\delta Q^*$.
If we identify that $H \equiv (3 k_\mathrm{B} T)/(2 L l_p)$, where $l_p$ is the polymer persistence length (discussed shortly) and $L \equiv \delta Q$ is the contour length, the $\sigma = 0$ limit of this force law is equivalent to that given by Marko and Siggia, the so-called MS-WLC spring \cite{marko1995statistical}.
This is shown in \fref{fig:FF force limits}, for both the $\sigma^* = 0$, $\delta Q^* = 6$ and $\sigma^* = 5$, $\delta Q^* = 8$ cases.
Note that compared to the FENE-Fraenkel force law, we see a different approach to the maximum-extensibility limit, as well as increased effective stiffness around $\sigma^*$ for $\sigma^* > 0$.
The MS-WLC-Fraenkel force law is considerably harder to deal with analytically, as its distribution function must be found numerically, but is used to show that results depend more on the extensibility, compressibility and average length than the fine-grained details of the force law in question.

\subsection{Bending potential, EV and HI}

One important feature of polymer chains which we wish to model is the semiflexibility, related to the energetic resistance to bending along the backbone.
This semiflexibility is represented by the persistence length, which can be thought of as the exponential decay constant for the autocorrelation of the tangent vector direction along the backbone curve \cite{Yamakawa2016}:
\begin{equation}
    \langle \bm{u}(s) \bm{u}(s') \rangle = e^{\frac{-|s-s'|}{l_p}}
\end{equation}
where $l_p$ is the persistence length, and $\bm{u}(s)$ is the tangent vector to the curve at position $s$, if the backbone is imagined as a continuous space-curve analogue of \fref{fig:angles schematic}.
This is also often expressed in terms of the measure of chain size more common for flexible chains, the number of Kuhn steps $N_k \equiv L/(2 l_p)$, where $L$ is the total polymer contour length.

In a continous chain, the inverse of the persistence length can be identified as a so-called stiffness parameter, essentially a flexural modulus which describes the energetic cost for chain bending.
For discrete models, we can use an equivalent potential which imposes an energetic cost based on the angle $\theta_\mu$.
In our case, the bending potential is given by:
\begin{equation}
\label{eq:bending potential}
    \phi_{\mathrm{b},\mu}/k_\mathrm{B} T = C (1-\cos{\theta_\mu})
\end{equation}
where $C$ is the bending stiffness, while $\theta_\mu$ and $\phi_{\mathrm{b},\mu}$ are the included angle and bending potential between vectors $\bm{Q}_\mu$ and $\bm{Q}_{\mu+1}$ respectively.
An expression for the force on bead $\mu$ due to the bending potential is given in \aref{Appendix Bending Force}, as well as an analytical expression for the angular distribution function.
For this form of the bending potential, Saadat and Khomami give a useful relation for the bending stiffness $C$ as a function of the ratio of contour length $L$ and persistence length $l_p$, represented by the number of Kuhn steps in each segment $N_\mathrm{K,s} = L/(2 N_\mathrm{s} l_p)$ \cite{Saadat2016}:
\begin{equation}
\label{bending potential expression SK}
    C = \frac{1+p_\mathrm{b,1}(2N_\mathrm{K,s}) + p_\mathrm{b,2}(2N_\mathrm{K,s})^2}{2N_\mathrm{K,s}+p_\mathrm{b,3}(2N_\mathrm{K,s})^2 + p_\mathrm{b,4}(2N_\mathrm{K,s})^3}
\end{equation}
where $p_\mathrm{b,i} = -1.237, 0.8105, -1.0243, 0.4595$ for $i=1,2,3,4$ respectively.
This is a Pade approximation chosen to exactly match the nearest neighbour correlation of a continuous wormlike chain at $N_\mathrm{K,s} = \{0,1,2,4,15,\infty\}$ \cite{Saadat2016}, while providing a good approximation for other $N_\mathrm{K,s}$.
Note that this form is technically only exact for Saadat and Khomami's specific force law \cite{Saadat2016}, but we find it gives the correct segment-segment correlation irrespective of the FENE-Fraenkel parameters used.
This allows us to express our chain semiflexibility in terms of the more physically relevant $l_p$, rather than simply as a function of the parameter $C$.

Experimentally, we measure the solvent quality by the equilibrium swelling, for example of the gyration radius \cite{Yamakawa1971}.
This is caused by the effective strength of polymer-solvent interactions, such that increased polymer-solvent attraction increases the equilibrium coil size.
This is a function of the so-called solvent quality $z$, which describes the universal swelling of a wide variety of experimental systems, based on renormalisation group calculations \cite{schafer2012excluded}.
In theory and simulations, the EV force is modelled through some effective bead-bead interaction strength, given by the EV potential.

In our simulations, the excluded volume force between beads is given by one of two potentials, the first of which is a truncated, purely repulsive LJ potential, what we will call the `hard-core' form as it does not allow bead overlap.
Specifically, we use the SDK potential \cite{Soddemann2001} with $\varepsilon = 0$, which has the exact form:
\begin{equation}
\label{SDK equation}
U_\mathrm{SDK}=\left\{\begin{array}{ll}
4\left[\left(\frac{d}{Q}\right)^{12}-\left(\frac{d}{Q}\right)^{6}+\frac{1}{4}\right]-\varepsilon, & Q \leq 2^{1 / 6} d \\
\frac{1}{2} \varepsilon\left[\cos \left(\alpha Q^{2}+\beta\right)-1\right], & 2^{1 / 6} d \leq Q \leq 1.82 d \\
0, & Q \geq 1.82 d
\end{array}\right.
\end{equation}
where $d$ is the range of the potential (similar to the well-known $\sigma$ of the LJ potential), $\varepsilon$ is the attractive well depth and $\alpha$ and $\beta$ are chosen such that the potential smoothly goes to zero at the cuttoff radius $1.82 d$ \cite{Soddemann2001}. 
Note that while we use $\varepsilon = 0$, which can only model a good solvent (similar to the WCA potential \cite{andersen1971relationship}), one can choose some $\varepsilon$ such that the attractive and repulsive forces balance, leading to a `hard-core' $\theta$-solvent with no net solvent-polymer interaction \cite{santra2019universality}, or even a poor solvent.
The second EV potential is the `soft-core' Gaussian potential, of the form:
\begin{equation}
\label{Gaussian potential}
    U_\mathrm{Gauss} = \frac{\nu_\mathrm{ev} k_\mathrm{B} T}{(2 \pi d_\mathrm{ev}^2)^{3/2}} \exp\left\{-\frac{1}{2} \frac{Q^2}{d_\mathrm{ev}^2}\right\}
\end{equation}
where $\nu_\mathrm{ev}$ is the strength of the excluded volume potential (with units of volume) and $d_\mathrm{ev}$ is the range of the potential \cite{Ottinger1996, prakash1999viscometric}.
In the limit of $d_\mathrm{ev} \rightarrow 0$, the Guassian potential approaches the delta-function potential.
This `soft' form of the excluded volume allows for bead overlap, but has the useful feature that the solvent quality, $z$, can be represented exactly in terms of the chain expansion caused by a particular choice of $\nu_\mathrm{ev}$ \cite{prakash1999viscometric}. 
This potential will generally be used in non-dimensional form, with:
\begin{equation}
    z^* = \nu_\mathrm{ev} \left(\frac{ k_\mathrm{B} T}{2 \pi H} \right)^{3/2}
\end{equation}
which allows the solvent quality $z$ to be expressed approximately as:
\begin{equation}
\label{z in terms of zstar}
    z^* = z \chi^3 / \sqrt{N}
\end{equation}
where the parameter $\chi$ is a scaled dimensionless spring length, which will be described shortly in \eref{eq: chi equation spring length}.
As $N \rightarrow \infty$ with $z^*$ corrected for $\chi$ as above, \eref{z in terms of zstar} is no longer an approximation but instead gives the exact universal swelling, which is a known function of $z$ from analytical renormalisation group theories \cite{Kumar2004}.

Hydrodynamic interactions are included via the RPY tensor, a regularisation of the Oseen-Burgers tensor, describing how the force on one bead influences the motion of the others:
\begin{equation}
    \bm{\Omega}(\bm{r}) = \frac{3 a}{4\zeta r} \left(A \bm{\delta} + B \frac{\bm{r} \bm{r}}{r^2} \right)
\label{eq:HI tensor}
\end{equation}
where the values of $A$ and $B$ depend on the bead separation:
\begin{subequations}
\begin{equation}
    A = 1 + \frac{2}{3} \left(\frac{a}{r}\right)^2, B = 1 - 2 \left(\frac{a}{r}\right)^2 \text{ for } r\ge 2a
\end{equation}
\begin{equation}
    A = \frac{4}{3}\left(\frac{r}{a}\right) - \frac{3}{8} \left(\frac{r}{a}\right)^2, B = \frac{1}{8} \left(\frac{r}{a}\right)^2 \text{ for } r < 2a
\end{equation}
\label{eq: AandB_RPY}%
\end{subequations}
where $a$ is the effective hydrodynamic bead radius, as in the definition of the bead friction $\zeta$. 
Note that we usually represent the strength of HI in terms of the parameter $h^*$, essentially a reduced bead radius.
This is given by:
\begin{equation}
    h^* = \sqrt{\frac{k_\mathrm{B} T}{H}} a \sqrt{\pi}
\end{equation}
the form of which comes from its use in the Zimm model with preaveraged HI \cite{bird1987dynamics}.

In general, calculations are performed and results are presented in the Hookean system of non-dimensionalisation, where we have length and force scales as described above, and time scale \newtxt{$\lambda_H$}.
\newtxt{Our full system of non-dimensionalisation is then:
\begin{equation}
    l_\text{H} \equiv \sqrt{\frac{k_\text{B} T}{H}}, \lambda_\text{H} \equiv \frac{\zeta}{4H}, F_\text{H} \equiv \sqrt{k_\text{B} T H}
\label{Hookean_system}
\end{equation}
and we denote non-dimensional properties with an asterisk, as in the following commonly-used examples:
\begin{equation}
    \dot{\gamma}^* = \dot{\gamma} \lambda_H
\end{equation}
\begin{equation}
    R_g^* = \frac{R_g}{l_H}
\end{equation}
\begin{equation}
    \eta_p^* \equiv (\eta - \eta_s)/n_p k_\mathrm{B} T \lambda_H
\end{equation}
where $n_p$ is the number density of polymers in solution, $\eta_p$ is the polymer contribution to viscosity defined below in \eref{eq: material functions eta, psi}, and $\eta_s$ is the solvent viscosity.}

Parameters are often expressed in terms of the quantity $\chi$, which is the ratio of the average length of a non-Hookean spring to that of a Hookean spring.
This quantity is useful as a natural way to express how parameters such as HI strength $h^*$ or EV radius $d$ should change as the spring force law is altered \cite{Sunthar2005parameterfree}.
It is best calculated in Hookean units, and is defined by:
\begin{equation}
\label{eq: chi equation spring length}
    \chi^2 = \frac{1}{3} \frac{\int Q^{* 4} e^{\phi^*}}{\int Q^{* 2} e^{\phi^*}}
\end{equation}
where $Q^*$ and $\phi^*$ are the non-dimensional spring length and spring potential respectively.
Although it is in principle possible to derive this quantity analytically for the FENE-Fraenkel spring, it has different forms depending on the values of $\sigma$ and $\delta Q$ (which cause the lower limit of integration to be either $\sigma-\delta Q$ or $0$). 
In practice, it is straightforward to calculate numerically by quadrature.
For the MS-WLC-Fraenkel spring force law, there is no analytical expression for $\chi$, and so the integrations must be performed numerically, with careful attention paid to avoid reaching floating-point infinities due to the exponentiation. 

\subsection{Brownian dynamics simulation methodology}

By including all of these physical effects in our equation of motion for the chain, we can derive the following Fokker-Planck equation for the evolution of the distribution function $\psi\left(\bm{r}_{1}, \ldots, \bm{r}_{N}\right)$ \cite{Ottinger1996, bird1987dynamics, Prabhakar2004Separation}:
\begin{widetext}
\begin{equation}
\label{eq: Fokker-Planck equation}
    \frac{\partial \psi^{*}}{\partial t^{*}}=-\sum_{\nu=1}^{N} \frac{\partial}{\partial \bm{r}_{\nu}^{*}} \cdot\left\{\bm{\kappa}^{*} \cdot \bm{r}_{\nu}^{*}+\frac{1}{4} \sum_{\mu} \bm{D}_{\nu \mu} \cdot \bm{F}_{\mu}^{\phi *}\right\} \psi^{*}+\frac{1}{4} \sum_{\nu, \mu=1}^{N} \frac{\partial}{\partial \bm{r}_{\nu}^{*}} \cdot \bm{D}_{\nu \mu} \cdot \frac{\partial \psi^{*}}{\partial \bm{r}_{\mu}^{*}}
\end{equation}
\end{widetext}
where $\bm{F}_{\mu}^{\phi *}$ is the total force on bead $\mu$ due to the sum of the spring, bending and EV forces, and the tensor $\bm{D}_{\nu \mu} = \delta_{\nu \mu} \bm{\delta} + \zeta \bm{\Omega}_{\nu \mu}$ takes into account hydrodynamic interactions between beads $\mu$ and $\nu$.

The numerical integration of \eref{eq: Fokker-Planck equation} is undertaken on the basis of the equivalent It\^{o} stochastic differential equation for the chain configuration \cite{Ottinger1996}, which we give in the same form as Prabhakar and Prakash \cite{Prabhakar2004Separation}:
\begin{equation}
\label{Ito SDE}
    \mathrm{d} \bm{R}=\left[\bm{K} \cdot \bm{R}+\frac{1}{4} \bm{D} \cdot \bm{F}^{\phi}\right] \mathrm{d} t^{*}+\frac{1}{\sqrt{2}} \bm{B} \cdot \mathrm{d} \bm{W}
\end{equation}
where $\bm{R}$ is a $3\times N$ matrix containing bead co-ordinates, $\bm{K}$ is a $3N \times 3N$ block matrix with the diagonal blocks containing $\bm{\kappa}^*$ and others equal to 0, $\bm{F}^{\phi}$ is a $3 \times N$ matrix containing total force vectors on each bead (due to spring, bending, and EV potentials), $\bm{D}$ is a $3N \times 3N$ block matrix where the $\nu \mu$ block contains the $\bm{D}_{\nu \mu}$ tensor components, $\bm{W}$ is a $3\times N$ dimensional Wiener process and $\bm{B}$ is a matrix such that $\bm{D} = \bm{B} \cdot \bm{B}^{\mathrm{T}}$.
The matrix $\bm{B}$ is not calculated directly, but instead the product $\bm{B} \cdot \mathrm{d} \bm{W}$ is evaluated using a Chebyshev approximation, as originally proposed by Fixman \cite{fixman1986construction, Prabhakar2004Separation}.
Additionally, the stochastic differential equation is integrated using a semi-implicit predictor-corrector method with a lookup table for the spring force law, the algorithm for which has been detailed extensively elsewhere \cite{Prabhakar2004Separation, Hsieh2006, somasi2002brownian, Ottinger1996, Hsieh2003}.

Simulations are generally run with $\mathcal{O}(10^3)$ trajectories for $50$ relaxation times or $5000$ strain units, whichever is shorter.
This ensures plenty of sampling at steady state for all runs besides the most extensible FENE and Hookean springs, which were run for $50$ relaxation times at all shear rates.

Several conformational, rheological and optical properties are extracted from our BD simulations.
The overall contribution of the polymers to the stress tensor is given by the Kramers expression \cite{bird1987dynamics}:
\begin{equation}
\label{eq: stress tensor expression}
    \bm{\tau}_p =  -n_p \sum_{\nu=1}^N \langle \bm{r}_\nu \bm{F}_{\nu}^{\phi} \rangle + n_p k_\mathrm{B} T \bm{\delta}
\end{equation}
where again $\bm{F}_{\nu}^{\phi}$ is the sum of the spring, EV and bending forces on each bead, and $n_p$ is the number density of polymers.
From this the following material functions can be extracted:
\begin{subequations}
\label{eq: material functions eta, psi}
\begin{equation}
    -\eta_p = \frac{\tau_{p,xy}}{\dot{\gamma}}
\end{equation}
\begin{equation}
    -\Psi_1 = \frac{\tau_{p,xx} - \tau_{p,yy}}{\dot{\gamma}^2}
\end{equation}
\begin{equation}
    -\Psi_2 = \frac{\tau_{p,yy} - \tau_{p,zz}}{\dot{\gamma}^2}
\end{equation}
\end{subequations}
namely the polymer contribution to the viscosity $\eta_p$, and the first and second normal stress coefficients $\Psi_{1}$ and $\Psi_2$ respectively.
Additionally, we measure the polymer extension in each direction using the gyration tensor, defined as:
\begin{equation}
    \bm{G} = \frac{1}{N} \left \langle \sum^N_{\nu=1} \bm{r}_\nu \bm{r}_\nu \right \rangle
\end{equation}
with the radius of gyration given by the trace of this tensor, and with the components:
\begin{equation}
\label{eq: gyration components}
    R_{g,\alpha} = G_{\alpha \alpha}
\end{equation}
where $\alpha = \{x, y, z\}$. 

Two orientation angles (often referred to as extinction angles) can be defined based on $\bm{G}$ and $\bm{\tau}_p$.
These are unfortunately also represented by $\chi$ in the literature, not to be confused with the equilibrium spring length from \eref{eq: chi equation spring length}.
They are denoted $\chi_G$ and $\chi_\tau$, and are essentially the orientation of the gyration tensor and stress tensor respectively, with the forms:
\begin{equation}
\label{eq: chi G definition}
    \chi_G = \frac{1}{2}\arctan{\frac{2 \langle G_{xy}\rangle}{\langle G_{xx} - G_{yy}\rangle}}
\end{equation}
\begin{equation}
\label{eq: chi tau definition}
    \chi_\tau = \frac{1}{2}\arctan{\frac{2 \tau_{xy}}{\tau_{xx} - \tau_{yy}}} = \frac{1}{2}\arctan{\frac{2 \eta_p}{\Psi_1 \dot{\gamma}}}
\end{equation}

Finally, we have the tumbling period (denoted $\tau_\mathrm{tumble}$), which we measure using two methods.
The first method is to count the total revolutions of the end-to-end vector of the polymer as a function of time, and so derive a kind of angular velocity.
This method was employed by Dalal et al. in a BD simulation study \cite{Dalal2012tumbling}, and also by Huber et al. in an experimental study directly imaging actin molecules \cite{Huber2014}.
The second method is to define a tumbling period related to the cross-correlation between conformational changes in flow and gradient directions \cite{Chen2013, teixeira2005shear, Huang2011}.
These quantities are reasonably straightforward to calculate in our simulations, but require some detailed explanation, and so we refer the reader to \aref{Appendix: Tumbling} for an in-depth description of our two methods.

Results are often plotted in terms of Wiessenberg number $Wi = \dot{\gamma^*} \eta_{p,0}^*$, also referred to as reduced shear rate $\beta$ in the literature \cite{bird1987dynamics}.
\newtxt{Note that both $\eta_{p}^*$ and $\dot{\gamma}^*$ have been defined in \eref{Hookean_system}.
In general, one can think of quantities in the `starred' non-dimensionalisation (such as $\dot{\gamma}^*$) as being scaled by the \textit{local bead/spring} relaxation time, while quantities presented in terms of reduced quantities $Wi$ or $\eta^*_p/\eta^*_{p,0}$ are scaled by the \textit{chain} relaxation time.
In other words, as $N$ is increased for a given bead-spring chain, $\dot{\gamma}^*$ remains the same, but $Wi$ increases for the same `true' shear rate.}
The zero-shear viscosity scales similarly to the end-to-end relaxation time up to a constant factor, and so if we instead scale shear rate by relaxation time rather than viscosity, qualitatively similar results are obtained \cite{bird1987dynamics}.
The zero-shear viscosity is determined predominately from the Newtonian plateau at low shear rates for models with HI, although initial estimates are derived using Green-Kubo relations over the stress autocorrelation at equilibrium.
We have compared several methods for determining zero-shear viscosity and relaxation times in \aref{Appendix TTCF}, such as autocorrelation, step strain procedures, and stretch-relaxation.
We also note that it is possible to calculate low-shear material properties using so-called transient time correlation functions (TTCF) \cite{todd_daivis_2017, Borzsak2002}, which to our knowledge have not been used in the context of BD previously, but which are used extensively in non-equilibrium molecular dynamics approaches
Our use of TTCF is also described in \aref{Appendix TTCF}.

Note that for models without HI, the zero-shear viscosity can be expressed directly in terms of the radius of gyration \cite{prakash2001}:
\begin{equation}
\label{eq: eta_0_analytical_equation}
    \eta_{p,0} = \frac{n_p \zeta}{6} N \langle R_g^2 \rangle_\mathrm{eq}
\end{equation}
and futher, for chains without EV, HI or a bending potential, the radius of gyration can be given analytically in terms of the number of beads and equilibrium spring length \cite{Sunthar2005parameterfree}:
\begin{equation}
    \langle R_g^2 \rangle_\mathrm{eq} = \frac{N^2-1}{6N} \langle Q^2 \rangle_\mathrm{eq}
\end{equation}
These expressions are also used to validate the BD predictions.

Finally, we employ variance reduction (VR) techniques at low shear rates to obtain more precise predictions \cite{Ottinger1996, wagner1997accurate, Kumar2003}.

\newtxt{\subsection{Describing a real polymer chain}}
\label{section: real polymer methods}

\begin{figure*}[!ht]
  \centerline{
  \includegraphics[width=17cm,height=!]{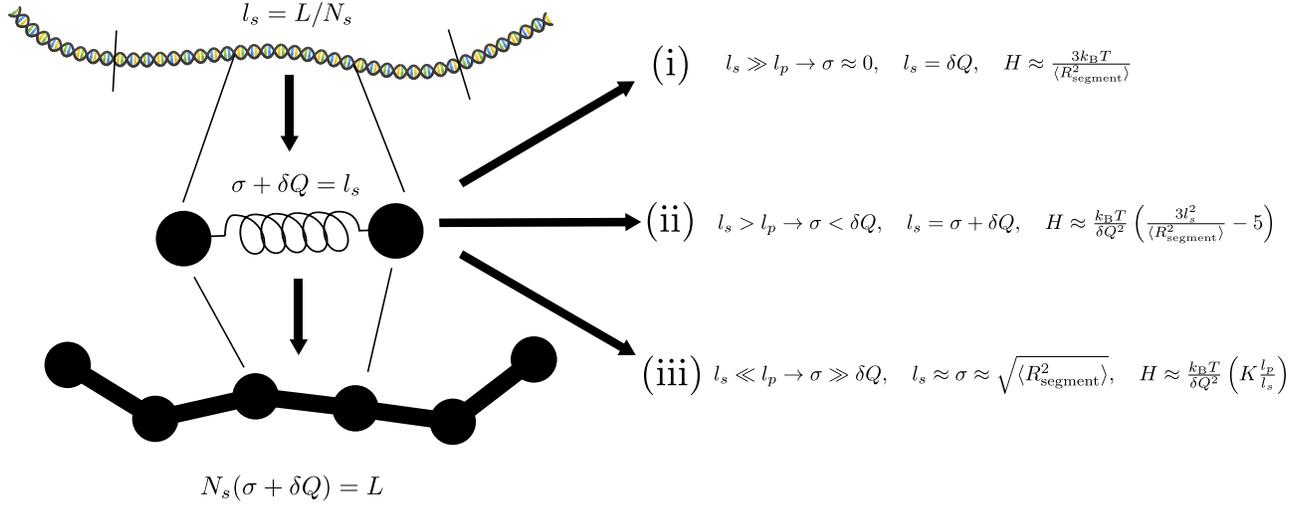}
  }
  \caption{\newtxt{Splitting chain into $N_s$ FENE-Fraenkel springs, with spring parameters chosen according to \eref{second H equation}. Top schematic represents a short strand of DNA, while the bottom schematic is a model with equivalent contour length and approximately equal end-to-end distribution function.}}
  \label{fig: methodology for real chain}
\end{figure*}

\newtxt{Finally, we wish to describe how we have generated the equilibrium configurations seen in \fref{fig: example eq configs FF} using our FENE-Fraenkel spring alongside a bending potential.
Although most of our results will be qualitative, we will use this model to generate semi-quantitative comparisons with PBLG data \cite{yang1958non} as a simple demonstration of its validity in the short-molecule regime.
We assume the semiflexible polymer chain (in this case, we assume DNA, but the method generalises to any polymer) can be characterised by two experimentally measured parameters, the total contour length $L$ and the persistence length $l_p$.
We give all lengths in units of DNA base pairs, for example describing a particular chain as having $L = 25$kbp and $l_p = 147$bp.
This choice for $l_p$ corresponds to the generally-accepted value of $l_p = 50$nm in excess salt, alongside a base pair length of $\approx 0.34$nm \cite{Pan2014zeroshear, Kubista1993, Rodger2009}.}

\newtxt{Assuming that the Kratky-Porod wormlike chain (KP WLC) is an accurate representation of the underlying chain with contour length $L$ and persistence length $l_p$, we seek to discretise the chain using $N_s$ FENE-Fraenkel springs and a bending potential such that we recover the correct end-to-end vector magnitude distribution function (given via a fit to the even moments of the KP chain derived by Hamprecht and Kleinert \cite{hamprecht2005end}).
This distribution function will be represented by $\psi(R)$, where $R$ is the end-to-end distance from the first to the last monomer in the entire real, physical chain, sometimes also characterised by the radius of gyration $R_g$.
We perform the discretisation (into $N_s$ segments) in two stages.
Firstly, for each of the $N_s$ segments with length $l_s = L/N_s$ (which is the same as choosing $N_\mathrm{K,s} = L/(2 N_\mathrm{s} l_p)$ from earlier), we set our FENE-Fraenkel spring parameters such that the total spring extensibility is equal to $l_s$, and also that the average spring length at equilibrium is equal to that for the underlying WLC segment.
The total spring extensibility condition is easy to satisfy by setting $\sigma + \delta Q = l_s$.
This procedure is illustrated in \fref{fig: methodology for real chain}.}

\begin{figure*}[t]
    \centerline{
    \begin{tabular}{c c}
        \includegraphics[width=8.5cm,height=!]{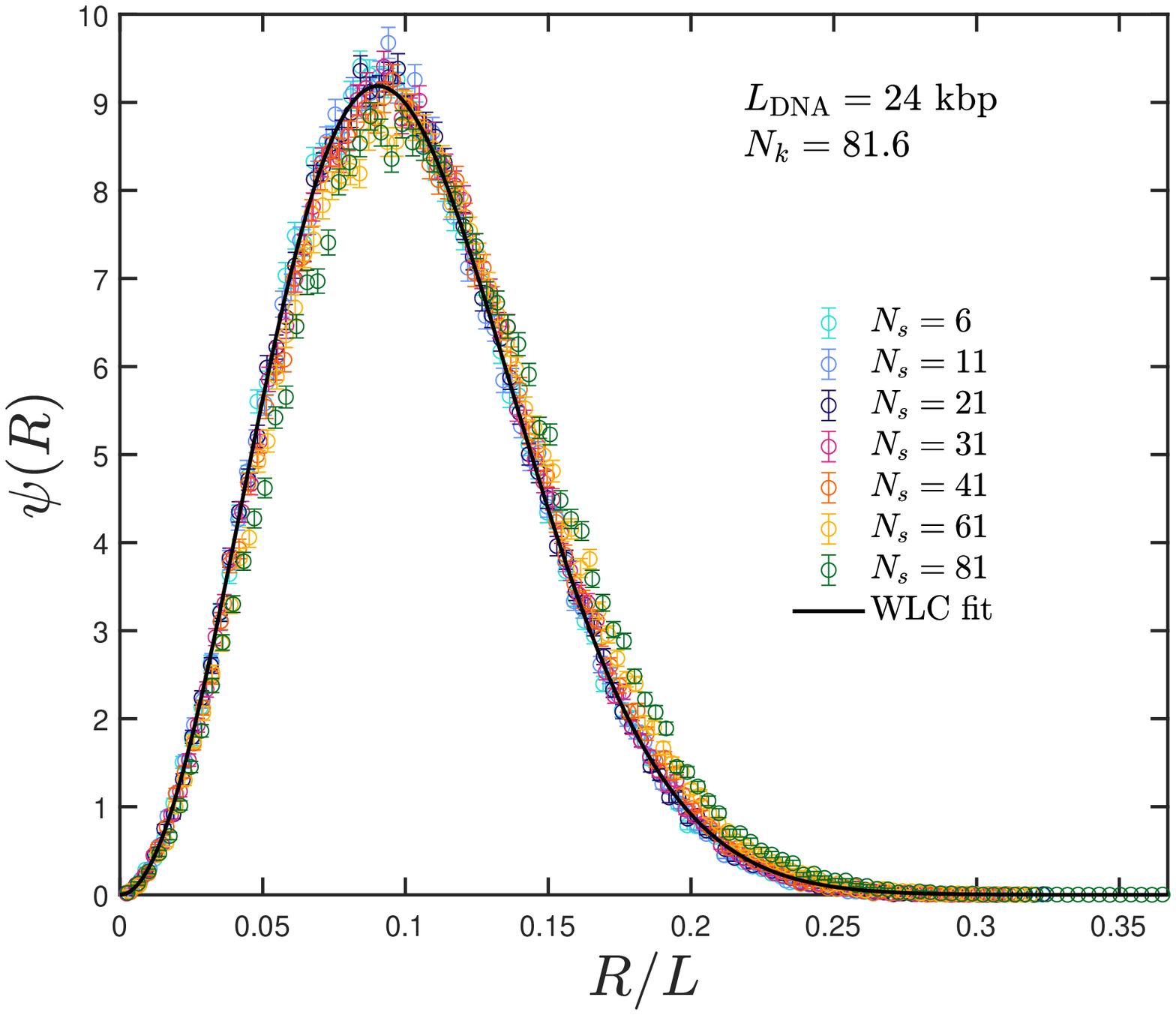}
        & \includegraphics[width=8.5cm,height=!]{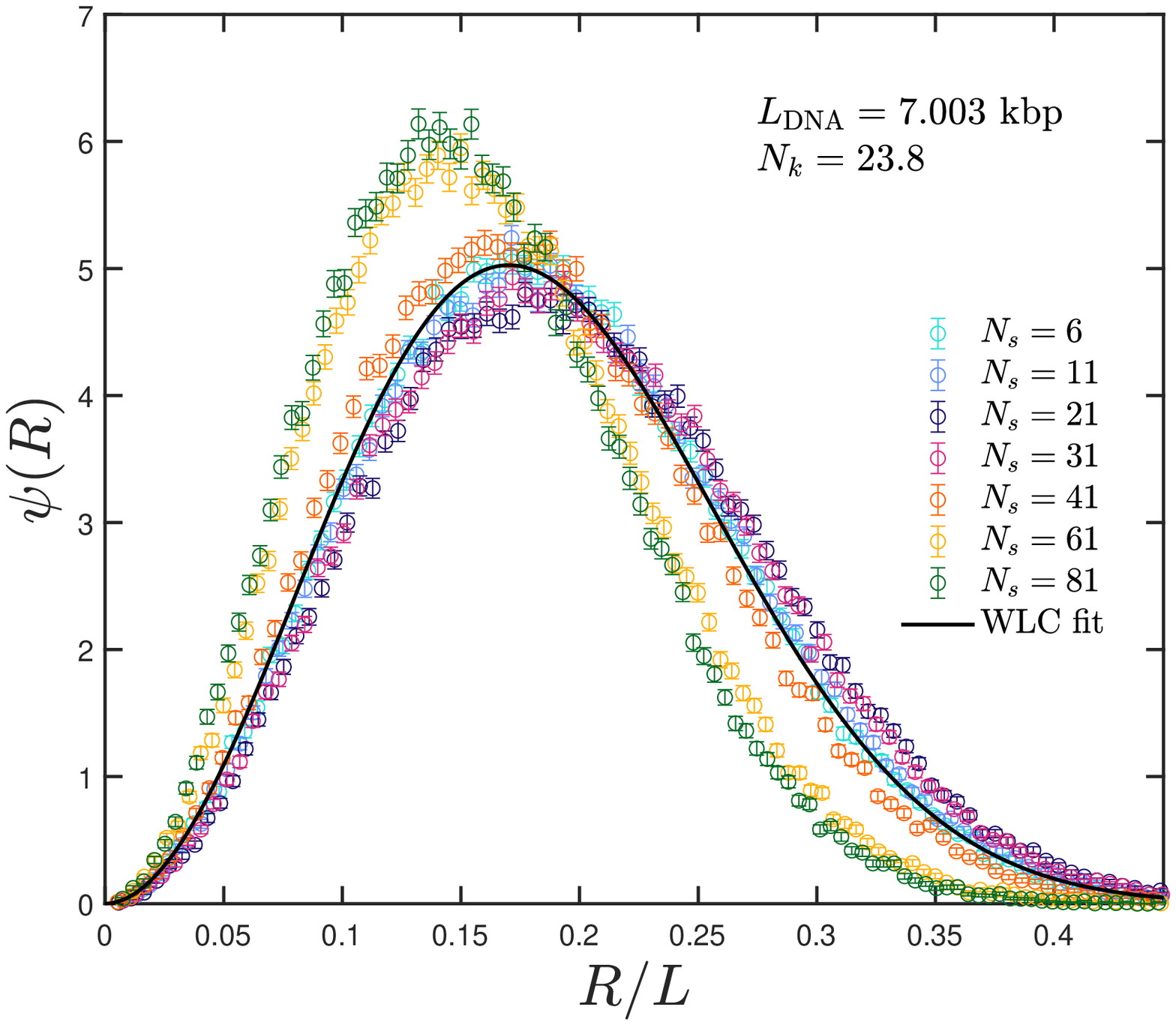} \\
        (a) & (b) \\
    \end{tabular}
    }
    \caption[End-to-end distribution function fits using our modelling scheme for FENE-Fraenkel springs and Saadat and Khomami's form of the bending potential]{End-to-end distribution function fits using our modelling scheme for FENE-Fraenkel springs and Saadat and Khomami's form of the bending potential. Analytical WLC distributions are given by the method of Hamprecht and Kleinert \cite{hamprecht2005end}. Here we have assumed $l_p = 147$ base pairs.}
    \label{fig: distribution function fits}
\end{figure*}

\newtxt{However, our system is still over-determined, since we have two chain properties, the segment length $l_s$ and the average segment end-to-end length $\langle R^2_\mathrm{segment} \rangle$, given as:
\begin{equation}
\label{end to end distance WLC}
    \langle R_\mathrm{segment}^2 \rangle = 2 l_s l_p - 2 l_p^2 \left(1 - e^{-l_s/l_p} \right)
\end{equation}
but three spring parameters, $\sigma$, $\delta Q$ and $H$.
Although there are likely many physically reasonable ways to resolve this this over-determination, we have chosen a somewhat ad-hoc method based upon the respective coil-like and rod-like limits.
First, when $l_s \gg l_p$, or equivalently $N_\mathrm{K,s} \gg 1$, it is well known that each spring should be approximately Hookean, as can be shown from simple physical arguments \cite{bird1987dynamics}.
In this limit, we can set $\sigma  = 0$, assume $\delta Q$ is large, and then find that the spring stiffness which gives an average $\langle Q^2 \rangle = \langle R_\mathrm{segment}^2 \rangle$ is:
\begin{equation}
    H  = \frac{3 k_\mathrm{B} T}{\langle R_\mathrm{segment}^2 \rangle}
\end{equation}
as shown as case (i) in \fref{fig: methodology for real chain}.
Secondly, as $l_s/l_p$ decreases (due to higher model $N_s$, a shorter underlying chain, or a stiffer underlying chain), the finite extensibility of the chain begins to influence its equilibrium distribution.
In this case, $\sigma$ is still considerably smaller than $\delta Q$ (so that $\delta Q \approx l_s$), but the FENE force law means that the spring stiffness must be changed to obtain the correct $\langle Q^2 \rangle = \langle R_\mathrm{segment}^2 \rangle$.
It has been shown by Sunthar and Prakash that in this limit, we have \cite{Sunthar2005parameterfree}:
\begin{equation}
\label{first H equation}
    H = \frac{k_\mathrm{B} T}{\delta Q^2} \left( \frac{3 l_s^2}{\langle R_\mathrm{segment}^2 \rangle} - 5\right)
\end{equation}
as is also displayed in case (ii) in \fref{fig: methodology for real chain}.
This equation works well when $l_s/l_p \gtrapprox 10$, but runs into serious issues as $\langle R_\mathrm{segment}^2 \rangle \rightarrow l_s^2$, with the spring constant eventually turning negative, which is clearly nonphysical.
This is because a FENE spring, having no `natural' length $\sigma$, cannot possibly represent a rigid rodlike molecule.
To correct this, we must increase $\sigma$ beyond $0$, and hence add an additional contribution to the spring constant $H$ which accounts directly for the bending rigidity of the underlying chain.
In fact, due to the finite extensibility about $\sigma$ of our FENE-Fraenkel spring due to finite $\delta Q$, the exact physical form of this extra contribution is less important than that it must increase with $l_p/l_s$, and be sufficiently large to ensure the spring potential does not turn negative.
We have chosen the form:
\begin{equation}
\label{second H equation}
    H = \frac{k_\mathrm{B} T}{\delta Q^2} \left( \frac{3 l_s^2}{\langle R_\mathrm{segment}^2 \rangle} - 5 + K \frac{l_p}{l_s}\right)
\end{equation}
for some arbitrary constant $K$, where $K = 5$ has been used in this paper.
This is demonstrated for case (iii) with $l_p \gg l_s$ in \fref{fig: methodology for real chain}, and importantly allows $\sigma$ to be set to approximately equal $l_s$ without disobeying the $\langle Q^2 \rangle = \langle R_\mathrm{segment}^2 \rangle$ constraint.}

\newtxt{Although this is a fairly ad hoc methodology, it correctly gives a Gaussian spring in case (i) for a very long chain with $l_s \gg l_p$, a FENE spring in case (ii), and a very stiff `rodlike' spring in case (iii) with $l_p \gg l_s$.
It also guarantees that $\langle Q^2 \rangle = \langle R_\mathrm{segment}^2 \rangle$, and that the chain never extends beyond $l_s$ since $\sigma + \delta Q = l_s$.
It is certainly possible to generate a model which more faithfully reproduces the energy of an elastic rod in conjunction with a bending potential.
However, one should not be too concerned over getting the spring potential exactly correct in this limit, since even the commonly used Kramers bead-rod chain has little fundamental physical meaning \cite{VanKampen1984}.
The FENE spring is also, of course, an approximation to the true force-extension behaviour, but one which seems to make little difference to rheological predictions if equlibrium chain properties are correctly reproduced \cite{Sunthar2005parameterfree,Pham2008}.}

\begin{figure*}[ht]
    \centerline{
    \begin{tabular}{c c}
        \includegraphics[width=8.5cm,height=!]{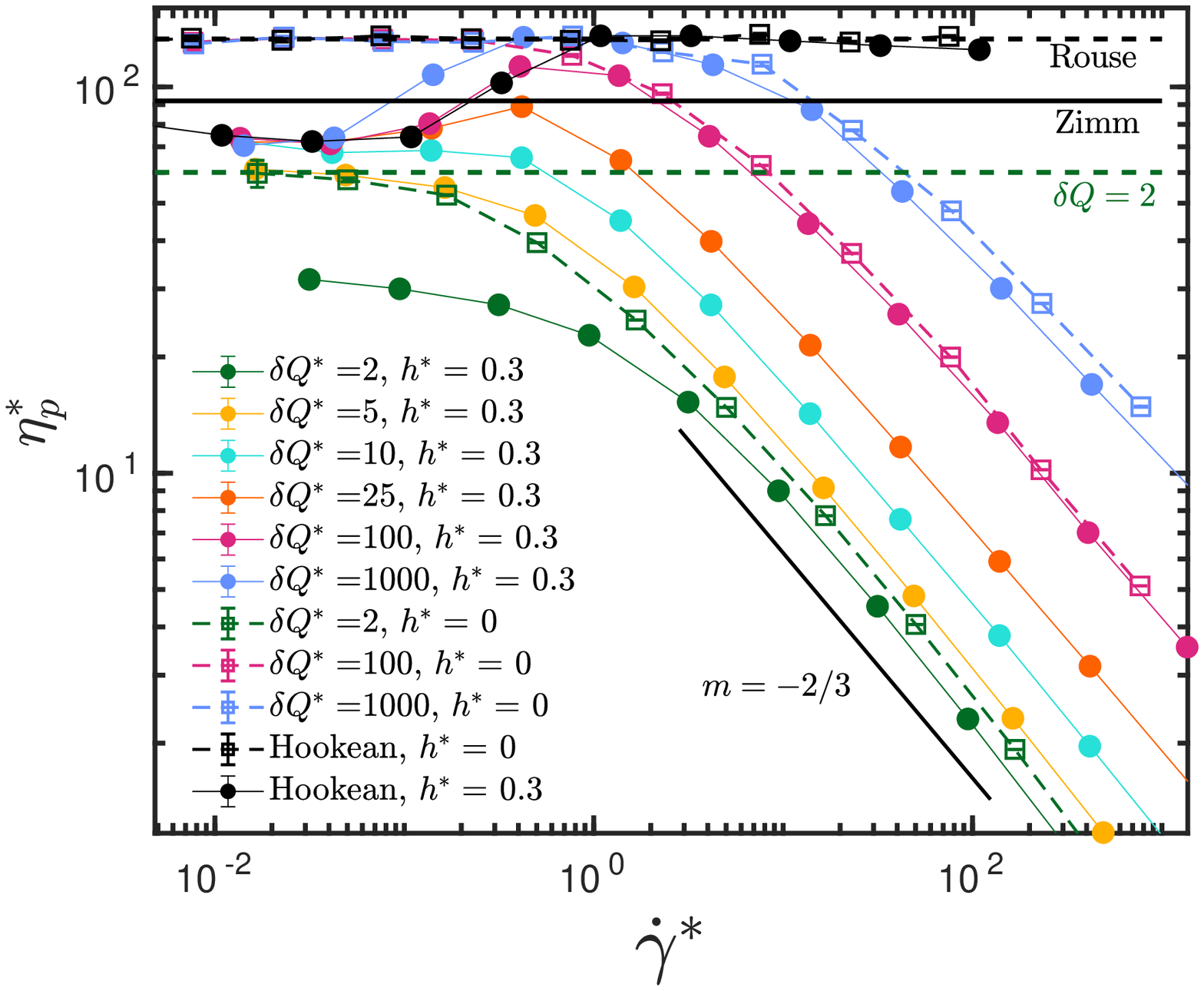}
        & \includegraphics[width=8.5cm,height=!]{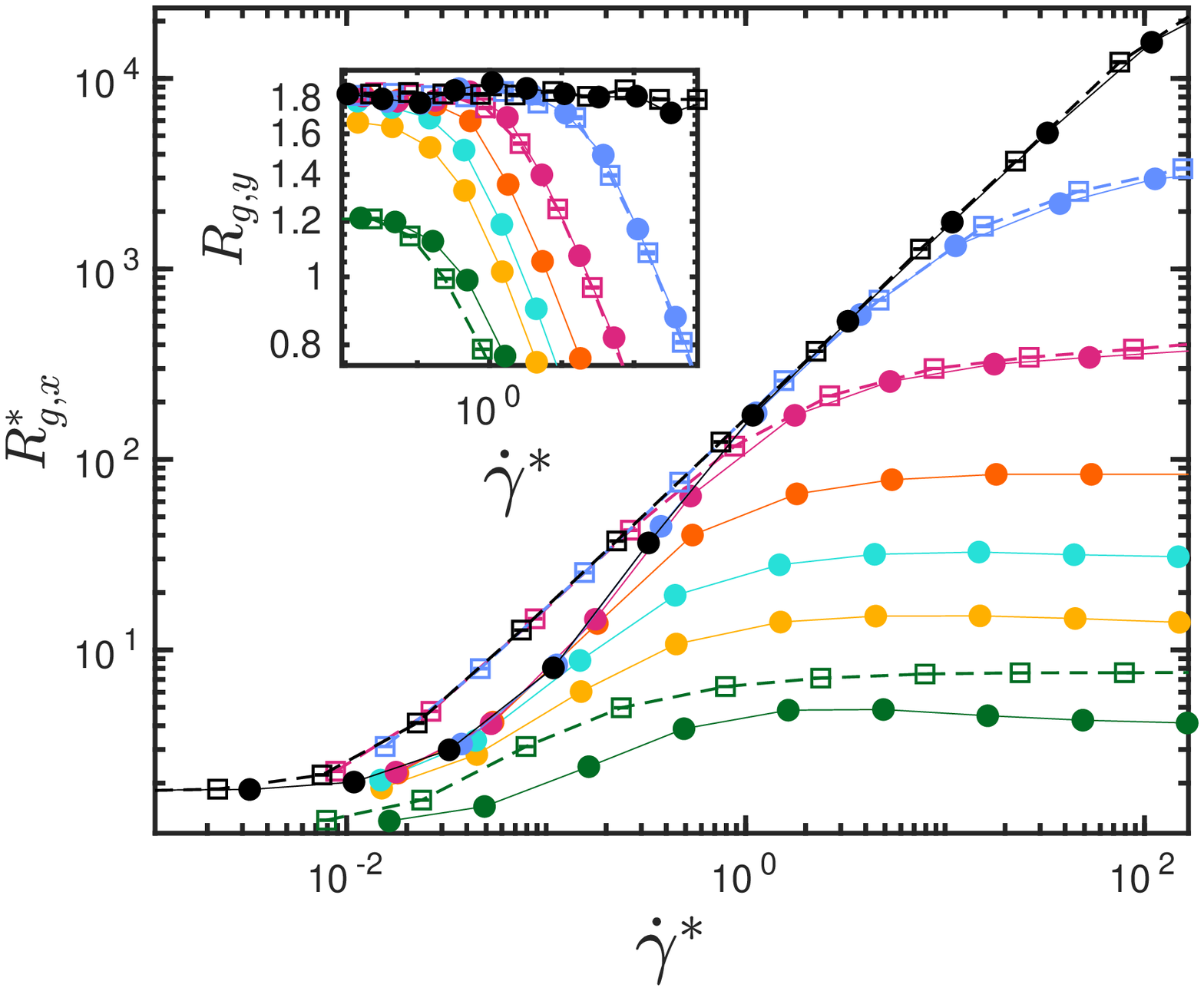} \\
        (a) & (b) \\
        \includegraphics[width=8.5cm,height=!]{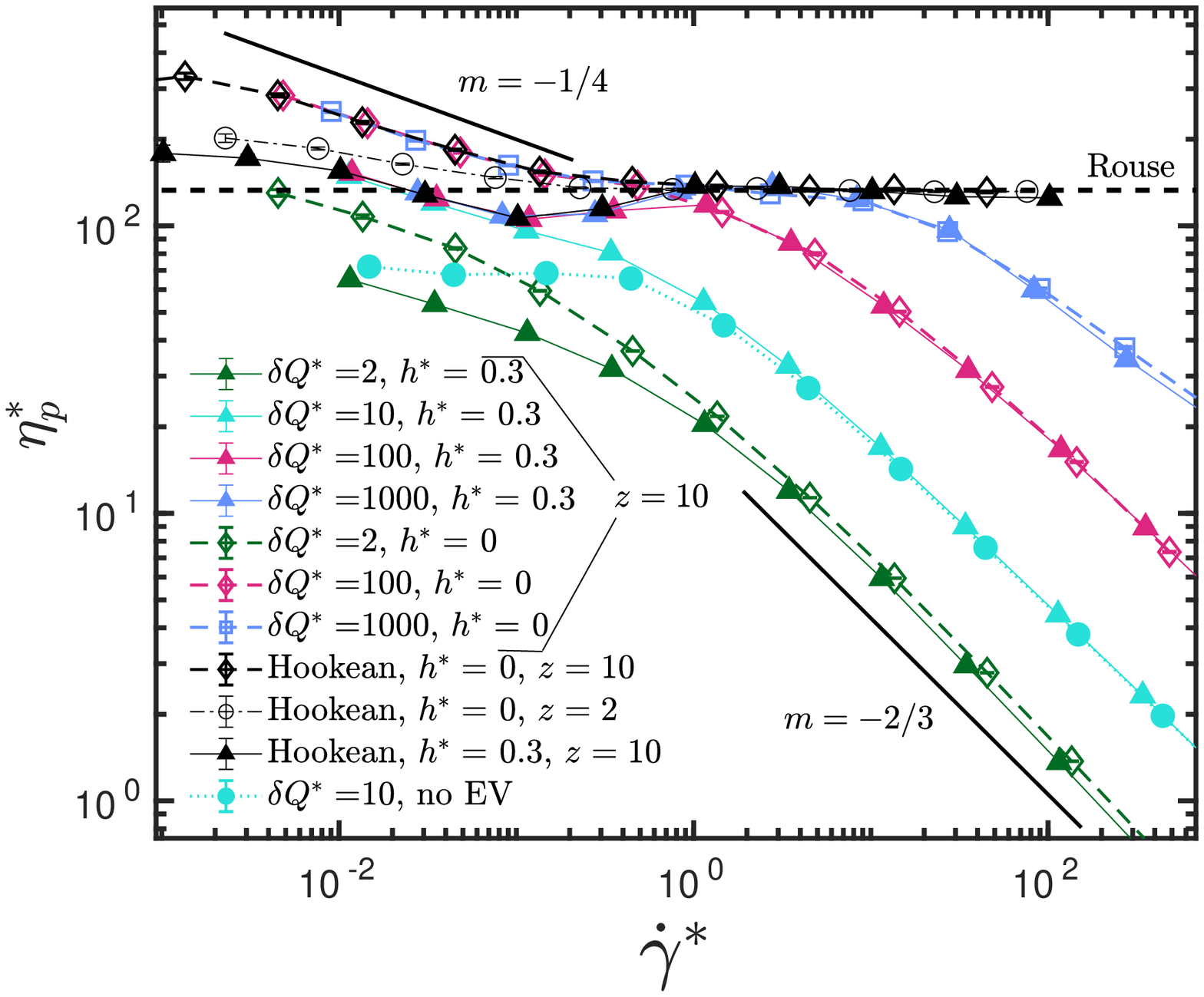}
        & \includegraphics[width=8.5cm,height=!]{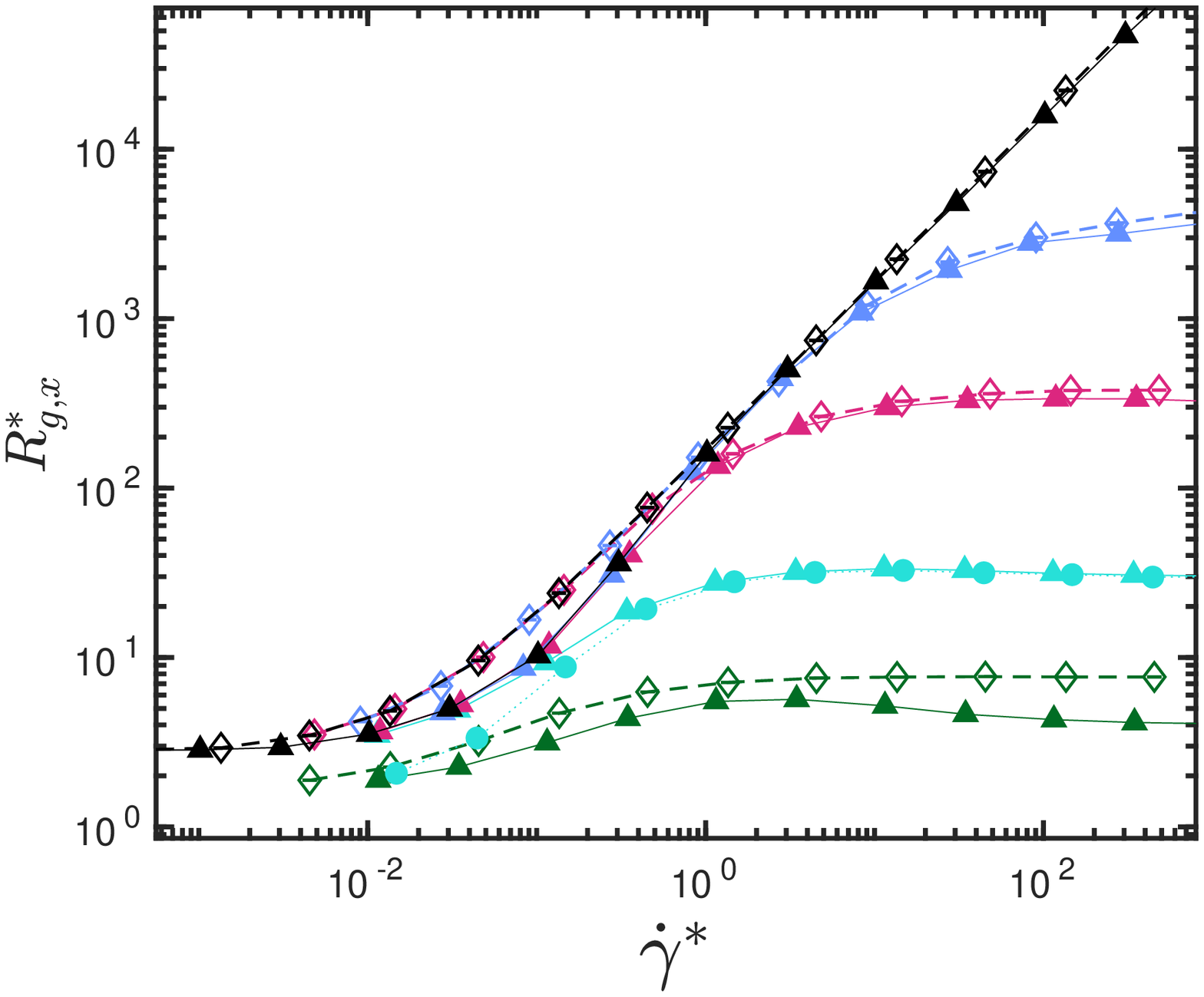} \\
        (c) & (d) \\
    \end{tabular}
    }
    \caption[Some results for N=20]{FENE springs of varying $\delta Q^*$, displaying both shear viscosity (a and c) and the $xx$-component of the gyration tensor (b and d) for 20-bead chains. Inset in (b) shows the scaling of the $yy$-component of the gyration tensor. Horizontal lines in (a) and (c) correspond to analytical results based on \eref{eq: eta_0_analytical_equation}, which can be calculated analytically with no HI, or using pre-averaged HI for the Zimm result. Labelled slopes ($m=-2/3$ and $m=-1/4$) are guides to the eye and do not imply exact terminal scaling with shear rate. For (c) and (d), all FENE chains include EV, with $z = 10$, with $z^*$ calculated via \eref{z in terms of zstar} and $d^* = {z^*}^{1/5}$. Where not visible, error bars are smaller than symbol size.}
    \label{fig:N20_Hook_FENE}
\end{figure*}

\newtxt{The second step is straightforward, namely to utilise the expression of Saadat and Khomami in \eref{bending potential expression SK} with our choice of $N_\mathrm{K,s}$ to determine the bending potential strength $C$.
Two examples of end-to-end distribution functions calculated using this procedure for $24$kbp and $7$kbp DNA are displayed in \fref{fig: distribution function fits} for several $N_s$.
Here we have assumed that the persistence length of DNA is 147 base pairs.
Note that the distribution functions seem to get worse for higher $N_s$ when modelling the shorter DNA fragment.
This is apparently due to a crudely chosen bending potential, such that matching only nearest-neighbor correlations is not sufficient to capture the full end-to-end distribution function when $N_{k,s} \approx 1$ (but it again becomes accurate in the $N_{k,s} \rightarrow 0$ limit \cite{Saadat2016}).
Since there is no simple analytical expression to capture this behaviour over the full range of $L$ and $l_p$, we will nevertheless use this scheme despite its shortcomings, noting that it would be possible to construct an iterative scheme to choose $C$ such that the correct end-to-end distance distribution is captured.}

\begin{table}[!ht]
\caption{FENE-Fraenkel spring parameters and bending potential constants for several DNA fragment lengths as shown in \fref{fig: methodology for real chain}. Parameters are calculated as described in text.}
\label{tab:my-table}
\begin{tabular}{|l|l|l|l|l|l|l|l|}
\hline
$L_\mathrm{DNA}$ (bp) & $N_s$ & C    & $\sigma$ (bp) & $\delta Q$ (bp) & $H \delta Q^2/k_\mathrm{B} T$ & $\sigma^*$ & $\delta Q^*$ \\ \hline
100                   & 5     & 7.2  & 20            & 0.44            & 35.8                          & 373        & 8.5          \\ \hline
100                   & 20    & 29.2 & 5             & 0.03            & 146                           & 3010       & 17           \\ \hline
100                   & 80    & 117  & 1.2           & 0.002           & 587                           & 24000      & 34           \\ \hline
1000                  & 20    & 2.8  & 47            & 2.7             & 13.9                          & 92.6       & 5.3          \\ \hline
10000                 & 20    & 0.64 & 270           & 230             & 2.5                           & 2.7        & 2.3          \\ \hline
100000                & 20    & 0.05 & 13            & 4987            & 23.9                          & 0.02       & 6.9          \\ \hline
100000                & 80    & 0.22 & 163           & 1097            & 5.31                          & 0.46       & 3.3          \\ \hline
\end{tabular}
\end{table}

\newtxt{We also wish to briefly touch upon the physical interpretation of the non-dimensional variables $\sigma^*$ and $\delta Q^*$, which will be used extensively in the rest of this paper.
From \tref{tab:my-table}, we can see that as the chain length is increased at $N_s = 20$, there is a decrease in $\sigma^*$ and an increase in $\delta Q^*$.
Roughly speaking, $\sigma^*$ corresponds to the rod-like nature of the spring, with a large $\sigma^*$ representing a stiff rod with high spring constant $H$. 
The value of $\delta Q^*$ then represents the total extensibility of the spring, such that a large $\delta Q^*$ (particularly relative to $\sigma^*$) represents a long, extensible chain.
In future figures in this paper, we will generally keep $\sigma^* + \delta Q^*$ fixed at some constant value for constant $N_s$, and change the relative values of $\sigma^*$ and $\delta Q^*$.
Although the analogy is not perfect, this is roughly equivalent to the transition between a short, stiff, rod-like chain for large $\sigma^*$ and small $\delta Q^*$, and a long, extensible chain for small $\sigma^*$ and large $\delta Q^*$.
However, here we caution against over-interpretation, as our primary aim is to investigate the qualitative features of various spring force law limits, rather than directly simulate a real polymer molecule.
For this reason, we have not used parameter sets representing a real chain in exploring the qualitative features, but instead kept $\sigma^* + \delta Q^*$ constant and explored the `spring-to-rod' parameter space, as will be seen below.}

\section{Results}
\label{Results}

We begin by summarising previous results for Hookean and FENE springs using our model with HI and EV, systematically displaying the effects of each piece of physics, as seen in \fref{fig:N20_Hook_FENE}.
Although these are certainly not novel findings, having been detailed for example by Ahn et al. in 1993 \cite{ahn1993bead}, they represent a wider range of parameter space than is currently in the literature, and are furthermore useful to inform later results.

\fref{fig:N20_Hook_FENE}~(a) and (b) give $\eta_p^*$ and $R_{g,x}^*$ curves for 20-bead Hookean and FENE chains without EV, and with and without HI, for a variety of FENE $b$-parameters (here identified as the total extensibility, $\delta Q^* \equiv \sqrt{b}$).
The Hookean chain without HI shows no deviation from the Rouse zero-shear viscosity with shear rate, as expected for an infinitely extensible chain.
Adding HI causes slight shear-thinning away from the Zimm viscosity, then shear-thickening towards the Rouse viscosity, due to HI being effectively weakened as the chain is stretched.

A highly extensible ($\delta Q^* > 100$) FENE chain with HI also displays this behaviour, following the Hookean $+$ HI result up until shear rate $\dot{\gamma}^* \approx 1$, after which the finite extensibility of the FENE chain begins to be felt, and the model displays a terminal shear-thinning slope of approximately $-0.6$.
This corresponds to a plateau in the chain extension in \fref{fig:N20_Hook_FENE}~(b), showing the relationship between extension and viscosity, caused by orientation and stretching of each link in the chain.
The inset to \fref{fig:N20_Hook_FENE}~(b) displays the $(-2/3)$ slope in $R^*_{g,y}$ for the FENE chains (not shown), notably the same as the shear-thinning exponent.
This can be intuitively understood in terms of the Giesekus expression for the stress tensor of a chain without HI, where the stress tensor is essentially proportional to the averaged gyration tensor \cite{bird1987dynamics}.
As $\delta Q^*$ is decreased, the shear-thickening vanishes, apparently due to the simultaneous decrease in the shear rate for onset of shear-thinning - in other words, the shear thinning kicks in before the HI has a chance to cause shear thickening.
Alternatively, one could reason that the chain can't stretch enough to reach the Rouse viscosity before the finite extensibility of the chain is reached.
Once $\delta Q^*$ is small enough, the finite extensibility causes a decrease in the coil size at equilibrium, leading to a lower zero-shear viscosity \cite{prakash2001}, as seen in the $\delta Q^* = 2$ and $\delta Q^* = 5$ cases both with and without HI.

Once we switch on EV (through a soft Gaussian potential), the low and intermediate shear behaviour changes, as shown in \fref{fig:N20_Hook_FENE}~(c) and (d).
This potential is set with $z^*$ as in \eref{z in terms of zstar}, and $d^* = {z^*}^{1/5}$, as per previous suggestions \cite{Kumar2004, Sunthar2005parameterfree}, giving a particular solvent quality $z$.
This EV potential causes a swelling at zero shear both in the viscosity and gyration radius, as can be seen by comparing the Hookean EV chain with the Rouse viscosity, and also the cyan dotted-line circles and cyan full-line triangles ($\delta Q^* = 10$, $z = 0$ and $10$) in \fref{fig:N20_Hook_FENE}~(c) and (d).
It is this swelling which defines the solvent quality, such that a smaller $z$ leads to less swelling as expected.

Beyond equilibrium, as the effective solvent quality is increased from $z=0 \rightarrow 2 \rightarrow 10$, we see some shear thinning prior to the $(-2/3)$ terminal exponent from the FENE springs.
This `intermediate' shear-thinning approaches a power-law slope of $-1/4$ as the solvent quality approaches infinity, which previous work has demonstrated exactly in the long-chain limit using BD and also renormalisation group approaches \cite{Kumar2004, Prabhakar2004SFG, ottinger1989renormalization}.
For the Hookean and highly extensible FENE chains ($\delta Q^* > 100$), there is something of a second Newtonian plateau at high shear rates, which is caused by the effect of EV lessening as beads are pulled apart due to flow.
At higher shear rates, the FENE springs show the expected $\approx (-2/3)$ power-law slope in viscosity, essentially unchanged by the addition of EV.
This can be seen particularly in the behaviour of the $\delta Q^* = 10$ FENE chains with and without EV (cyan symbols), where \fref{fig:N20_Hook_FENE}~(c) and (d) show a difference in zero-shear behaviour, but identical high-shear properties.

\newtxt{Some shear thickening also remains in the presence of EV, when HI is also switched on.
This shear-thickening is sensitive to EV in the sense that strong EV with high $z$ will cause the shear-thickening to disappear.
However, the shear-thickening is not caused by the EV, but instead, EV is an additional effect which may eliminate the impact of HI.
For example, consider \fref{fig:N20_Hook_FENE}(c) for the Hookean chain with $z = 10$ and HI.
Essentially what is occurring is that HI reduces the zero-shear viscosity compared to no-HI, while EV increases it.
However, both of these effects diminish at higher shear rates for a Hookean chain, since the beads move far apart and no longer feel a strong EV or HI force.
Therefore, the viscosity moves towards the Rouse viscosity - this is an increase for a chain with only HI, and a decrease for a chain with only EV.
That is to say, HI and EV affect the chain in opposite ways at low shear rates, but it is only HI which directly causes shear-thickening - EV is simply an effect over the top of it which counteracts this shear-thickening.}

Overall, our results match with previous theoretical and computational findings for FENE and Hookean springs both with and without HI and EV.
In summary, the key features are an $\approx (-2/3)$ slope in viscosity at high shear rates for finite extensibility, and an $\approx (-1/4)$ power law slope in viscosity at intermediate shear rates due to EV (which plateaus at high shear rates for Hookean chains), slight shear-thickening due to HI, and differences in onset of shear thinning due to finite extensibility.

\subsection{FENE-Fraenkel spring results}

\begin{figure}[t]
    \centering
    \includegraphics[width=8.5cm,height=!]{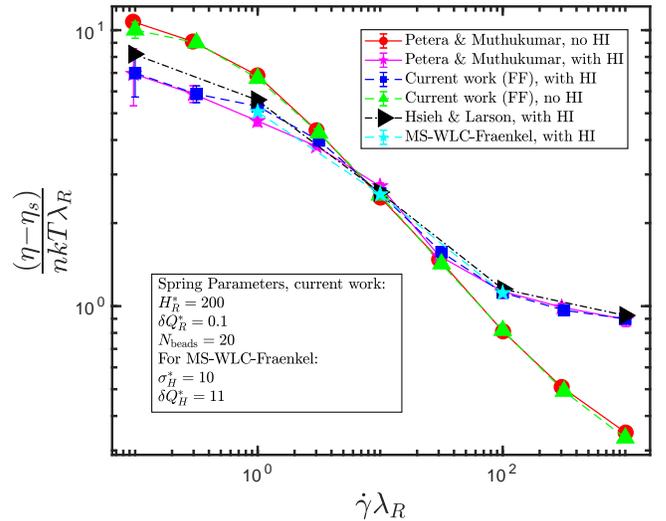}
    \caption{Comparisons between the FENE-Fraenkel spring in the current work and the bead-rod simulations of Petera and Muthukumar \cite{Petera1999}, as well as the previous FENE-Fraenkel simulations of Hsieh and Larson \cite{Hsieh2006}. The timestep used was $\Delta t^*_H = 0.01$. Note that results are given in the `rodlike' system of non-dimensionalisation, where $\lambda_R = \sigma^2 \zeta/k_\text{B} T$, $H^*_R = H \sigma^2/k_\text{B} T$, and $\delta Q^*_R = \delta Q/\sigma$. In these units, the equivalent spring parameters used by Hsieh and Larson were $H^*_R = 1000$, $\delta Q^*_R = 0.01$, and $\Delta t^*_H = 4$. The MS-WLC-Fraenkel spring parameters are chosen to have the same relative extensibility around $\sigma$ as our FENE-Fraenkel chain. Where not visible, error bars are smaller than symbol size.}
    \label{fig: Muthukumar comparisons}
\end{figure}

\begin{figure*}[t]
    \centerline{
    \begin{tabular}{c c}
        \includegraphics[width=8.5cm,height=!]{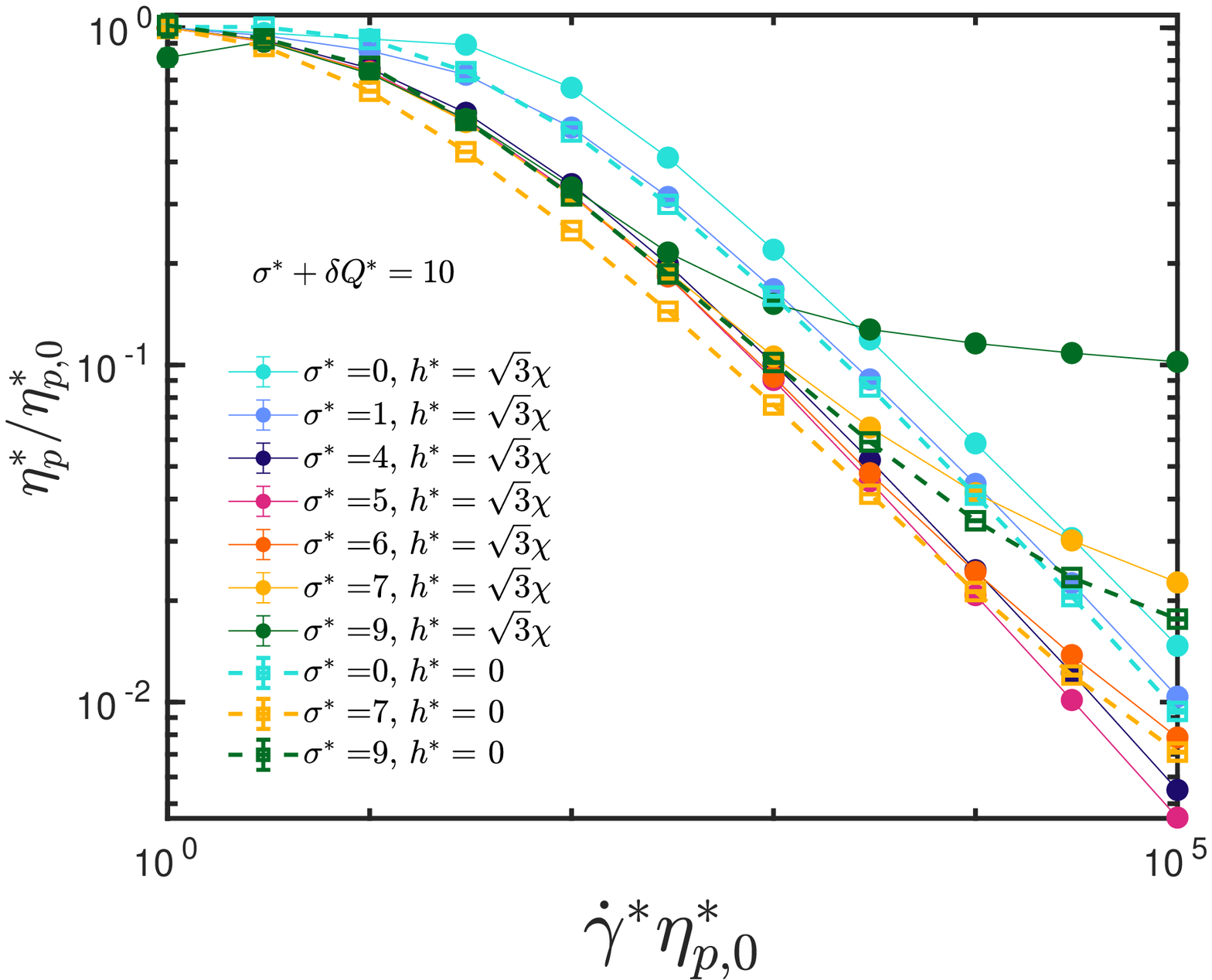}
        & \includegraphics[width=8.5cm,height=!]{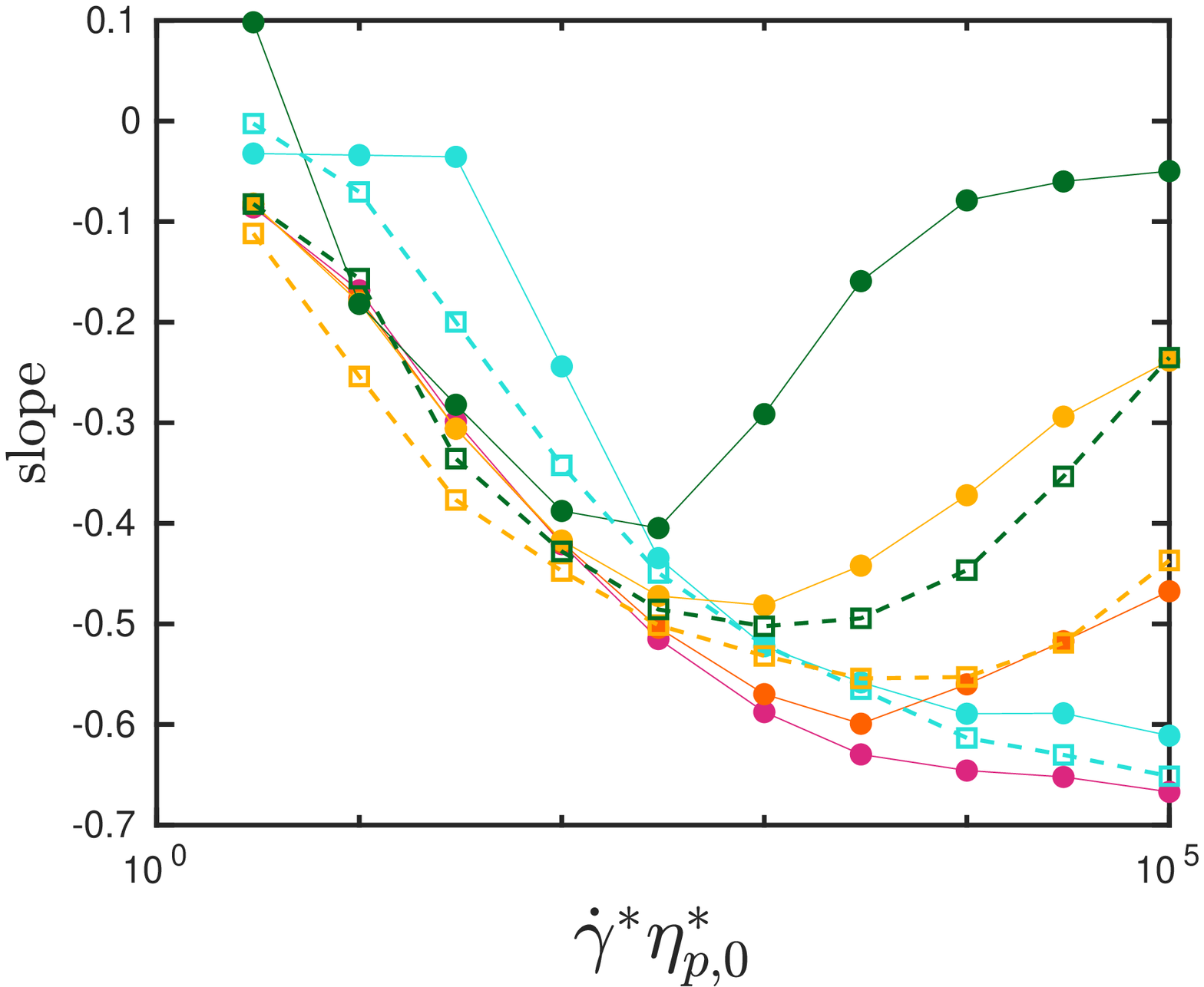} \\
        (a) & (b) \\
        \includegraphics[width=8.5cm,height=!]{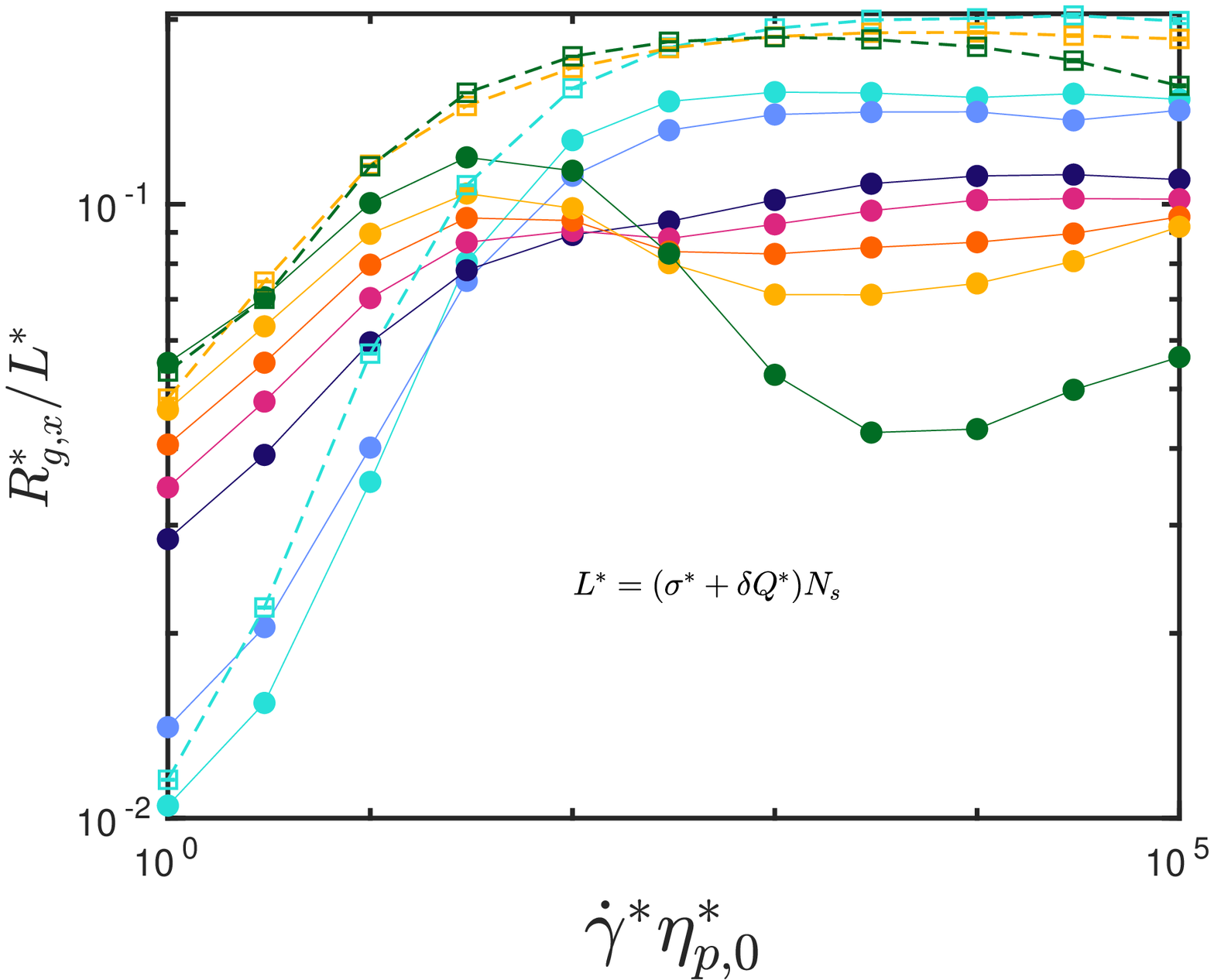}
        & \includegraphics[width=8.5cm,height=!]{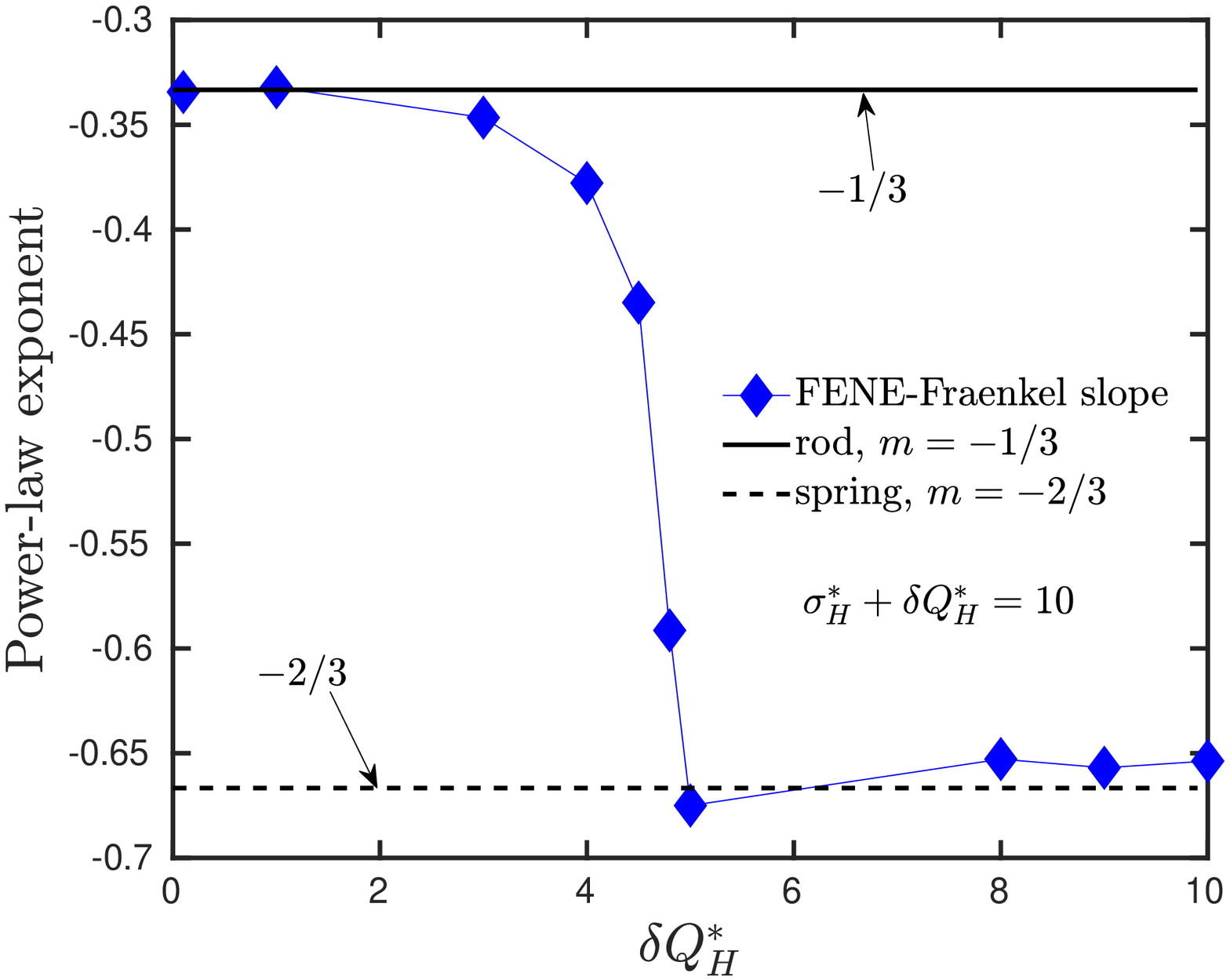} \\
        (c) & (d) \\
    \end{tabular}
    }
    \caption{Calculated properties for FENE-Fraenkel springs with $\sigma^* + \delta Q^* =10$, changing the value of $\sigma^*$. Results with HI have $h^*$ scaled by $\chi$ calculated from \eref{eq: chi equation spring length}, such that the beads are on average osculating at equilibrium. The values of $\chi$ for $\sigma^* = 0$ to $9$ are $\chi = \{0.9759,1.2856,2.6145,3.1303,3.6591,4.1922,5.2360\}$ respectively. Note that qualitatively similar results are seen when we set the strength of HI to a constant $h^* = 0.3$ for all springs irrespective of $\chi$. Zero-shear viscosity is found from low-shear results, which agrees with Green-Kubo calculations to within error bars (see \aref{Appendix TTCF}). Properties for curves (a) and (c) are defined as in \eref{eq: material functions eta, psi} and \eref{eq: gyration components}, while `slope' in (c) is given by the log-log gradient of (a) at each shear rate. The total contour length $L^*$ used to normalise $R^*_{g,x}$ in plot (c) is given by $L^* = (\sigma^* + \delta Q^*) N_s$. Plot (d) is reproduced from a previous paper \cite{pincus2020viscometric}. Where not visible, error bars are smaller than symbol size.}
    \label{fig:FENE-Fraenkel N20}
\end{figure*}

Now that we have outlined the behaviour of FENE and Hookean springs with HI and EV, we can `stiffen' our FENE-Fraenkel springs by increasing $\sigma^*$ to head towards the bead-rod limit.
We first note that our model is indeed able to reproduce the bead-rod results of Petera and Muthukumar \cite{Petera1999} in the `rodlike' limit, even when HI is included (see \fref{fig: Muthukumar comparisons}), in contrast to the previous results of Hsieh and Larson \cite{Hsieh2006}.
We suspect that this difference is due to their unsuitably large timestep \cite{pincus2020viscometric}, noting that our model is able to reproduce the bead-rod results with both a $5\times$ smaller spring constant, as well as a $10\times$ larger extensibility.
As we will see, this finding is not surprising given the range of crossover between bead-spring and bead-rod behaviour.
We also note that a recent paper by Kumar and Dalal has demonstrated that a Fraenkel spring is indeed able to reproduce bead-rod behaviour when sufficiently stiff \cite{kumarfraenkel}.

\begin{figure}[t]
    \centerline{
    \begin{tabular}{c}
        \includegraphics[width=8.5cm,height=!]{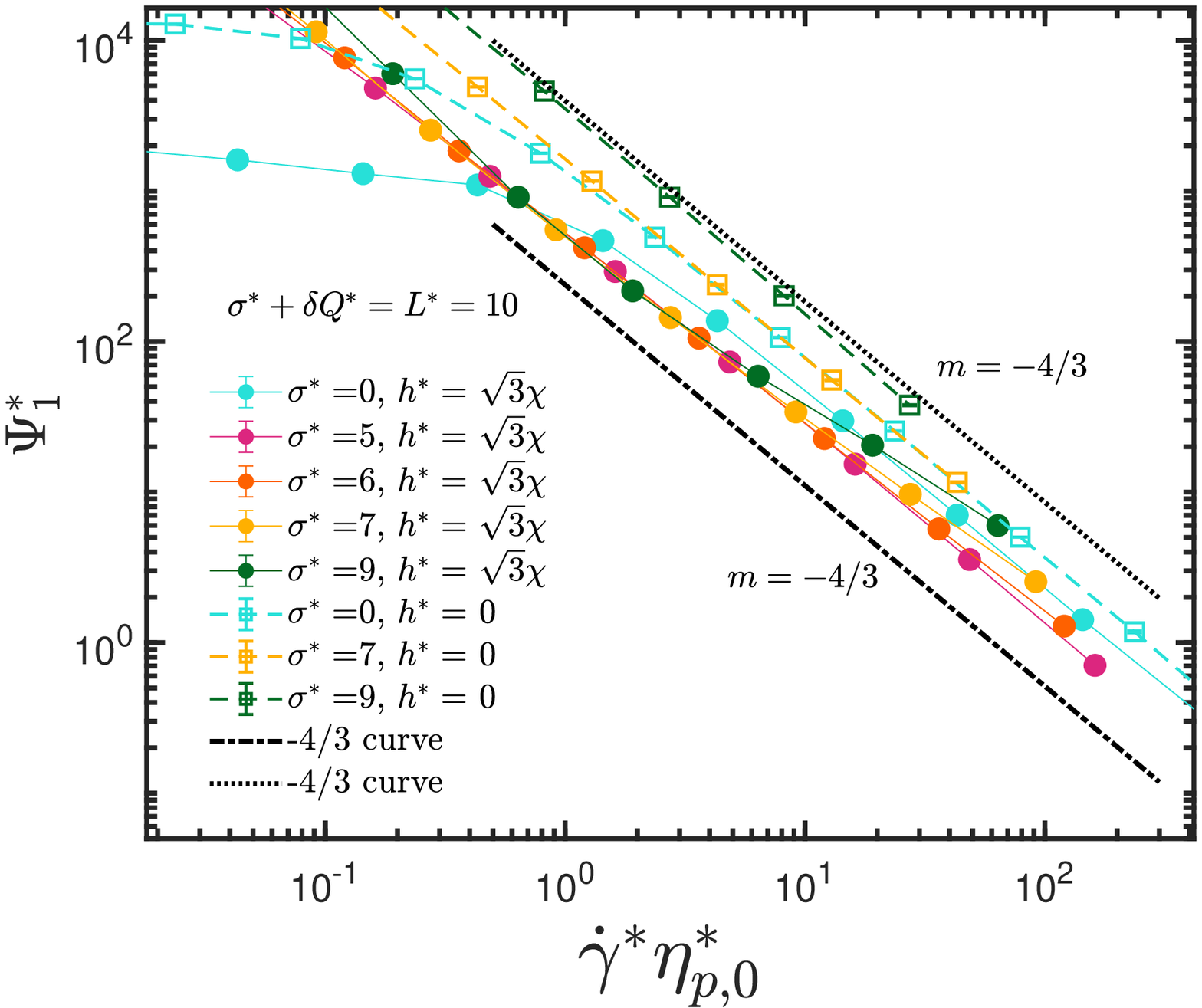} \\
        (a) \\
        \includegraphics[width=8.5cm,height=!]{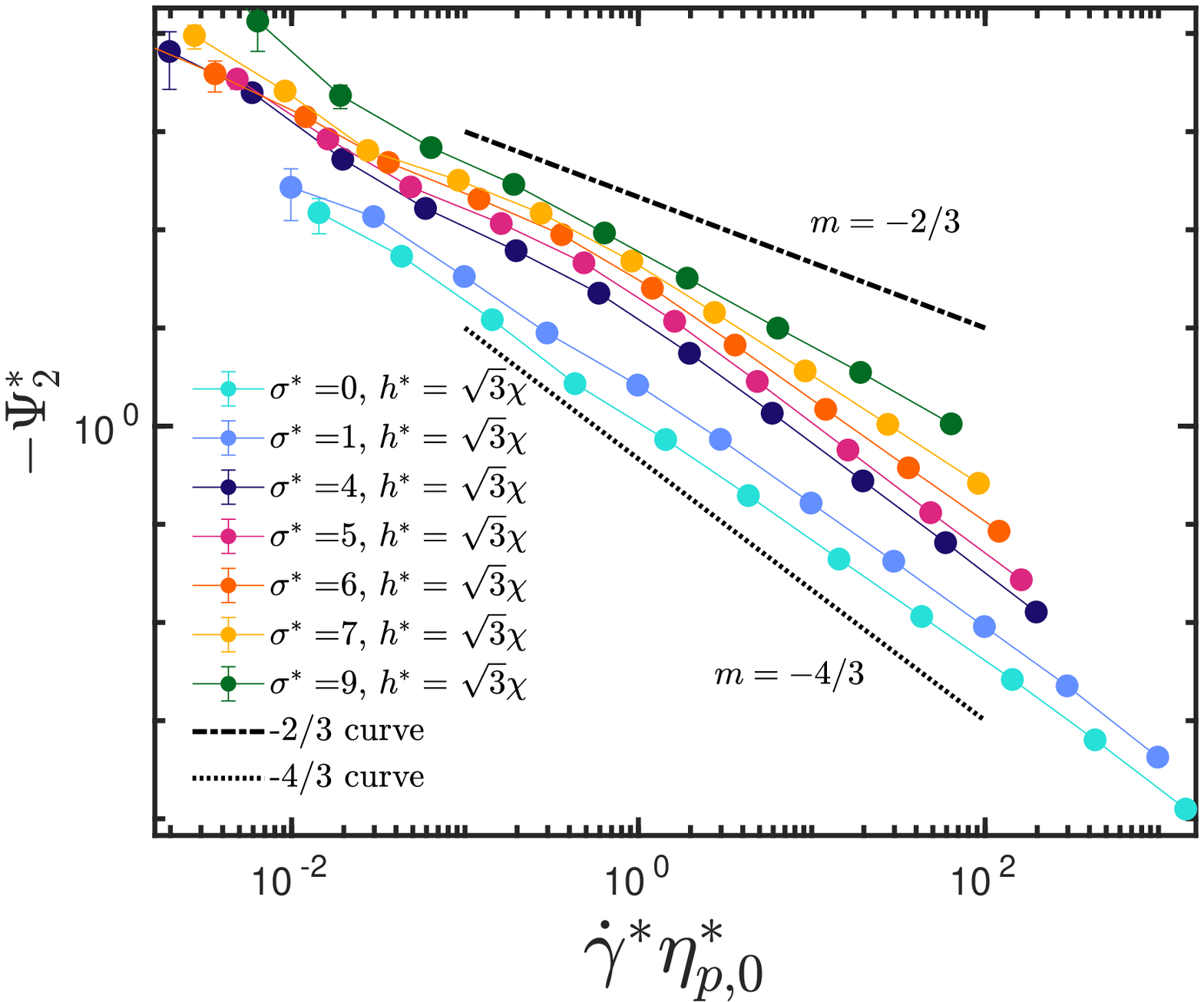} \\
        (b) \\
    \end{tabular}
    }
    \caption{Normal stress coefficients (a) $\Psi_1^*$ and (b) $\Psi_2^*$ as a function of shear rate. 20-bead FENE-Fraenkel springs are used, with parameters chosen such that $\sigma^*+\delta Q^* = 10$. Error bars are unfortunately not small enough to accurately display results for $\Psi_2^*$ over a range of shear rates for models with smaller $N$, or for the case of $h^* = 0$. Where not visible, error bars are smaller than symbol size. Dotted lines are guides for the eye and do not imply exact terminal slopes.}
    \label{fig:FF_Psi1_Psi2}
\end{figure}

\begin{figure*}[t]
    \centerline{
    \begin{tabular}{c c}
        \includegraphics[width=8.5cm,height=!]{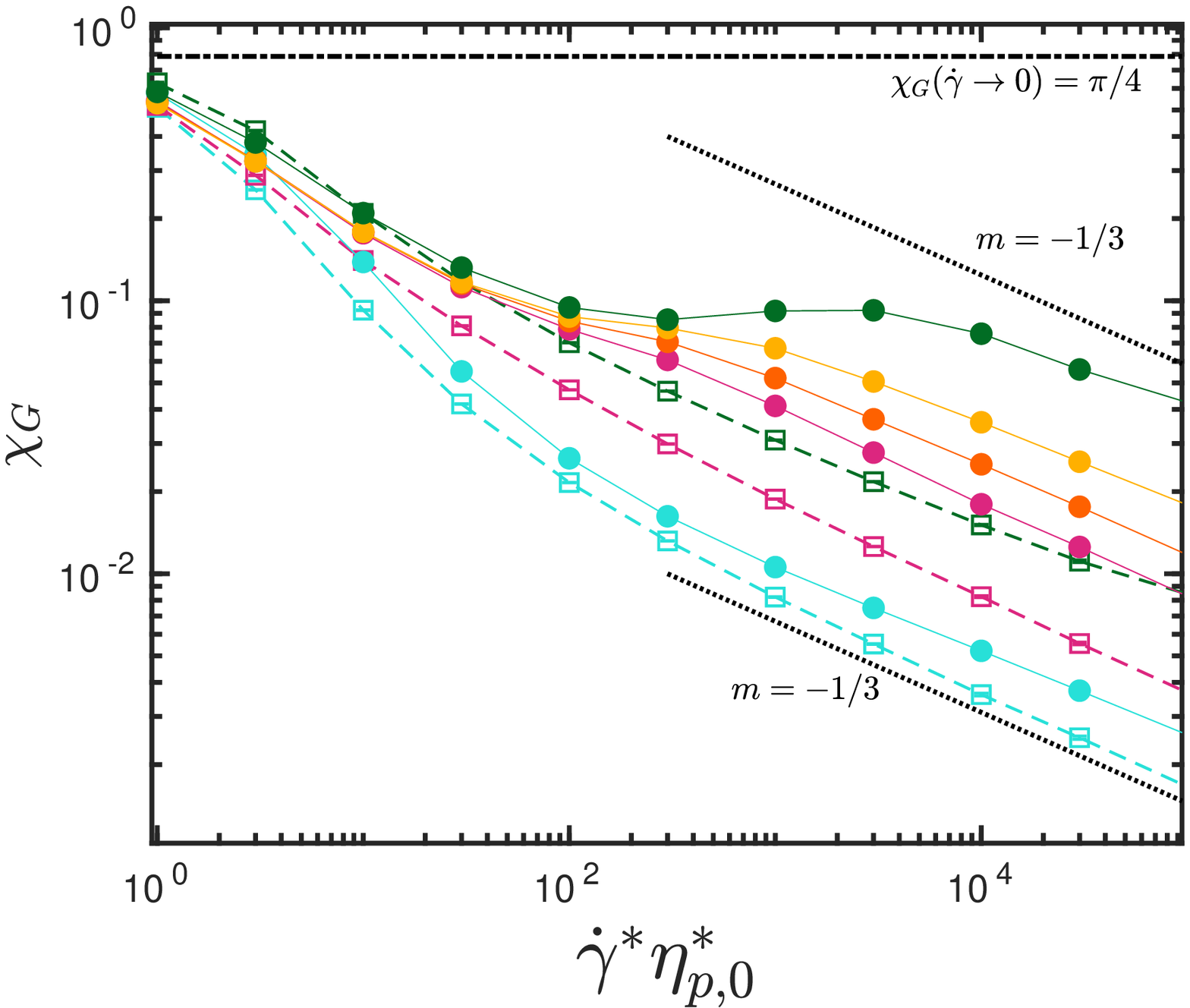} 
        &
        \includegraphics[width=8.5cm,height=!]{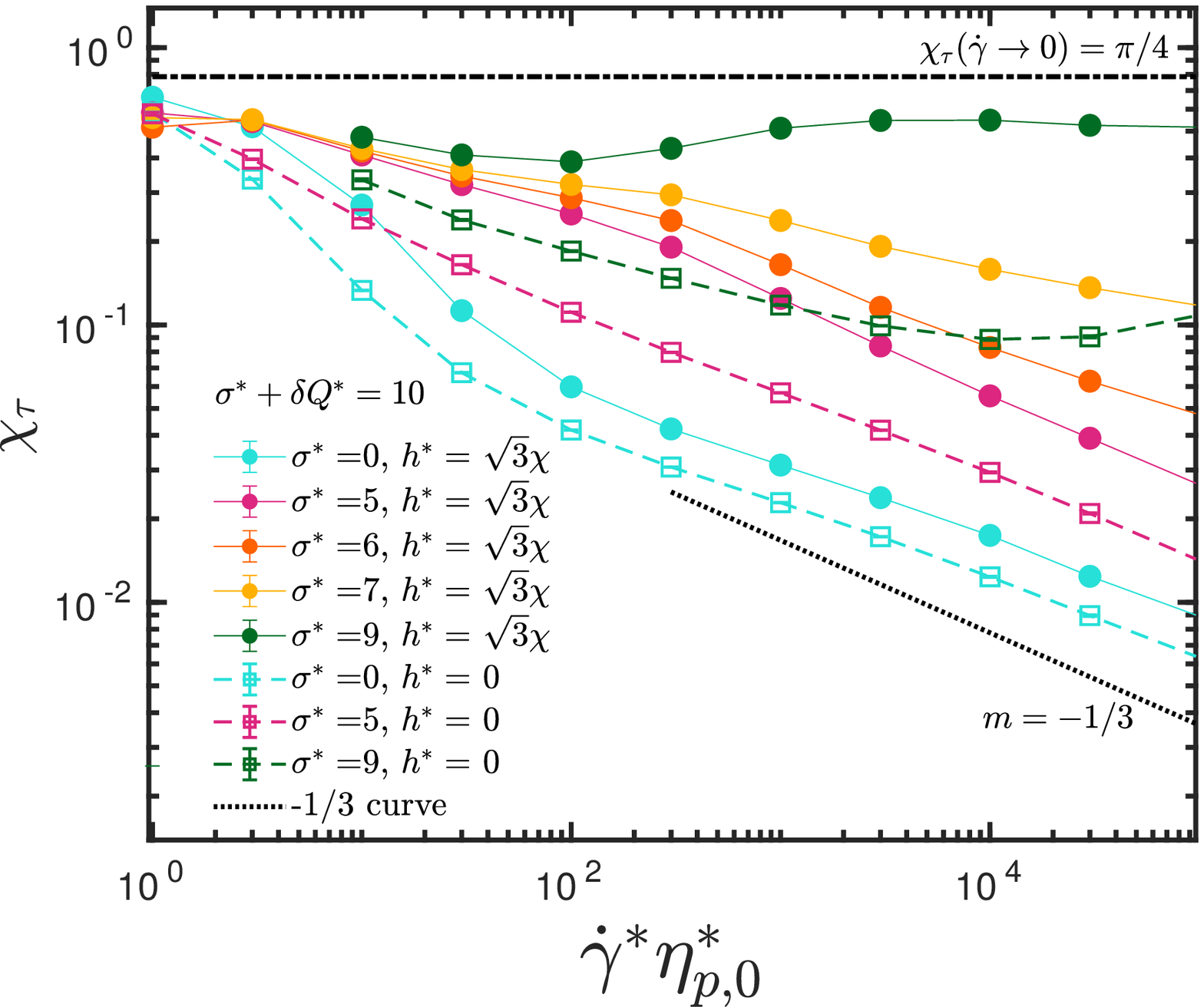} \\
        (a) & (b) \\
    \end{tabular}
    }
    \caption{Orientation angles (a) $\chi_G$ and (b) $\chi_\tau$ as a function of shear rate. 20-bead FENE-Fraenkel springs are used, with parameters chosen such that $\sigma^*+\delta Q^* = 10$. The zero-shear value for both $\chi_G$ and $\chi_\tau$ of $\pi/4$ (since $1/2 \arctan{\infty} = \pi/4$) is displayed as a horizontal dotted line. First two $\sigma^* = 9$ data points for $\chi_\tau$ at low shear have not been included due to very large error bars. Where not visible, error bars are smaller than symbol size. Dotted lines are guides for the eye and do not imply exact terminal slopes.}
    \label{fig:FF_chi_G_chi_tau}
\end{figure*}

\begin{figure}
    \centering
    \includegraphics[width=8.5cm,height=!]{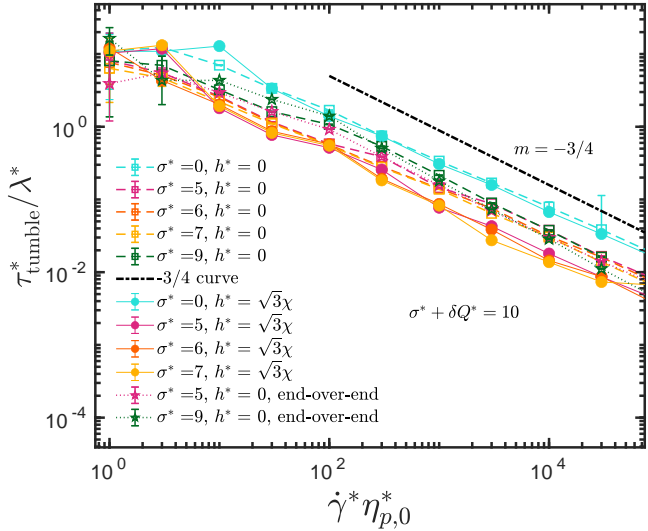}
    \caption{Tumbling calculations, using cross-correlation method as detailed in \eref{tumbling cross-correlation}. Results are given as the tumbling period $\tau_\mathrm{tumble}$ divided by the relaxation time $\lambda$ calculated from the end-to-end auto-correlation of the radius of gyration at equilibrium. Also given are two results with the `end-over-end' tumbling time calculations, showing nearly identical, and certainly qualitatively similar behaviour. $m=-3/4$ line is a guide for the eye and does not imply exact terminal slopes.}
        \label{fig:Tumbling FENE-Fraenkel N20}
\end{figure}

\begin{figure}[t]
    \centerline{
    \begin{tabular}{c}
        \includegraphics[width=8.5cm,height=!]{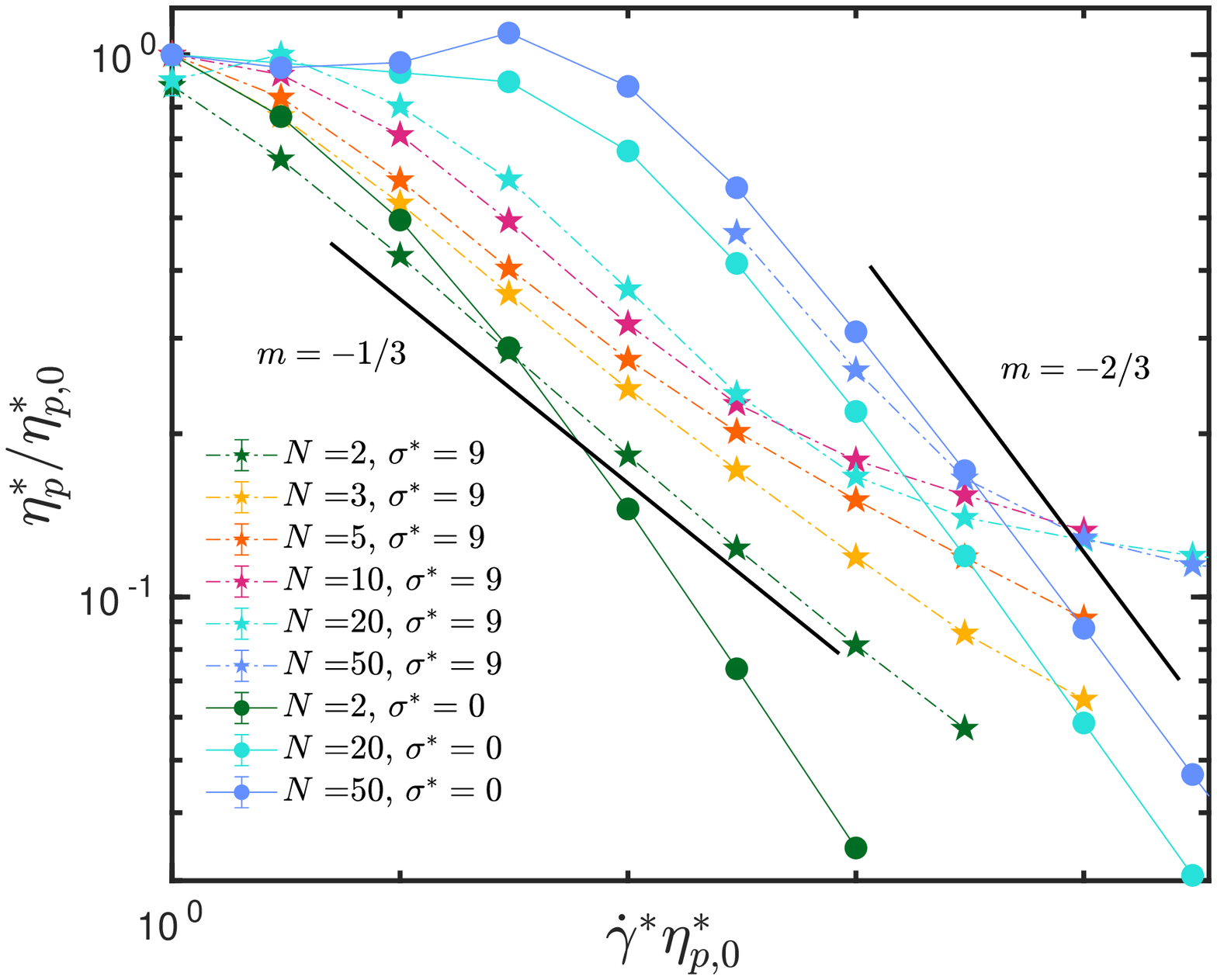} \\
        (a) \\
        \includegraphics[width=8.5cm,height=!]{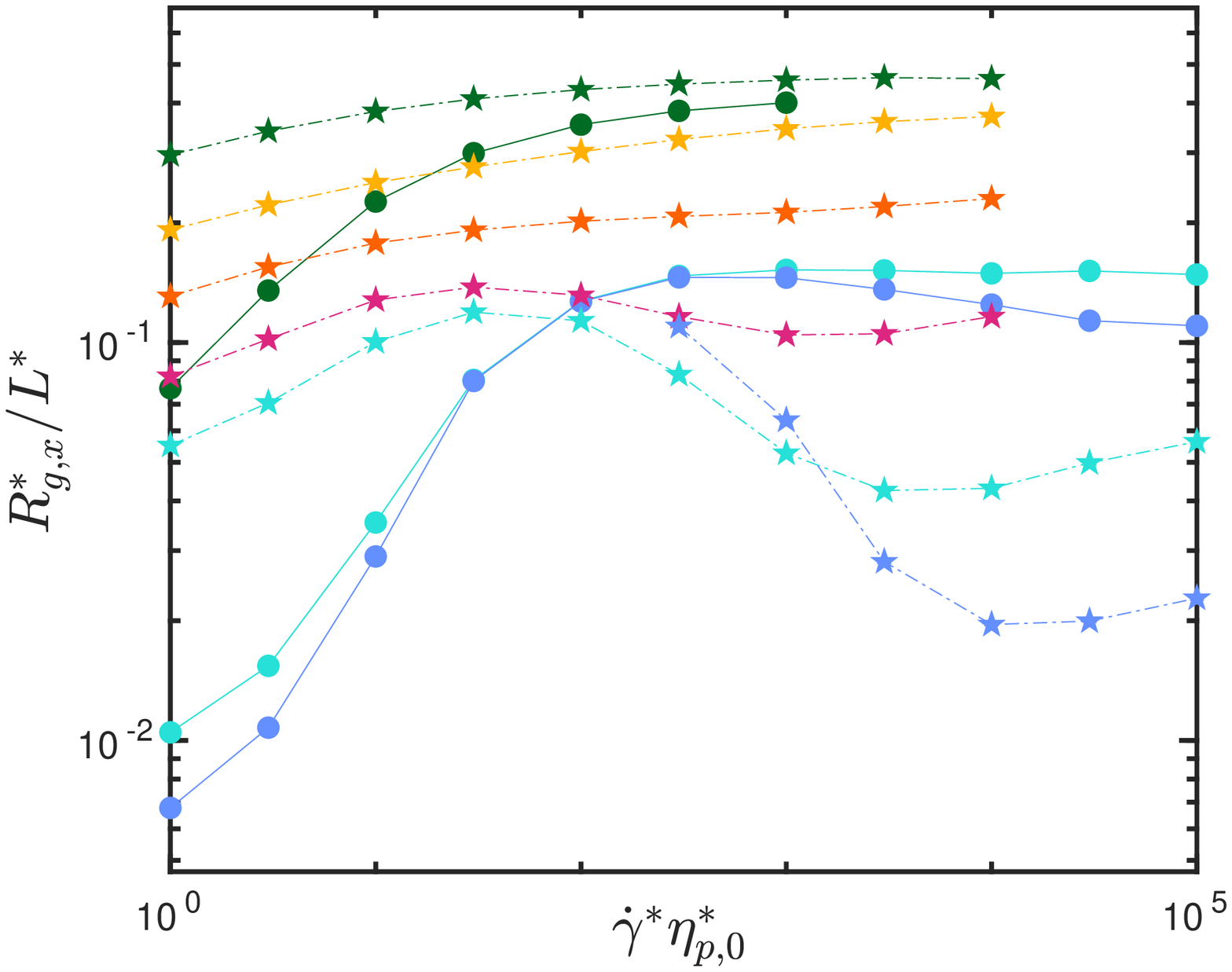} \\
        (b) \\
    \end{tabular}
    }
    \caption{FENE-Frankel springs with HI but no EV and $\sigma^* + \delta Q^* = 10$, presented as a function of bead number. Plot (a) gives the viscosity divided by the zero-shear viscosity versus Weissenberg number, while plot (b) gives the $xx$-direction gyration tensor component divided by total contour length of the chain $L^* = (\sigma^* + \delta Q^*) N_s$. Note that $\sigma^* = 0$ corresponds to a FENE spring with $\delta Q^* = 10$, while $\sigma^* = 9$ is a `rod-like' FENE-Fraenkel spring. $m=-1/3$ and $m=-2/3$ lines are guides for the eye and do not imply exact terminal slopes.}
    \label{fig:eta_sig9_sig0_N_sweep}
\end{figure}

Guided by previous results for FENE-Fraenkel dumbbells \cite{pincus2020viscometric}, we will principally investigate the change in power-law slope in viscosity with shear rate, as well as the compression in gyration radius at high shear seen in bead-rod models.
For the following sets of figures, namely Fig.~\ref{fig:FENE-Fraenkel N20} and Figs.~\ref{fig:FF_Psi1_Psi2} to \ref{fig:bending potential results}, we go from bead-spring to bead-rod behaviour by keeping $\sigma^* + \delta Q^* = 10$ fixed, while changing $\sigma^*$.
For example, the cyan symbols in \fref{fig:FENE-Fraenkel N20}~(a) with $\sigma^*=0$ and $\delta Q^* = 10$ are exactly the same as our cyan-coloured FENE spring results in \fref{fig:N20_Hook_FENE}, while the green-coloured symbols have $\sigma^* = 9$ and $\delta Q^* = 1$.
This is following the procedure from our previous paper on FENE-Fraenkel dumbbells \cite{pincus2020viscometric}, which showed a smooth crossover from bead-spring to bead-rod behaviour using this arrangement.

\begin{figure}[t]
    \centerline{
    \begin{tabular}{c} 
        \includegraphics[width=6.8cm,height=!]{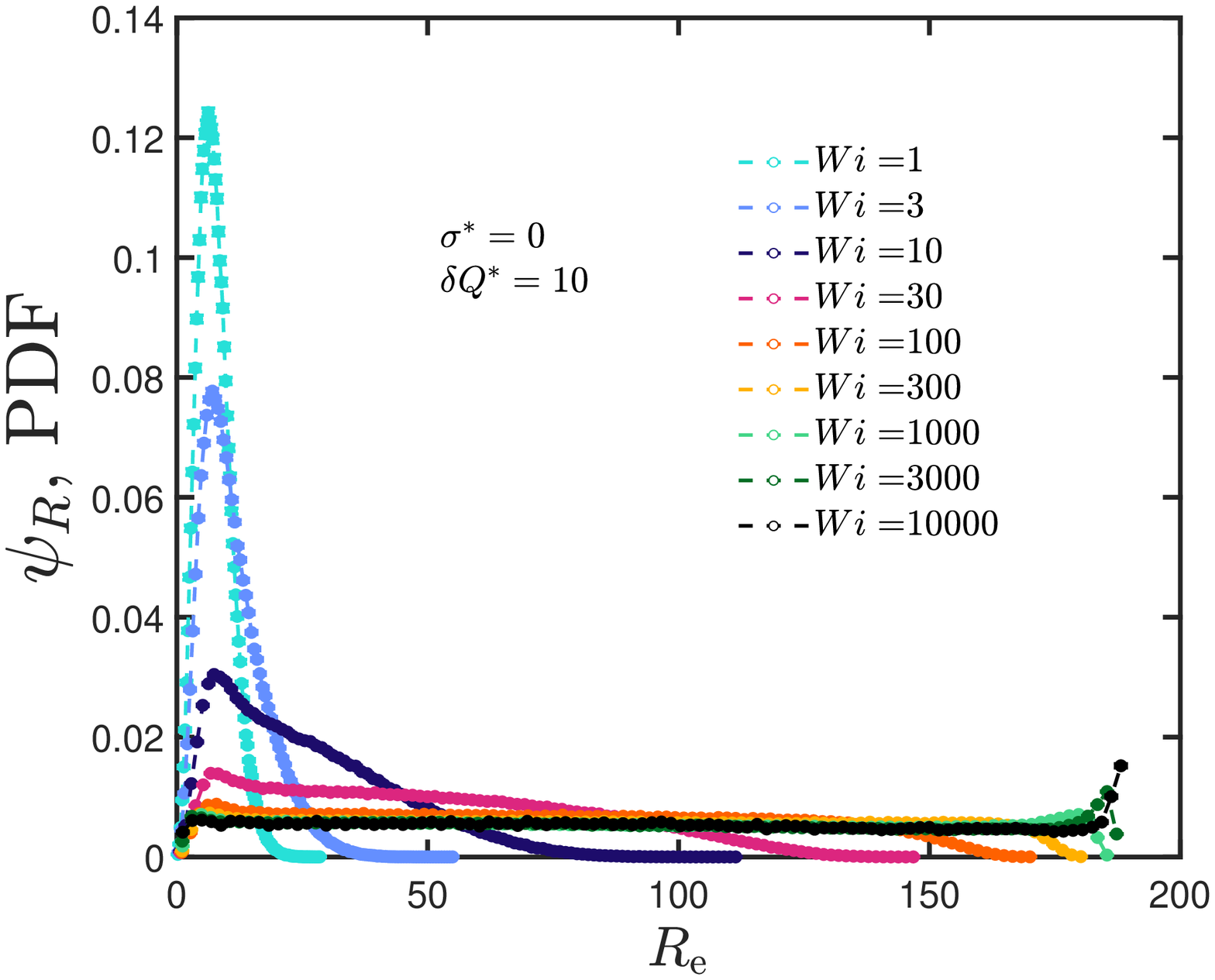} \\
     (a) \\[10pt]
        \includegraphics[width=6.8cm,height=!]{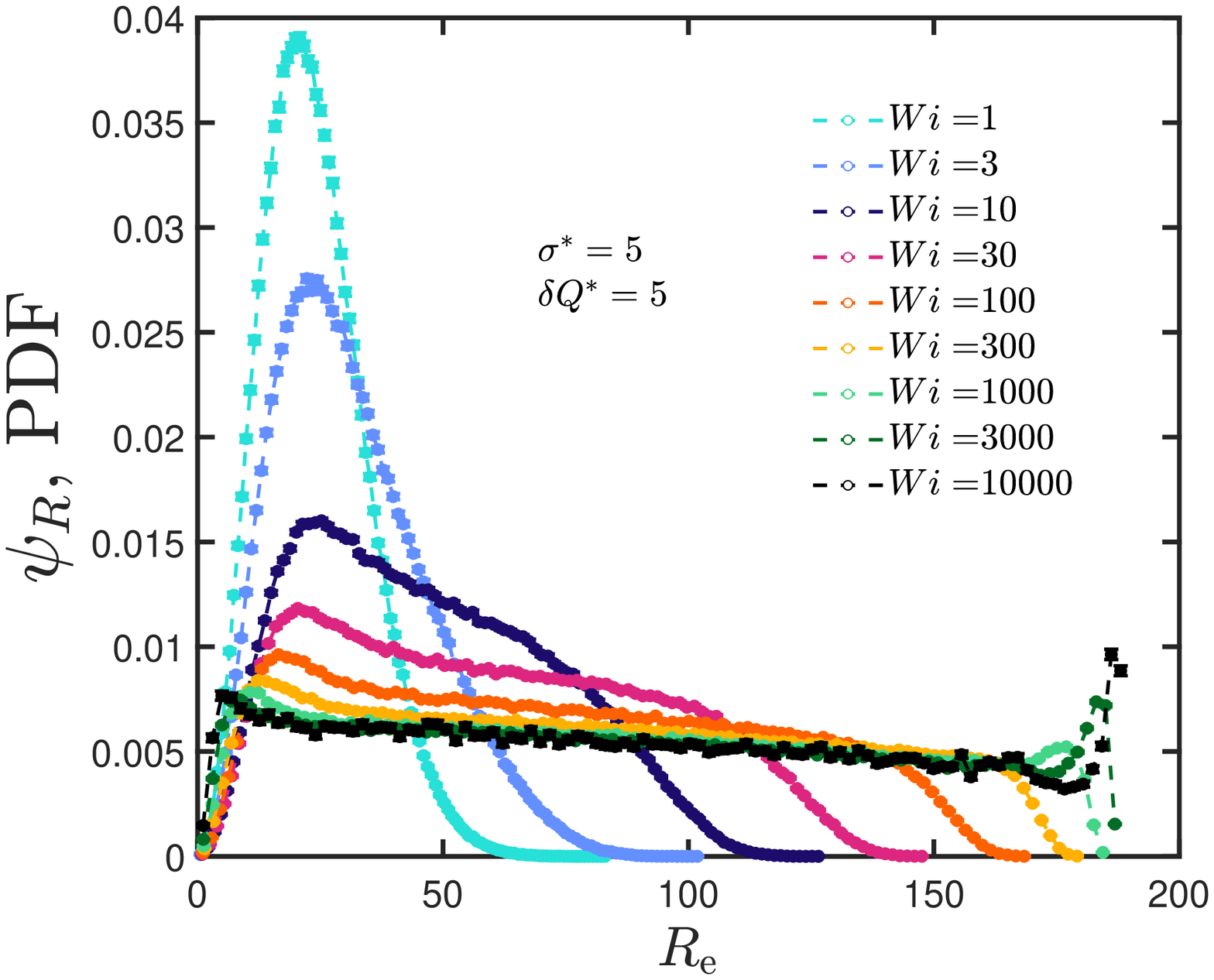} \\ 
    (b) \\[10pt]
        \includegraphics[width=6.8cm,height=!]{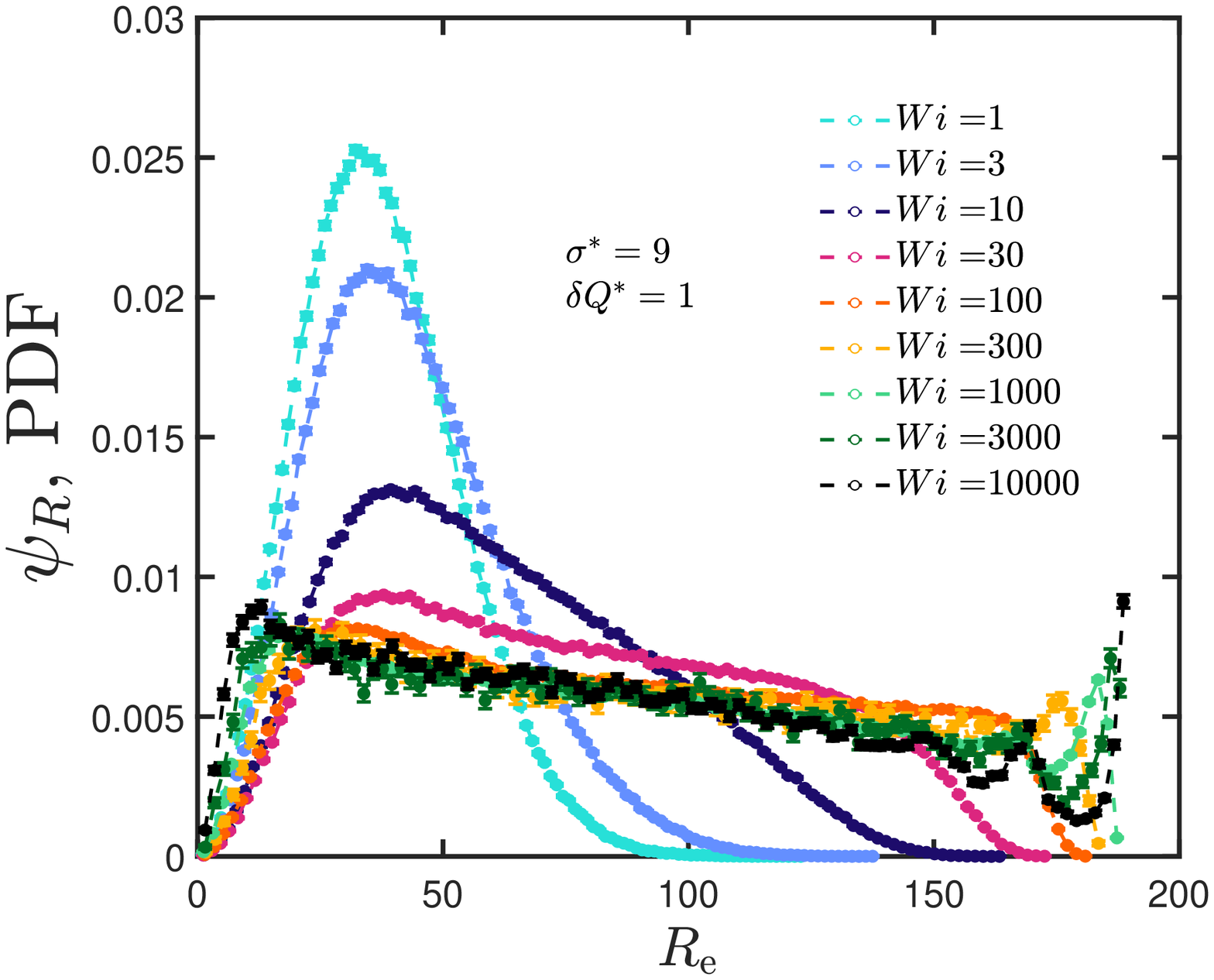} \\
     (c) \\
    \end{tabular}
    }
    \caption{End-to-end ($R_e$) distribution function ($\psi_R$) plots for different FENE-Fraenkel springs with $N=20$ at different $Wi = \dot{\gamma} \eta_{p,0}^*$ numbers. all simulations have $h^* = \sqrt{3} \chi$ as in \fref{fig:FENE-Fraenkel N20}.}
    \label{fig: dist functions with Wi}
\end{figure}

\begin{figure*}
    \centerline{
    \begin{tabular}{c c}
        \includegraphics[width=8.5cm,height=!]{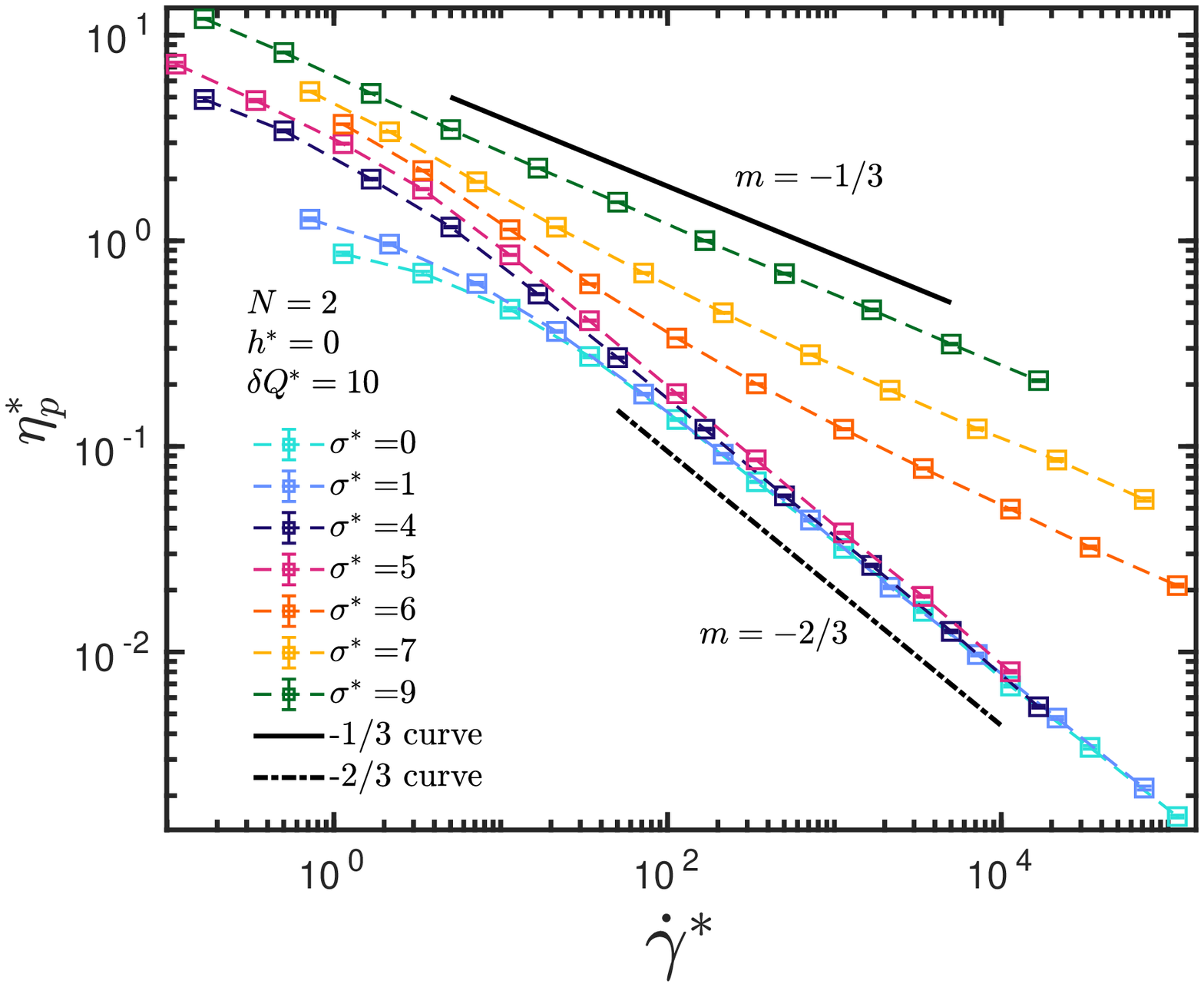}
        & \includegraphics[width=8.5cm,height=!]{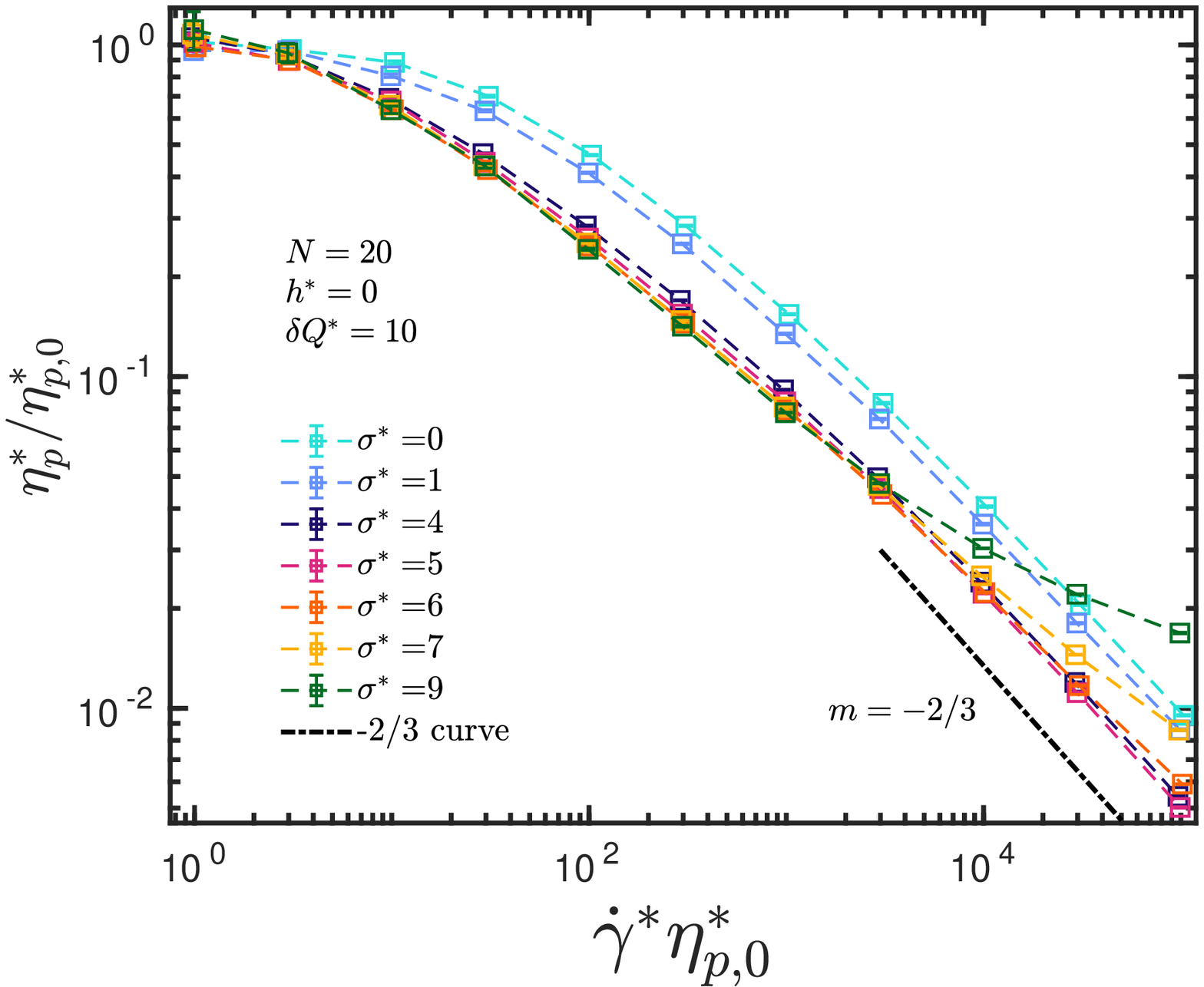} \\
        (a) & (b) \\
    \end{tabular}
    }
    \caption{Viscosity scaling without HI or EV for MS-WLC-Fraenkel spring chains, as described in \eref{MS-MOD non dim eqn}. In this case, $\delta Q^* = 10$, giving the same total extensibility at a particular $\sigma^*$ as for the FENE-Fraenkel spring (although not necessarily the same effective spring constant). Plot (a) gives the non-dimensional viscosity as a function of shear rate for MS-WLC dumbbells without HI. Plot (b) gives the viscosity normalised by zero-shear viscosity against Weissenberg number for 20-bead MS-WLC spring chains, also without HI. $m=-1/3$ and $m=-2/3$ lines are guides for the eye and do not imply exact terminal slopes.}
    \label{fig:MS WLC visc examples}
\end{figure*}

We first examine \fref{fig:FENE-Fraenkel N20}~(a), which gives the normalised viscosity as a function of Weissenberg number for $N=20$.
Alongside it, \fref{fig:FENE-Fraenkel N20}~(b) displays the log-log gradient of the lines in (a).
Pure FENE springs (cyan symbols) give the expected $(-2/3)$ slope in viscosity at high shear rates, as seen in \ref{fig:N20_Hook_FENE}.
This same `spring-like' terminal slope is observed for $\sigma^* = 1$, $4$, and $5$.
However, at $\sigma^* = 6$ and beyond, we see an increase in the power-law exponent, leading to a plateauing of the viscosity at high shear rates.
This is observed both with and without HI, although the effect is more pronounced with HI.
The same scaling can be seen in \fref{fig:FENE-Fraenkel N20}~(c), particularly for models with HI, where a compression in the flow direction at high shear rates begins roughly when $\sigma^* = 6$, and is more pronounced for higher $\sigma^*$.
This was previously noted by several authors using both stiff Fraenkel springs \cite{Sendner2009, Dalal2012regimes, Dalal2014} and true rigid constraints \cite{Liu1989, Petera1999}.

Interestingly, we note that since $\sigma^* + \delta Q^* = 10$, this change in behaviour occurs when $\sigma^* > \delta Q^*$, or in other words when the spring is no longer infinitely compressible.
To see this, note the behaviour of the lower bound in \fref{fig:FF force limits}, in which the force goes to infinity at $\sigma^* - \delta Q^*$ when $\sigma^* > \delta Q^*$.
This is exactly the behaviour observed for FENE-Fraenkel dumbbells \cite{pincus2020viscometric}, where the terminal slope showed a change from $(-2/3)$ to $(-1/3)$ when $\sigma^* > 5$.
For reference, we have reproduced a key figure from Ref.~[9] in \fref{fig:FENE-Fraenkel N20}~(d), which gives the terminal slope at high shear for FENE-Fraenkel dumbbells as a function of $\delta Q^*$.
These results taken together imply that even for chains, the crossover from bead-spring to bead-rod behaviour is related to the compressibility of the underlying segment, whether in the form of a rigid rod, very stiff Fraenkel spring, or the current FENE-Fraenkel spring.

To test this idea for another form of the spring potential, we use the so-called MS-WLC-Fraenkel spring, introduced in \eref{MS-MOD non dim eqn}.
Note that for this form of the spring force law, setting a constant $\delta Q^* = 10$ and changing $\sigma^*$ from  $0 \rightarrow 10$ implies the same behaviour as setting $\sigma^* + \delta Q^* = 10$ for a FENE-Fraenkel spring.
Additionally, unlike the FENE-Fraenkel spring, the effective spring constant (or the linear relationship between force and extension about $\sigma^*$) changes as $\sigma^* \rightarrow \delta Q^*$, such that one both `stiffens' the spring and makes it less extensible for higher $\sigma^*$.
Results are shown in \fref{fig:MS WLC visc examples}, without HI or EV and with constant $\delta Q^* = 10$ and variable $\sigma^*$. 

For dumbbells in \fref{fig:MS WLC visc examples}~(a), we again see a clear $(-2/3)$ power law slope at high shear rates when $\sigma^* < 5$, as expected for `spring-like' force laws.
Note that in the $\sigma^* = 0$ limit, the traditional MS-WLC force law is recovered, as has been used by many other authors \cite{schroeder2005dynamics, Sunthar2005parameterfree, li2000comparison, stoltz2006concentration}, which again gives a $(-2/3)$ power law slope.
However, the crossover in behaviour from $\sigma^* = 5$ to $\sigma^* = 6$ is even more pronounced than for the FF spring, possibly due to the increased effective spring stiffness $H$ as $\sigma^* \rightarrow \delta Q^*$.
The crossover from bead-spring to bead-rod behaviour again occurs when the spring can no longer compress to zero length, reinforcing the conclusion that `rod-like' behaviour is intimately linked to spring compressibility.
For the bead-spring-chain results in \fref{fig:MS WLC visc examples}~(b), we see a power-law slope of $\approx -0.6$ for $\sigma^* < 5$, with a gradual plateau in viscosity at high shear rates for $\sigma^* > 5$, qualitatively identical to the behaviour for FF springs.
The straightforward conclusion is that for shear flow, the precise form of the spring force law seems to be less qualitatively relevant than the average spring length, spring stiffness, extensibility and compressibility.

\newtxt{FENE-Fraenkel bead-spring chains also show the same flattening of the end-to-end distribution function seen in prior work \cite{hur2000brownian, li2000comparison, schroeder2005dynamics} irrespective of the value of $\sigma$ and $\delta Q$, as shown in \fref{fig: dist functions with Wi}.
It is somewhat remarkable that the flattening seems to be almost identical in a broad qualitative sense, with all the distributions largely collapsing by $Wi \sim \mathcal{O}(300)$.
This is evidently a universal feature of polymer models which does not depend upon the particular form of the spring potential.}

\begin{figure}[t]

    \centerline{
    \begin{tabular}{c}
        \includegraphics[width=10cm,height=!]{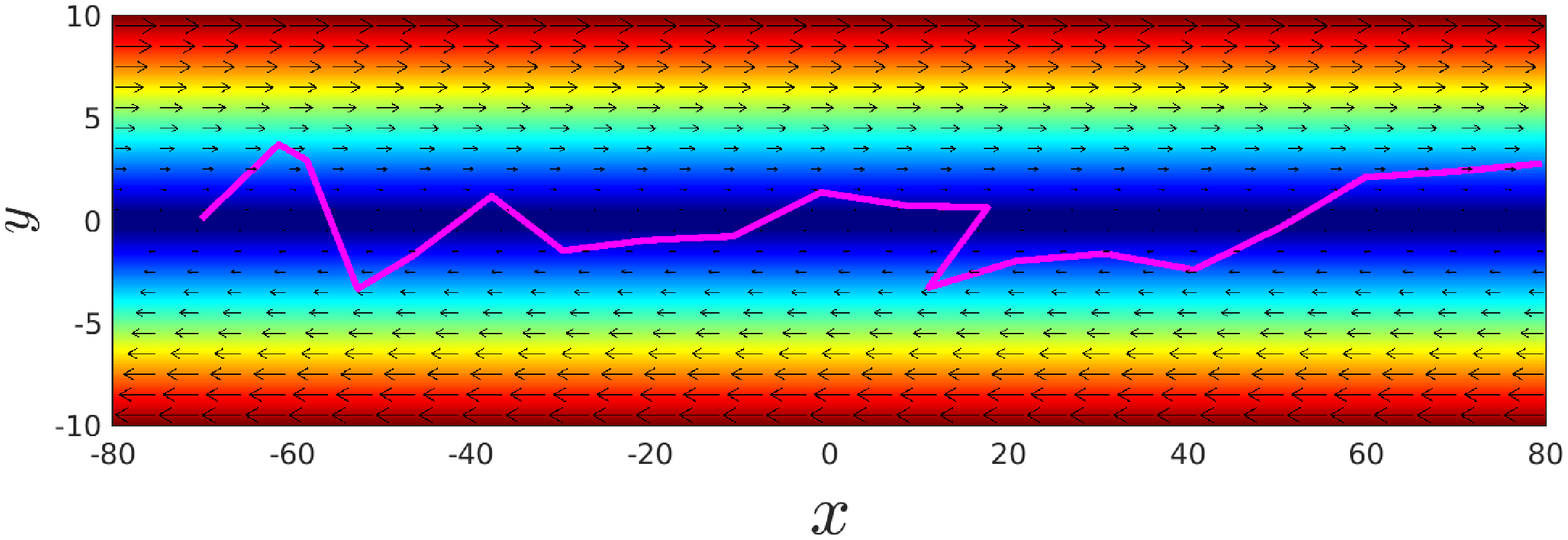}\\
        (a) \\
        \includegraphics[width=10cm,height=!]{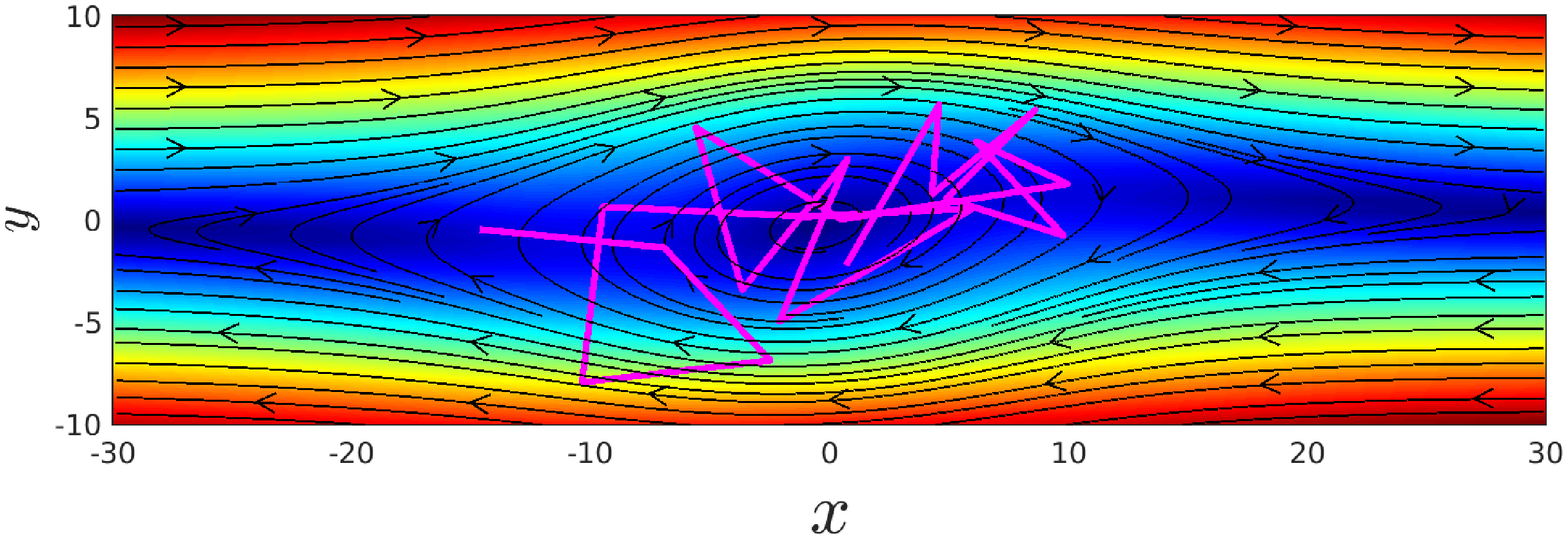}\\
        (b) \\
        \includegraphics[width=10cm,height=!]{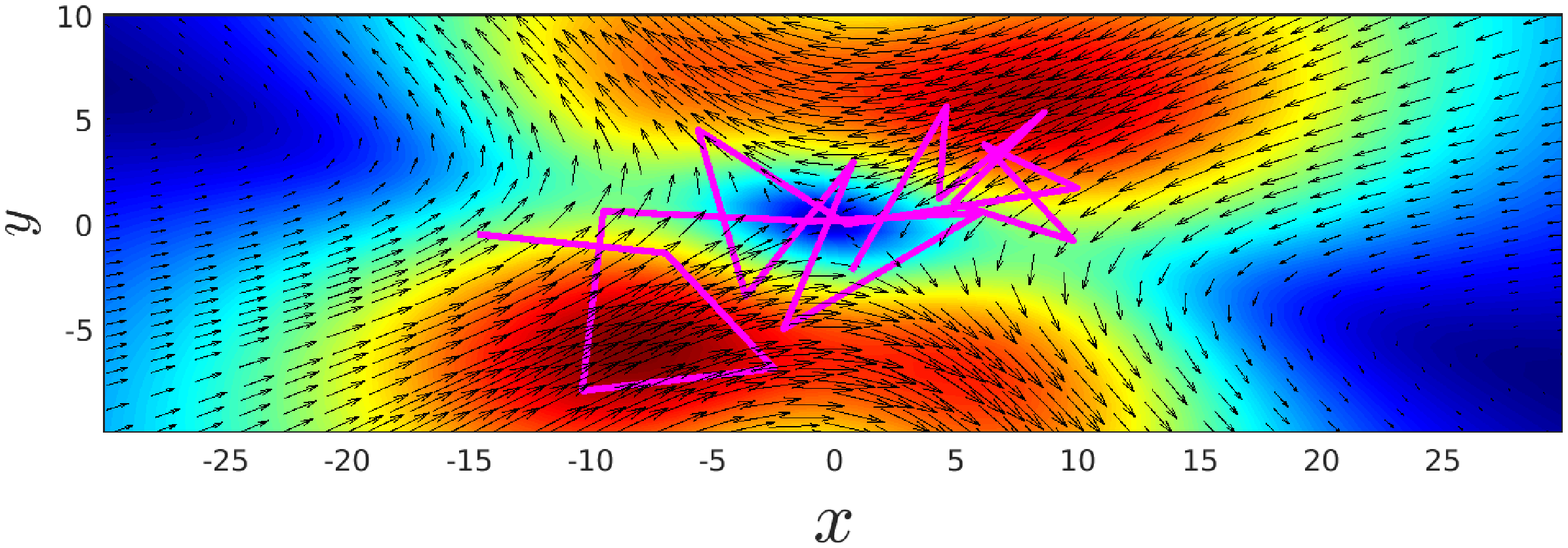}\\
        (c) \\
    \end{tabular}
    }
    \caption{Velocity field due to shear flow and HI for $N=20$, $\sigma^* = 9$, $\delta Q^* = 1$ and $Wi=3000$. (a) is with $h^* = 0$ (no HI), (b) and (c) are with $h^* = \sqrt{3} \chi$. (c) shows the flow only due to HI, whereas (a) and (b) show the total flow. Colour scales are not the same between figures. HI perturbation is averaged over several timesteps and trajectories.}
    \label{fig:HI visualisation}
\end{figure}

Returning briefly to \fref{fig:FENE-Fraenkel N20}, one obvious feature is in the stark difference in both $R^*_{g,x}$ compression and high-shear plateau as one switches on HI.
To investigate this effect, we have visualised the flow field caused by polymer deformation in \fref{fig:HI visualisation}~(a), (b) and (c).
First examining \fref{fig:HI visualisation}~(a), a $20$-bead chain with $\sigma^*=9$, $\delta Q^* = 1$ at $Wi=3000$ with $h^* = 0$, one can see the stretch of an example polymer trajectory in the flow direction caused by the background shear flow.
The direction of shear flow is shown by the streamlines, while the colour represents the magnitude of velocity at each point.
While the background shear flow does have a rotational component, causing tumbling, the elongational stretch leads to the classic increase in $R^*_{g,x}$.

We then turn to \fref{fig:HI visualisation}~(b) and (c), where HI has been switched on.
HI causes a change in the flow field at each point corresponding to:
\begin{equation}
    \bm{v}' = \left[ \bm{\Omega} \cdot \bm{F} \right]
\end{equation}
where $\bm{v}'$ is the velocity perturbation due to HI, and $\bm{\Omega}$ is the RPY tensor described in \eref{eq:HI tensor}.
In \fref{fig:HI visualisation}~(b), we have plotted the streamlines of $\bm{v}_\mathrm{shear flow} + \bm{v}'$, again with colour representing velocity magnitude at each point (and averaged over several relaxation times and trajectories).
The drastic change in flow field around the center of mass of the chain is immediately obvious, demonstrating clearly why HI is often referred to as `backflow'.
This effect is highlighted in \fref{fig:HI visualisation}~(c), which plots just $\bm{v}'$ without the background shear flow.
As the flow stretches out the chain, the spring and entropic forces pull it back towards its center of mass, causing a velocity disturbance which opposes the background shear flow.
It is apparently this effect which leads to a compression of bead-rod chains at high shear and with $h^*\gg0$, as the backflow disrupts elongation.
This process does not occur in bead-spring models, likely due to their stretchability, which causes the effective force of HI on each bead to be diminished due to increased distance between beads.
Also, as we will see in \fref{fig:FENE-Fraenkel EV changes}, the addition of a hard-core repulsion between beads disrupts this compression, likely due to increased bead-bead separation and less of a `coiled' shape for the chain at high shear.

The behaviour of the other material functions in shear flow, the normal stress coefficients $\Psi_1^*$ and $\Psi_2^*$, are displayed in \fref{fig:FF_Psi1_Psi2}.
The first normal stress coefficient $\Psi_1^*$ at high shear rates is given in \fref{fig:FF_Psi1_Psi2}~(a), where all $\sigma^*$ values display the expected $(-4/3)$ power-law slope without HI.
However, when HI is switched on, there is a slight increase in the power-law slope, but only, as we have come to expect, for the cases of $\sigma^* > 5$.
While a slope of $-4/3$ has been widely reported for a variety of models \cite{Fan:1985jk, Stewart:1972gt, Liu1989, Doyle1997, lyulin1999brownian, schroeder2005dynamics}, several bead-rod simulations show an $\approx -1.1$ power law slope with HI \cite{Liu2004, Aust1999, Moghani2017}.
This again suggests that the change in behaviour is linked to the rodlike characteristics of the underlying model.
We also briefly report results for the second normal stress coefficient $\Psi_2^*$ in \fref{fig:FF_Psi1_Psi2}~(b).
We do not report results without HI, since they are considerably smaller in magnitude and error bars overlap with $0$ for a wide range of shear rates.
Interestingly, $\Psi_2^*$ is negative for all $\sigma^*$, in contrast with the results for dumbbells where a crossover from positive to negative $\Psi_2^*$ was seen for sufficiently extensible springs \cite{pincus2020viscometric}.
We have attempted to calculate the magnitude of $\Psi_2^*$ for intermediate bead numbers, but results are difficult to interpret due to the large error bars (one needs significantly more sampling to observe a difference in $\Psi_2^*$ than $\Psi_1^*$).
It appears that either a sufficiently `spring-like' dumbbell, or a bead-rod chain with sufficient beads, gives a negative $\Psi_2$, while a bead-rod dumbbell ($N=2$) gives a positive $\Psi_2$.
We have discussed the history of $\Psi_2$ calculations in our previous paper \cite{pincus2020viscometric} - to summarise, there is still no clear experimental consensus on the correct sign of $\Psi_2$.

The orientation or extinction angles $\chi_G$ and $\chi_\tau$ from \eref{eq: chi G definition} and \eref{eq: chi tau definition} are plotted in \fref{fig:FF_chi_G_chi_tau}.
For $\chi_G$, all models show a roughly $(-1/3)$ power law slope at high shear rate shear rate irrespective of $\sigma^*$ or the presence of HI.
However, the intermediate-shear behaviour is quite varied, with low $\sigma^*$ and no-HI curves displaying a fairly monotonic decrease, while the $\sigma^* > 5$ cases with HI show an initial $\approx (-1/3)$ slope, then a slight leveling off before the final terminal slope.
This is likely related to the behaviour of the components of the gyration tensor - while we have not displayed this behaviour in figures, $R^*_{g,y}$ tends to decrease monotonically for all non-Hookean springs irrespective of HI or EV, so a decrease in $R^*_{g,x}$ at intermediate shear rates is reflected in a change in behaviour of $\chi_G$ for certain FENE-Fraenkel springs.
At higher shear rates, $R^*_{g,x}$ levels off as seen in \fref{fig:FENE-Fraenkel N20}~(c), but the continuing decrease of $R^*_{g,y}$ leads to a further decrease in $\chi_G$.

The stress tensor orientation $\chi_\tau$ displays similar behaviour, particularly at low and intermediate shear rates.
At high shear rates, the $\sigma^* > 5$ cases deviate from the other curves, with a plateau in $\chi_\tau$ for $\sigma^* = 9$.
This plateau implies that the terminal slopes of $\eta_p$ and $\Psi_1$ differ by only $\dot{\gamma}^{-1}$, meaning that $\eta_p/\Psi_1^* \dot{\gamma}^*$ is a constant.
For FENE-Fraenkel dumbbells, the divergence of $\chi_G$ and $\chi_\tau$ was a clear marker of the change from bead-spring to bead-rod behaviour, and this seems to hold somewhat for chains.
While several other authors have calculated $\chi_G$ and $\chi_\tau$ \cite{Doyle1997, Aust1999, li2000comparison, Kumar2004, Winkler2010}, the only direct calculation of the power-law slope appears to be from Schroeder et al. \cite{schroeder2005dynamics}, who found a $-0.43$ slope in $\chi_G$ from both bead-spring simulations and direct imaging of DNA chains, albeit with rather large error bars.

Calculations of the tumbling times of our FENE-Fraenkel chains was performed using the two methods detail in \aref{Appendix: Tumbling}.
Interestingly, we do not find any difference in the scaling of tumbling times for all FENE-Fraenkel springs in \fref{fig:FENE-Fraenkel N20}.
This is seen in \fref{fig:Tumbling FENE-Fraenkel N20}, which shows a nearly $-3/4$ power-law slope in tumbling time with shear rate irrespective of the inclusion of HI, or the stiffness of the springs.
The implication is that tumbling time is a universal function of the shear rate and overall polymer relaxation time, in contrast to what we have seen for other solution properties where the included non-linear physics can lead to drastic changes in behaviour.
Our finding of a $(-3/4)$ power-law slope agrees with the findings of Dalal and coworkers \cite{Dalal2014} for 100-bead Fraenkel-spring chains without HI, but their model changed to a $-1.1$ power law slope at high shear when HI (but not both EV and HI) was included.
They used only the end-on-end method for calculating tumbling times, and so it would be enlightening to re-run the analysis for their system using the cross-correlation method.

\newtxt{We also note that there are several measures of tumbling time which we have not explored in the current work, and which may hold significant insights into the behaviour.
Besides the rotation of the end-to-end vector \cite{Harasim2013} and the autocorrelation of the gyration tensor components \cite{Chen2013}, one can also determine the peak in the power spectral density \cite{Schroeder2005_characteristic}, the peak in the mean squared displacement of beads in the shear direction \cite{Usabiaga2011}, the autocorrelation function of Rouse vectors \cite{Usabiaga2011}, and time between successive conformations where the $x$- or $y$-component of the end-to-end vector is zero \cite{Winkler2006, Das2008}.
Theoretical results generally derive a scaling expression for one or more of these properties based upon an advection-diffusion balance of the components of the chain above and below the flow gradient plane \cite{teixeira2005shear, Usabiaga2011, Schroeder2005_characteristic, Dalal2012tumbling}, with findings of an expected $(-2/3)$ slope with HI.
Das and Sabhapandit \cite{Das2008} have in fact performed a full analytical calculation of the time between successive zero-crossings of the $x$-component of the end-to-end vector for a Rouse chain (still a very difficult problem despite the relative analytical tractability of the Rouse model), but their findings are difficult to compare with other calculations.
Therefore, while the literature on polymer tumbling in shear flow is relatively well-developed, it is unfortunately challenging to directly relate the findings of different authors.
This would likely be a fruitful avenue for future investigation, particularly if the findings could be linked to macroscopic rheological properties.}

We also wish to check how the qualitative behaviour of our FF springs changes with number of beads.
This is shown in \fref{fig:eta_sig9_sig0_N_sweep}, for both $\sigma^* = 0$ and $\sigma^* = 9$. 
For the FENE spring, $\sigma^* = 0$, we see an $\approx (-2/3)$ slope in viscosity with shear rate for all chain lengths at high shear.
However, the onset of shear thinning occurs at successively higher shear rates for longer chains, which matches with the later onset of shear thinning for more extensible FENE springs (larger $\delta Q^*$).
Additionally, the $N=50$ FENE chain with HI displays slight shear-thickening at intermediate shear rates.
Overall, this suggests that for FENE springs, qualitatively one finds that changing the extensibility is equivalent in many ways to changing the number of springs, once there are enough beads (in this case, $>20$) for sufficient degrees of freedom.

However, behaviour is somewhat different for bead-rod chains.
The $\sigma^* = 9$ dumbbell of course gives the expected $-1/3$rd power law slope in viscosity, and a monotonic increase in gyration tensor component in the flow direction due to alignment.
At higher bead numbers, we again have a delayed onset of shear thinning, but also the appearance of a high-shear plateau in viscosity, which becomes more pronounced as $N$ is increased. 
It's clear that the chain conformation observed in \fref{fig:HI visualisation} can only occur once there are sufficient numbers of chain links for significant compression to occur, and likely for significant backflow to be felt.
The compression in the flow direction appears more pronounced at higher bead numbers, as seen in \fref{fig:eta_sig9_sig0_N_sweep}~(b). 
Additionally, we do see a slight plateauing of the viscosity for $N=5$, even though there is no apparent flow-direction compression - this compression is not a necessary condition for a change in the shear-thinning exponent.
Finally, there appears to be no shear-thickening for bead-rod chains, although this may require far larger numbers of beads.
For example, a $350$-bead chain with HI and EV simulated using an extrememly efficient algorithm by Moghani and Khomami \cite{Moghani2017} shows slight hints of intermediate-shear thickening before  the terminal shear-thinning slope is reached, however it's unclear whether this is a real result or due to lack of sufficient sampling at low shear.

\subsection{Addition of EV and bending potentials to FF bead-spring chains}

As previously mentioned, EV potentials can be given in either `soft' or 'hard' forms.
The `hard' potential is the purely repulsive SDK (in \eref{SDK equation}, with $\varepsilon = 0$), where the force diverges to infinity at small bead separation, preventing overlap.
The `soft' potential is the Gaussian potential (given in \eref{Gaussian potential}, with $d^* = {z^*}^{1/5}$), which has a finite force at low bead separations, pushing beads apart but not completely preventing overlap.
For the Gaussian potential, there are well-developed theories of polymer swelling based on so-called two-parameter theory \cite{schafer2012excluded}, which says that the value of some property away from the theta-point (for simulations, this implies no EV potential) can be written as the value at the theta point multiplied by some function of the universal scaling variable $z$, the solvent quality \cite{Prakash2019review, Kumar2003}.
Specifically, it has previously be shown that by calculating $z^*$ as per \eref{z in terms of zstar}, one obtains universal predictions of swelling in the long chain limit (as $N \rightarrow \infty$), irrespective of the value of $d^*$ \cite{Sunthar2005parameterfree}.
Additionally, one finds a universal $-1/4$ power-law scaling in the viscosity with shear rate for $z\rightarrow \infty$ and $N \rightarrow \infty$ \cite{Kumar2004}, a result in alignment with renormalisation group calculations \cite{ottinger1989renormalization}. 
In summary, we generally define our EV potential not in terms of the direct microscopic details of the polymer, but instead the measured static or dynamic swelling at equilibrium, which should be independent of fine-grained details such as the exact form of the potential in the long-chain limit.

For purely repulsive hard-core potentials, as we have used here for the case of the SDK potential with $\varepsilon = 0$, correspond to the athermal limit, where $z \rightarrow \infty$ \cite{Hayward1999}.
As has been mentioned, a highly repulsive potential appears to remove the high-shear plateau in viscosity for bead-rod models \cite{Petera1999, Liu2004, Moghani2017}, as well as the compression at high shear \cite{Dalal2014}.
We wish to determine whether the `soft' and `hard' potentials are equivalent in shear flow, particularly for the less extensible FENE-Fraenkel springs.

\begin{figure}[t]
    \centerline{
    \begin{tabular}{c}
        \includegraphics[width=8.5cm,height=!]{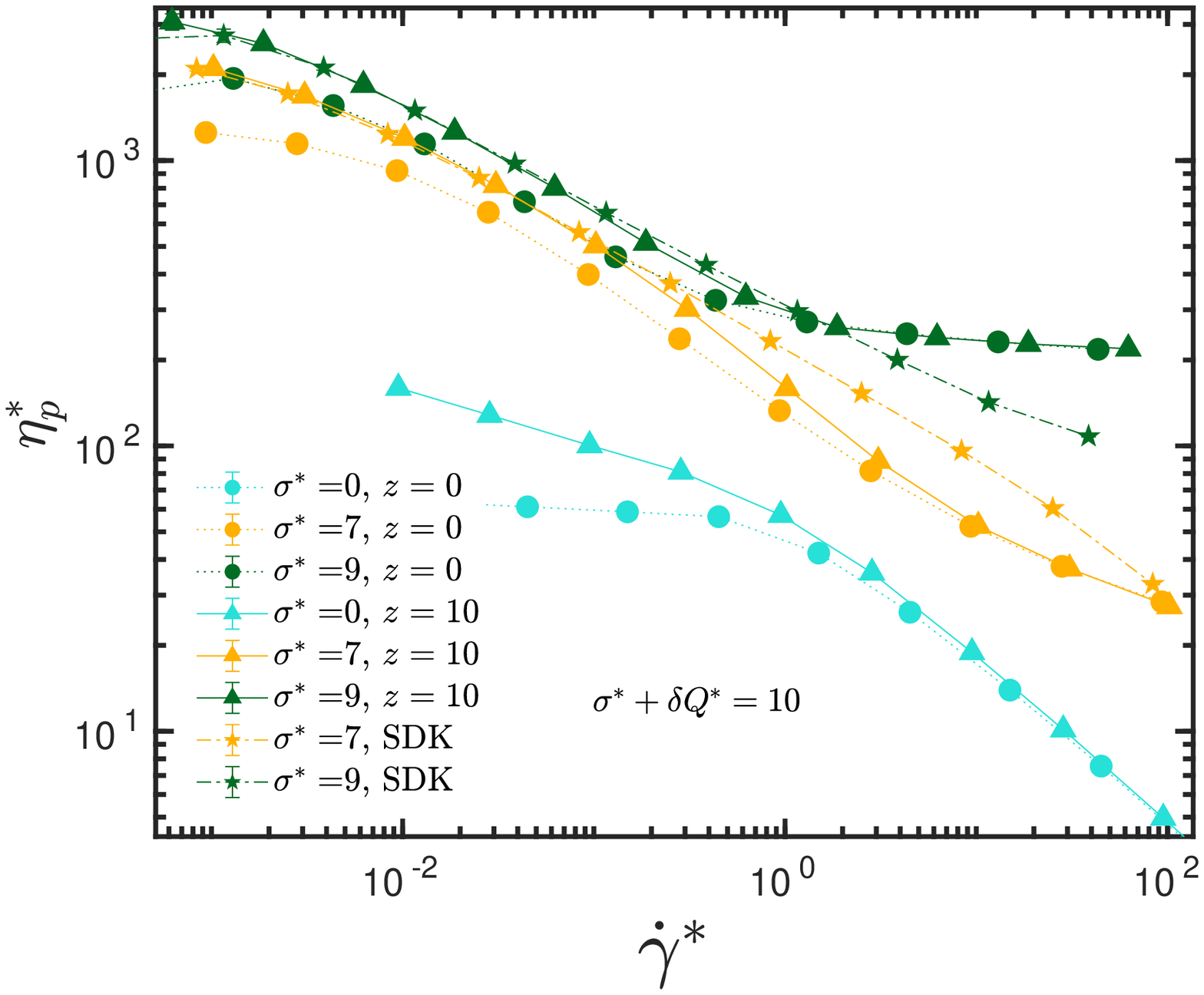}\\
        (a) \\
        \includegraphics[width=8.5cm,height=!]{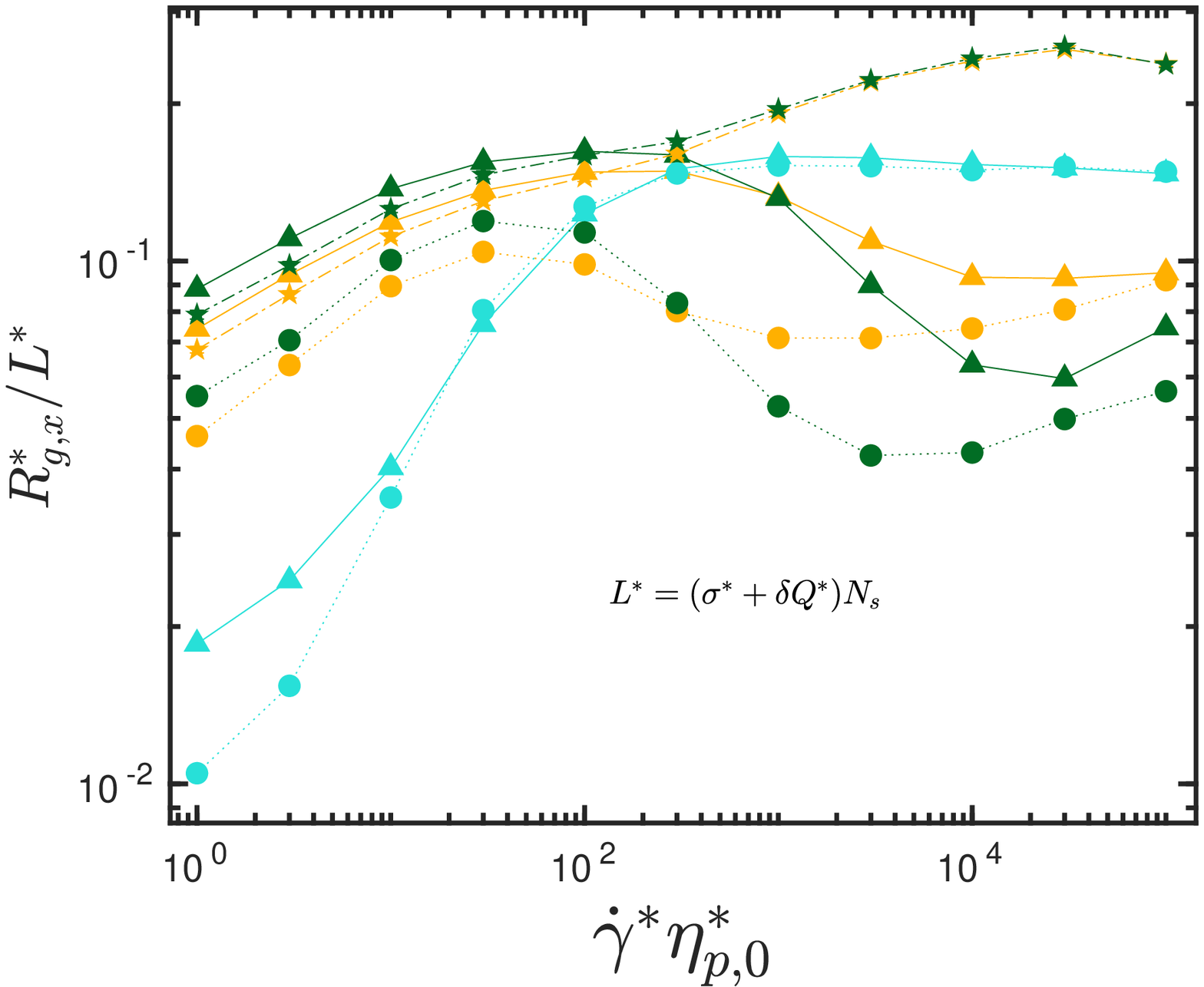} \\
        (b) \\
    \end{tabular}
    }
    \caption{Effects of inclusion of excluded volume potentials on the shear-behaviour of various FENE-Fraenkel springs. Plot (a) gives the non-dimensional viscosity against the non-dimensional shear rate. Plot (b) gives the $xx$-component of the radius of gyration tensor normalised by the total contour length $L^* = (\sigma^* + \delta Q^*) N_s$. All examples have $\sigma^* + \delta Q^* = 10$, $N=20$ and $h^* = \sqrt{3} \chi$. Circle symbols with dotted lines have no EV. Triangle symbols with solid lines have Gaussian potentials using parameters $z=10$, $z^* = z \chi^3/\sqrt{N}$ and $d^* = {z^*}^{1/5}$. Star symbols with dashed lines have an SDK potential using parameters $\varepsilon = 0$ and $d^* = 0.8 \sqrt{3} \chi$.}
    \label{fig:FENE-Fraenkel EV changes}
\end{figure}

Here we present results for $N=20$ bead chains with both Gaussian and purely repulsive SDK potentials, in order to study the differences for more rod-like models.
Both of these potentials cause swelling at equilibrium, seen in the zero-shear viscosity of \fref{fig:FENE-Fraenkel EV changes}~(a) and the equilibrium gyration radius in \fref{fig:FENE-Fraenkel EV changes}~(b).
In fact, the Gaussian potential leads to slightly more equilibrium swelling than the repulsive SDK potential.

\begin{figure*}[t]
    \centerline{
    \begin{tabular}{c c}
        \includegraphics[width=8.5cm,height=!]{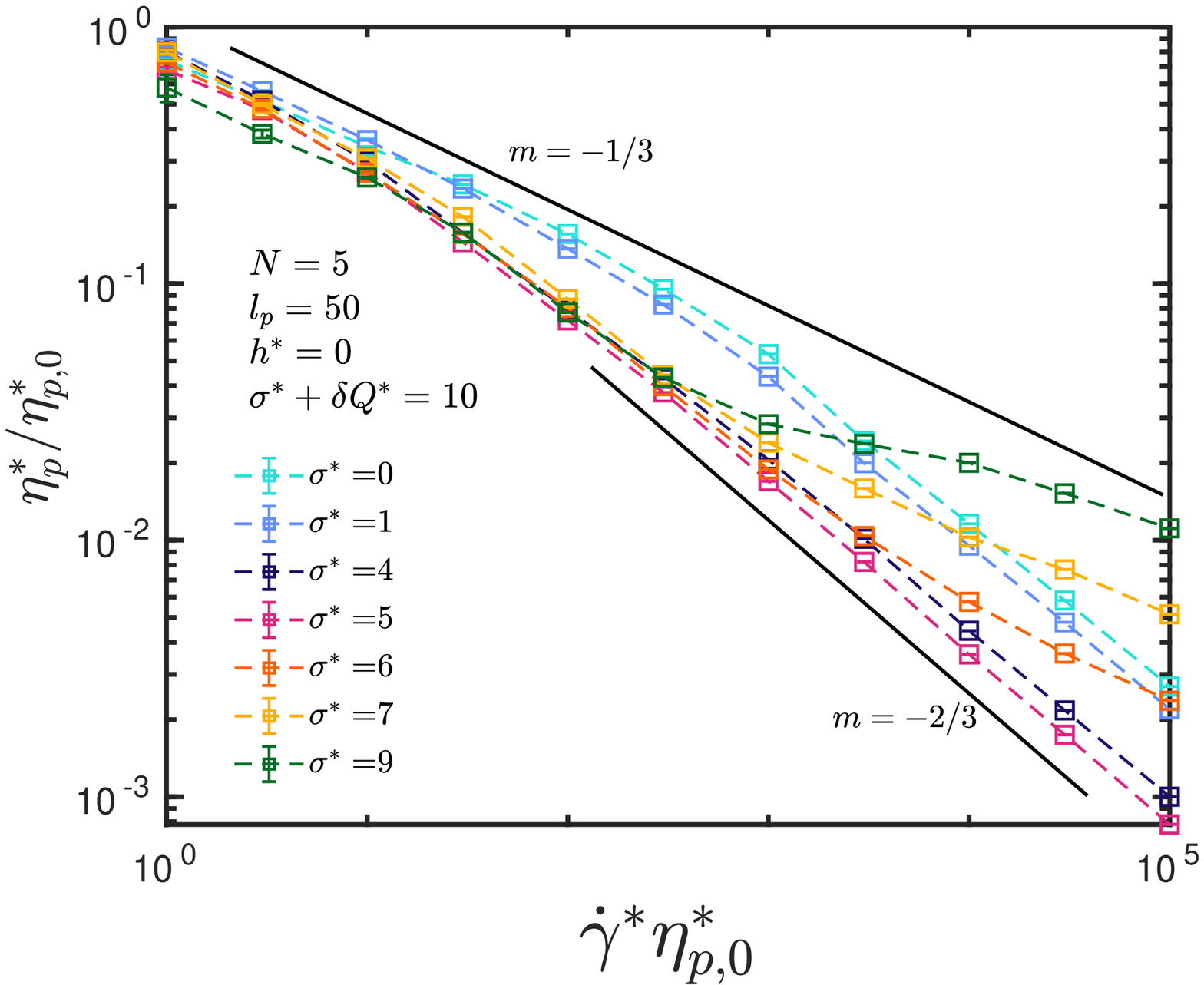}
        & \includegraphics[width=8.5cm,height=!]{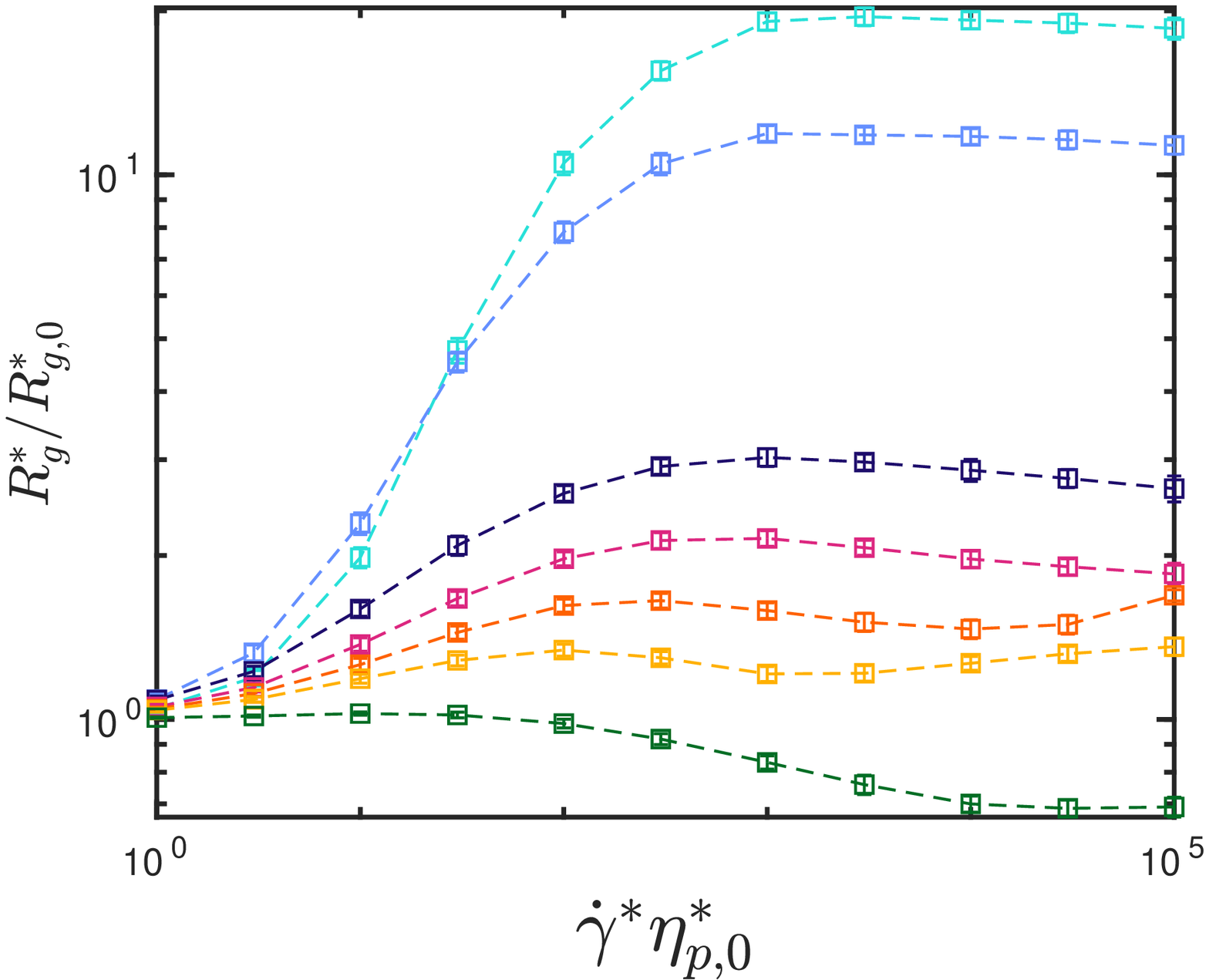} \\
        (a) & (b) \\
        \includegraphics[width=8.5cm,height=!]{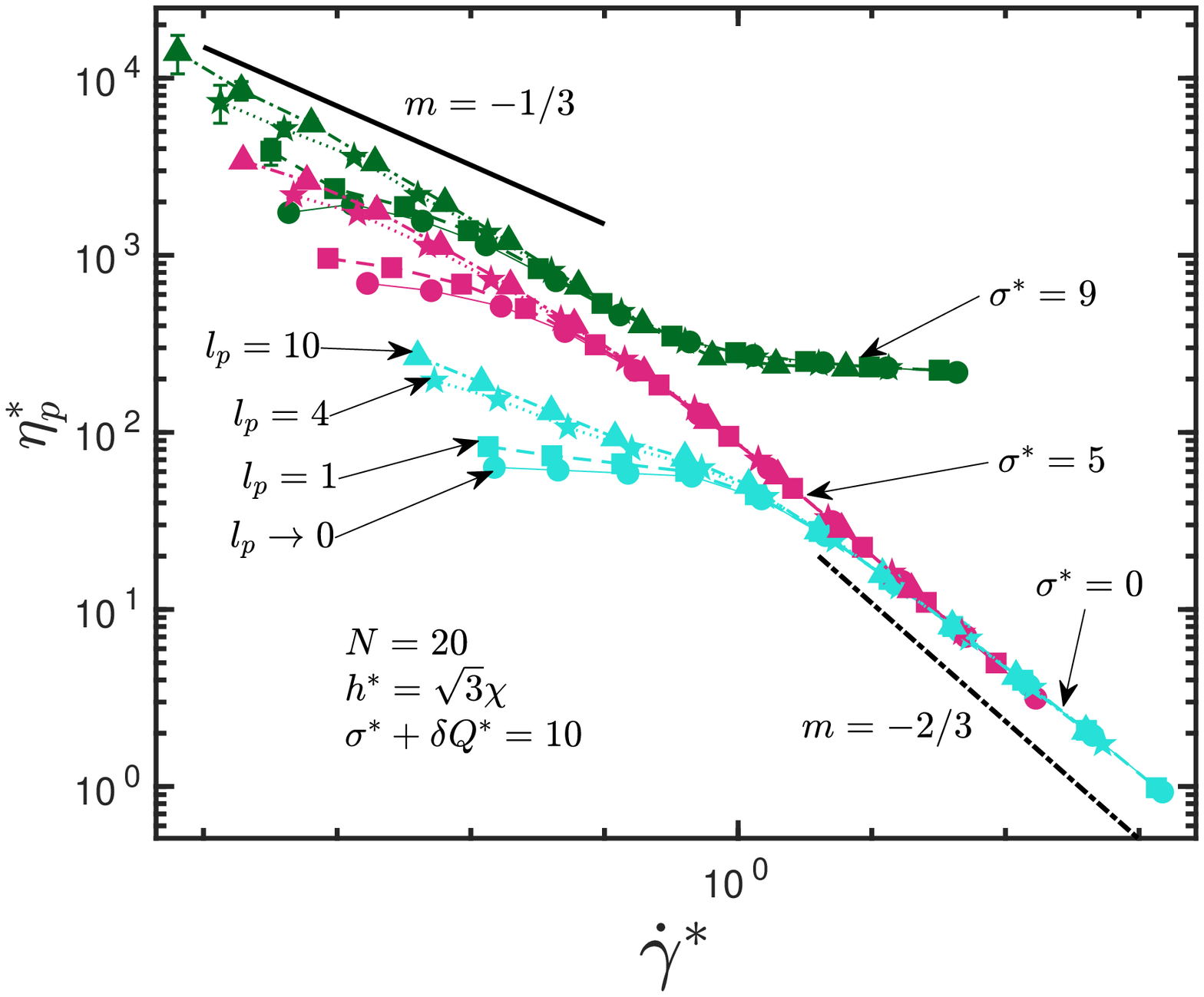} 
        & \includegraphics[width=8.5cm,height=!]{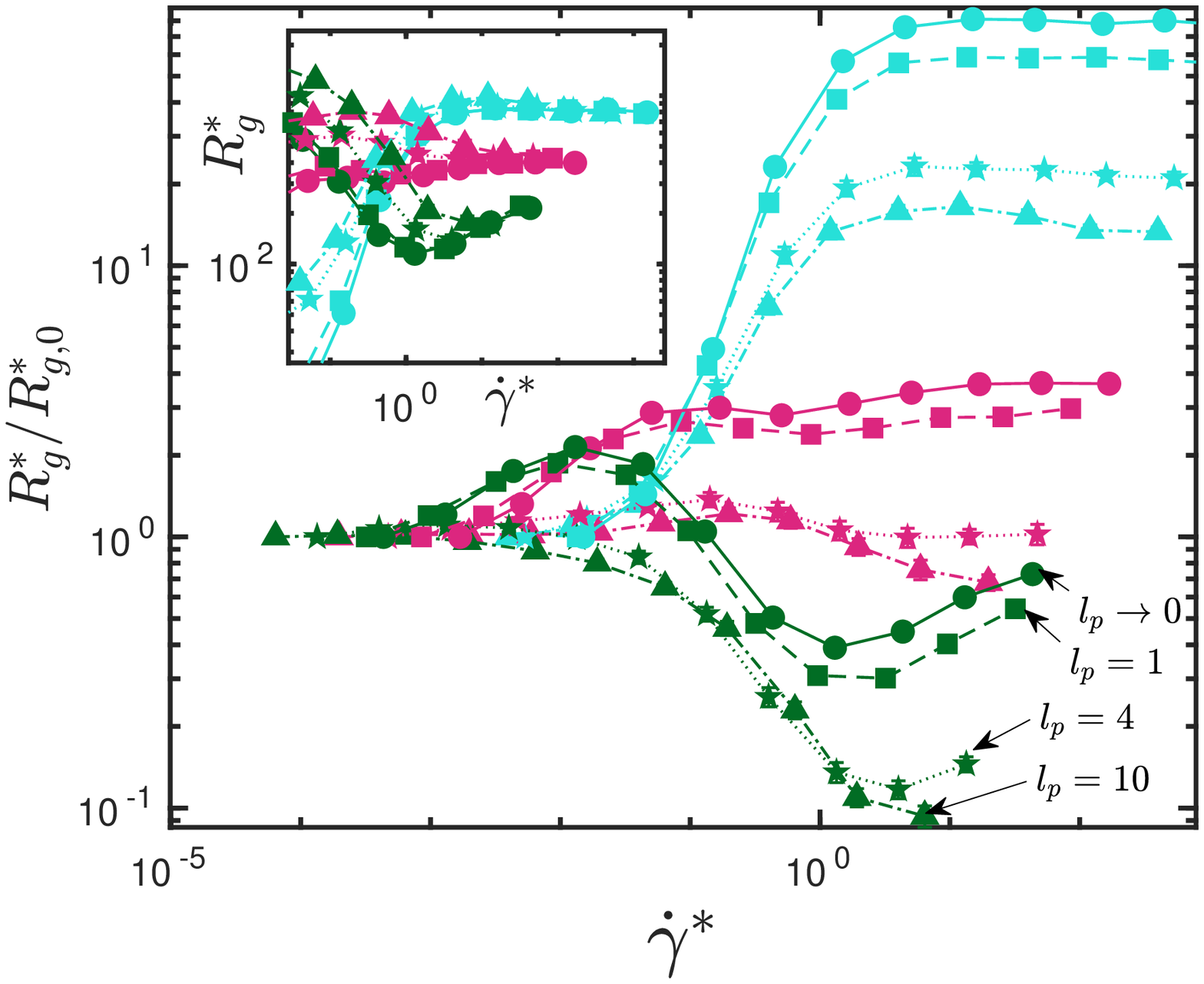} \\
        (c) & (d) \\
    \end{tabular}
    }
    \caption{FENE-Fraenkel springs of bead number $N=5$ for (a) and (b), as well as $N=20$ for (c) and (d), with a bending potential. In all cases, $l_p$ is defined in terms of the number of links, i.e. for $l_p = 50$ in (a), we must travel 50 springs along the chain in order for the correlations in segmental unit vectors to have decayed by $63.2\%$. As before, we have $\sigma^* + \delta Q^* = 10$. For (a) and (b), which display the normalised polymer viscosity and normalised gyration radius respectively against Weissenberg number, the zero-shear viscosity is calculated via \eref{eq: eta_0_analytical_equation}. Plots (c) and (d) give the dimensionless polymer viscosity and normalised gyration radius respectively against the dimensionless viscosity. Inset to plot (d) instead displays the radius of gyration without the zero-shear normalisation. $m=-1/3$ and $m=-2/3$ lines are guides for the eye and do not imply exact terminal slopes. Where not shown, error bars are smaller than symbol size.}
    \label{fig:bending potential results}
\end{figure*}

While the two EV potentials have a similar qualitative effect near equilibrium, as the shear rate is increased there are considerable differences.
The most obvious effect is the convergence in viscosity for the no-EV and Gaussian-EV cases, leading to a similar terminal shear-thinning slope for the $\sigma^* = 0$ (FENE) case, or a high-shear plateau for the $\sigma^* = 7$ and $\sigma^* = 9$ cases.
This is also somewhat apparent in the $R^*_{g,x}$ scaling, which shows similar qualitative behaviour with shear rate for all three $\sigma^*$ values with and without Gaussian EV.

However, the SDK potential, which prevents bead overlap, leads to entirely different qualitative results.
The high-shear plateau is entirely absent, and $R^*_{g,x}$ monotonically increases to a similar value for both $\sigma^* = 7$ and $\sigma^* = 9$.
Although these results are given for only $N=20$, other authors find similar results for longer chains \cite{Dalal2014}, with the longest available in the literature being Moghani and Khomami's $350$-bead-rod chains with full HI and repulsive hardcore EV \cite{Moghani2017} (which shows no compression at high shear rates, and an $\approx -0.28$ power-law slope in viscosity).
There are several tentative conclusions which can be drawn from this result.
Firstly, the swelling at equilibrium due to EV does not necessarily predict the shear-flow behaviour for finite chains (for which universal scaling results don't necessarily apply).
There is a clear distinction between potentials which cause chain swelling but allow bead overlap, and those which cause the same swelling but do NOT allow bead overlap.
Secondly, as seen by the similar terminal shear-thinning slopes for $\sigma^* = 7$ and $\sigma^* = 9$, and particularly the convergence of $R^*_{g,x}$, implies that a hard EV potential in some sense `takes over' from the spring potential at high shear rates.
An SDK potential with $d^* = 0.8 \sqrt{3} \chi$, as in \fref{fig:FENE-Fraenkel EV changes}, almost represents a `pearl-necklace' model, where beads exclude each other on roughly the range given by half their average spring length.
Finally, although not shown here, we note that these effects diminish as $d^*_\mathrm{SDK}$ is reduced (the effective range of interactions), even for 100-bead chains as previously demonstrated by Dalal et al. \cite{Dalal2014}.
A more detailed study could use an SDK or LJ potential with attractive and repulsive components, carefully determine the $\varepsilon$ which represents a $\theta$-solvent \cite{santra2019universality}, and then compare results with Gaussian EV in the long-chain limit at the same solvent quality $z$.

\newtxt{This shows that at low shear rates, there is universal behaviour, but beyond a certain point it does matter what form of potential is used. 
Simple blob scaling arguments suggest that the critical Weissenberg number when flow penetrates the `Pincus blob' \cite{Pincus1976} is approximately \cite{sasmal2017parameter} $Wi_\mathrm{c} = N_k^{3 \nu}$, where in this case $N_k \approx 20$ for a 20 bead-rod chain and $\nu \approx 0.6$.
Therefore, $Wi_\mathrm{c} \sim \mathcal{O}(10^2)$, which is approximately the $\dot{\gamma}^* \eta_{p,0}^*$ at which the two potentials diverge in \fref{fig:FENE-Fraenkel EV changes}.
In other words, at low shear rates, the flow has not penetrated the Pincus blob, and hence the exact form of the potential is not important, as molecular details do not matter.
However, at higher shear rates, universality breaks down and the form of the potential is important - one must be more careful to make a physically relevant choice, for which soft-core potentials are rarely appropriate.}

\newtxt{In terms of the reason for the divergence of behaviour, we suspect the difference is due to a `soft' versus a `hard' potential.
This is similar to the difference between a Hookean and a FENE spring, where the Hookean spring, while it does cause a larger force for higher shear rates, can be endlessly stretched, leading to a constant viscosity.
A FENE spring has a finite extensibility, which leads to a decrease in viscosity.
It seems likely that a similar effect is at play with the two potentials, where at a certain shear rate the flow `overcomes' the softer Gaussian potential and still pushes beads together, while the `hard' potential cannot be endlessly compressed irrespective of the flow strength.}

We now arrive at our final piece of qualitative physics, the bending potential, which represents polymer semiflexibility.
This has been done by other authors in BD simulations \cite{lyulin1999brownian, Ryder2006, Jain2009, Sendner2009, Dalal2012regimes, Dalal2014}, often alongside an additional torsional potential and rodlike bead-bead links.
Generally, this leads to a $(-1/3)$ slope in the viscosity at high shear rates for very strong bending potentials, as expected for `stiff' polymers.
Additionally, the high-shear compression in the flow direction appears to be lessened through introducing semiflexibility \cite{Dalal2012regimes, Dalal2014}.

On the other hand, semi-analytical models enforce the semiflexibility directly through an averaged constraint on the segmental (or tangent vector) correlation along the backbone \cite{Winkler1994, harnau1995dynamic}.
However, to be analytically tractable, these models often relax the constraints on the segmental stretch and total contour length, leading to a chain which can extend and contract in response to external forces.
For example, the model of Winkler \cite{Winkler2006} gives a $(-2/3)$ power law slope in viscosity at high shear irrespective of the underlying semiflexibility of the chain, somewhat in contrast to expectations for highly inflexible polymers.
\newtxt{Therefore, although the spring potential and bending potential are often changed in proportion to each other when simulating a real polymer chain \cite{Saadat2016}, it is insightful to study the independent effects of the two potentials to clarify the accuracy of previously-applied models.}

In \fref{fig:bending potential results}, we present results with a very stiff bending potential, where $l_p$ is given in units of the spring length.
For example, for $N = 5$ and $l_p = 50$, we have $N_\mathrm{K,s} = 4/(100) = 0.04$, leading to $C = 49.8$ as per \eref{bending potential expression SK}.
Focusing firstly on the viscosity for \fref{fig:bending potential results}~(a) with $N=5$, $l_p = 50$ and no HI, we see a clear initial $(-1/3)$ power-law slope at intermediate shear rates, as one would expect for a highly `stiff' bending potential.
At higher shear rates, the more `spring-like' FENE-Fraenkel chains again give a $(-2/3)$ power-law slope in the viscosity, while the $\sigma^* > 5$ chains return to an approximately $(-1/3)$ power-law slope.
Interestingly, the chains all seem to display fairly similar qualitative behaviour in the extension $R_g^*$ in \fref{fig:bending potential results}~(b), and there is no clear difference in the behaviour from $\sigma^* =5$ to $\sigma^* = 6$.
The $\sigma^* = 9$ chain in fact compresses in shear flow relative to its equilibrium length, likely due to the shear flow deforming the semiflexible chain into a hairpin-like configuration observed in both experimental and simulation studies \cite{Harasim2013, Nikoubashman2017}.

The behaviour for a range of smaller $l_p$ values is displayed in \fref{fig:bending potential results}~(c) and (d), using $N=20$ bead chains and $h^* = \sqrt{3} \chi$.
At low shear rates, with $l_p \ge 4$, all chains display an $\approx -1/3$ power-law slope in viscosity.
However, at higher shear rates, the behaviour collapses onto the same curve irrespective of the bending stiffness, following the results without a bending potential almost exactly.
This occurs for both the viscosity and radius of gyration, as can be seen in the inset to \fref{fig:bending potential results}~(d).
This behaviour is remarkably similar to that previously observed for Gaussian EV, where the intermediate-shear behaviour is altered, but results collapse at a certain $\dot{\gamma}^*$.

\newtxt{\subsection{Semi-Quantitative Comparisons}}
\label{section: quantitative}

\begin{figure}
    \centering
    \includegraphics[width=8.5cm,height=!]{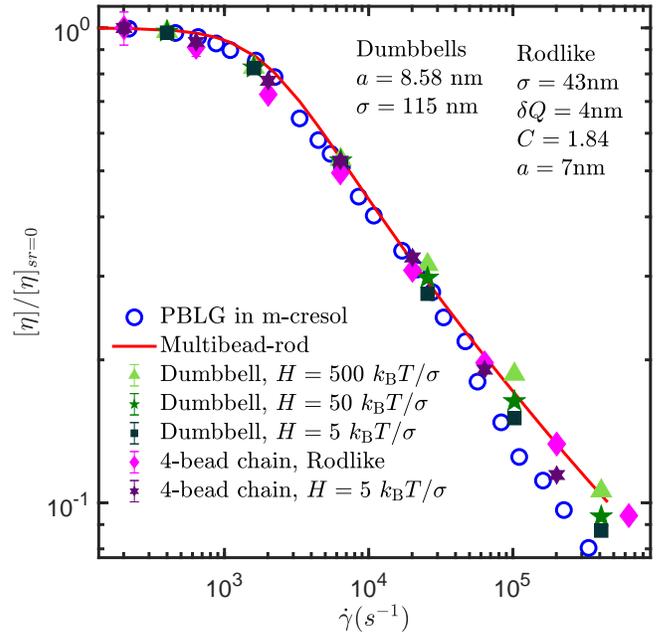}
    \caption{\newtxt{Semi-quantative comparisons with experimental data for PBLG from Yang \cite{yang1958non}, which has a contour length of $140$ nm and a persistence length of around $90$ nm \cite{yang1958non, Mead1990}. The dumbbell and multibead-rod data is generated as described in a previous paper \cite{pincus2020viscometric}. The `Rodlike' 4-bead chain has parameters as described in \sref{section: real polymer methods}, with the exact parameters given in the top-right of the figure. The other 4-bead chain uses the same parameter selection as for the dumbbells, resulting in somewhat lower effective $H$. The hydrodynamic bead radius of $a = 7$ nm has been chosen to match the experimental data.}}
    \label{fig: PBLG comparisons}
\end{figure}

\newtxt{As a final point, we wish to compare our model with data for a real polymer, PBLG, which has a contour length of $140$ nm and a persistence length of around $90$ nm \cite{yang1958non, Mead1990}.
We have compared results with those from a previous paper using FENE-Fraenkel dumbbell and multibead-rod models \cite{pincus2020viscometric}, as shown in \fref{fig: PBLG comparisons}.
In the current work, we have extended the dumbbell model to a 4-bead chain.
In all cases, we have selected the hydrodynamic bead radius to roughly match experimental results, hence the semi-quanitative moniker.
There are two different ways we have selected spring parameters for the 4-bead chain.
The first is to follow the method in \sref{section: real polymer methods}, where the exact parameters used for each spring are shown in the top right of \fref{fig: PBLG comparisons}.
The second is to simply apply the dumbbell parameters to the 4-bead model, setting $H = 5 k_\mathrm{B} T/\sigma$, while keeping $\sigma + \delta Q = l_s$.
All the 4-bead rod models have an effective bead radius $a = 7$ nm.}

\newtxt{As can be seen, the `rodlike' parameters selected using the method outlined in \sref{section: real polymer methods} quite closely follow the rigid multibead-rod results, and scale similarly to the PBLG results at low shear rates.
Interestingly, simply setting $H = 5 k_\mathrm{B} T/\sigma$ gives very similar results up to quite high shear rates, and seems to be a slightly superior description of the experimental results at high shear rates.
This is despite the $H = 5 k_\mathrm{B} T/\sigma$ being chosen in such a way that $\langle Q^2 \rangle \ne \langle R^2_\mathrm{segment} \rangle$ at equilibrium.
This is of course speculation based on limited experimental data, but it seems possible that the PBLG molecule is more flexible or extensible than either the WLC or multibead-rod model would suggest, such that the extensibility of the FENE-Fraenkel spring leads to a more faithful depiction of behaviour at high shear rates.
To summarise, the straightforward conclusion is that when $l_p > l_s$ and $\sigma \approx l_s$, the exact form or value of the spring constant is less important than the `rodlike' nature of the spring.}

\section{Conclusions}
\label{Conclusions}

We began by presenting the full range of behaviour seen in previous experimental and theoretical studies in \fref{fig:Experimental schematic} and \fref{fig:Simulations schematic} and describing a model which could supposedly reproduce this range of behaviour.
The changes in simulated polymer solution properties as a function of model parameters were then studied in detail, highlighting how each piece of physics affects the observed rheology.
To conclude, let us refer back to \fref{fig:Experimental schematic} and \fref{fig:Simulations schematic}, and demonstrate that by appropriately selecting parameters, we can indeed qualitatively recover the full range of expected changes in viscosity with shear rate using our model.
In other words, we wish to show that given some experimental or simulated measurements of dilute polymer solution rheology in shear flow (such as a curve of viscosity versus shear rate), as well as basic details about the polymer microstructure, we can choose our FENE-Fraenkel spring parameters, as well as bending potential, HI, and EV, in order to qualitatively match that behaviour. 

We begin with \fref{fig:Simulations schematic}, pointing out which of our simulations correspond to each curve.
Note that our list numbering refers to the curves in \fref{fig:Simulations schematic}, such that, for example, (a) is qualitative comparisons with FENE chains and HI: \cite{kishbaugh1990discussion}:
\begin{enumerate}[(a)]
    \item The shear-thickening is observed for sufficiently extensible FENE springs with HI, as shown in \fref{fig:N20_Hook_FENE} and \fref{fig:eta_sig9_sig0_N_sweep}. 
    \newtxt{Strong EV tends to counteract this effect, since it increases the zero-shear viscosity.}
    \item The high-shear plateau seen for bead-spring models is again recovered for sufficiently extensible FENE (or Hookean) springs with Gaussian EV, as seen in \fref{fig:N20_Hook_FENE}~(c).
    \item The $\approx (-2/3)$rds power-law slope in viscosity with shear rate is found for any sufficiently compressible bead-spring model, as demonstrated most clearly in \fref{fig:FENE-Fraenkel N20} and \fref{fig:MS WLC visc examples}. This may even be the case when a strong bending potential is used, as suggested by \fref{fig:bending potential results}.
    \item The characteristic $(-1/3)$rd power-law slope in viscosity found in rodlike models is approached for sufficiently incompressible springs, seen here quite clearly for the dumbbell results of \fref{fig:MS WLC visc examples}~(a). A strong bending potential also seems to give an intermediate $(-1/3)$rd slope in the viscosity as per \fref{fig:bending potential results}, however the terminal slope may instead correspond to the form of the spring potential rather than bending potential.
    \item The high-shear plateau for bead-rod chains is seen for sufficiently incompressible bead-spring chains, as in \fref{fig:FENE-Fraenkel N20}~(a) and \fref{fig:MS WLC visc examples}~(b).
\end{enumerate}

Furthermore, we show that moving from the bead-spring-chain to bead-rod-chain leads to a plateauing of the viscosity at high shear rates, as well as a compression in the flow direction.
This crossover occurs when $\sigma^* > \delta Q^*$ (or $\sigma^* > \delta Q^*/2$ for MS-WLC-Fraenkel spring), suggesting that the compressibility of a force law gives it either bead-rod or bead-spring-like behaviour.
We also found that `hard-core' and `soft-core' EV potentials give considerably different results at finite shear rates for finite chain lengths, and the effects of a strong bending potential depend heavily on the form of the force law used to link beads.

We also briefly mention some features of the experimental results which can be matched onto our models, roughly corresponding to the physics expected to be important in those real polymer solutions.
However, this is largely qualitative - ideally we would seek to develop a systematic method to obtain quantitative predictions of  experimental behaviour in future work.
The following features were present in \fref{fig:Experimental schematic} which can be seen in our FF-spring-chain simulations (again, letters represent curves in the original figure):
\begin{enumerate}[(a)]
    \item The extremely high-molecular-weight polystyrene in a theta solvent should in theory be modelled by a highly extensible bead-spring-chain with HI but no EV (or for a value of $\varepsilon$ corresponding to a $\theta$-solvent). This could correspond to the behaviour of Hookean or $\delta Q^* \gg 0$ FENE chains seen in \fref{fig:N20_Hook_FENE}, making the speculative assumption that the shear-thickening regime has not been reached.
    \item This polymer solution uses the same polystyrene molecule as in (a), but with a higher solvent quality, leading to a $\approx -0.1$ power-law slope in viscosity. This somewhat corresponds to a highly extensible FENE chain with finite $z$, as per \fref{fig:N20_Hook_FENE}~(c). For example, the Hookean $h^* = 0$, $z = 2$ curve has an initial gradient of $\approx -0.1$, which is seen at intermediate shear rates before the $(-2/3)$ slope due to finite extensibility.
    \item For a shorter polystyrene chain in a close-to-theta solvent, we expect a FENE spring-chain with some relatively small $\delta Q^*$ to be a reasonable model, leading directly to a $(-2/3)$ slope in viscosity at high shear rates, as seen in all of our highly-compressible FF springs.
    \item The three DNA chains of 24 kbp, 48.5 kbp and 165.6 kbp show slight differences in the shear-thinning exponent, as well as differences in the onset of shear thinning. Qualitatively, we have seen that all our models appear to have a later onset of shear-thinning as the chain extensibility is increased, as in \fref{fig:N20_Hook_FENE} and \fref{fig:eta_sig9_sig0_N_sweep}. The shear rates reached are also not particularly large - it is possible that the chains are still in the crossover region between zero-shear and high-shear behaviour, leading to differences in slopes. One might also speculate that the semiflexibility of DNA causes two different slopes at intermediate and high shear rates, as in \fref{fig:bending potential results}, although this is again purely speculative.
\end{enumerate}

Beyond presenting a unified model for examining the properties of previous bead-rod chains and spring-force laws in detail, we hope that this work will be useful in the future development of multiscale modelling approaches.
While several authors have developed models which are able to represent a section of polymer chain on many length scales at equilibrium \cite{Saadat2016, Koslover2014, underhill2006alternative, underhill2004coarse}, our current results suggest that this is not sufficient to ensure correct reproduction of properties at finite shear.
In future studies, we hope to present a multiscale modelling procedure based upon this FENE-Fraenkel spring \newtxt{which can represent both very short rodlike polymer segments as well as longer, coil-like polymers}.
In this way, one may be able to represent both a short section of semiflexible polymer chain, as well as a very large segment of a more flexible polymer, using the same continuous fine-graining procedure.

\begin{acknowledgments}

This research was supported by an Australian Government Research Training Program (RTP) Scholarship.
Simulations were performed on the MonARCH HPC Cluster through the Monash eResearch Centre and eSolutions-Research Support Services, as well as the MASSIVE HPC facility (www.massive.org.au), and the Gadi supercomputing cluster through resources and services from the National Computational Infrastructure (NCI), which is supported by the Australian Government.
The authors would also like to thank Aritra Santra, Dominic Robe and R. Kailasham for helpful advice on the efficient implementation of several algorithms used in the paper.

\end{acknowledgments}

\section*{Data Availability Statement}

The data that support the findings of this study are available from the corresponding author upon reasonable request.
The code used to generate the data in this paper can be retrieved at gitlab.erc.monash.edu.au/jagadeeshan-molecular-rheology/single-chain.

\appendix

\section{Bending Force}
\label{Appendix Bending Force}

\begin{figure}[!ht]
    \centering
    \includegraphics[width=8cm,height=!]{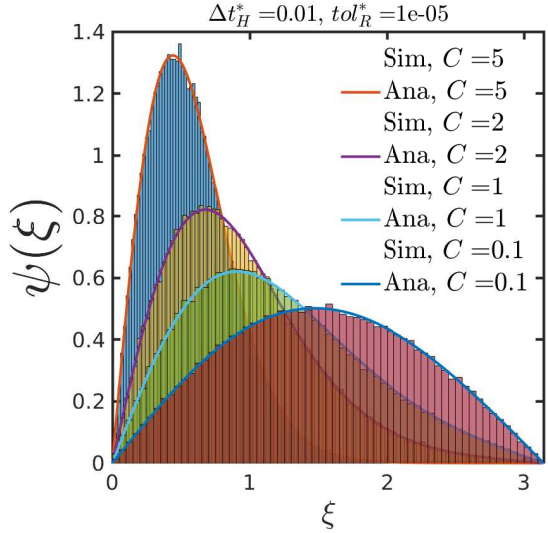}
    \caption{Comparison of analytical and simulated bending angles $\xi$ against bending stiffness value $C$.}
    \label{fig: C curves no HI 3 beads}
\end{figure}

If we denote the force on bead $\mu$ due to bending potential $\nu$ as $\bm{F}^\mathrm{b}_{\mu, \nu}$, then the total force on bead $\mu$ is:
\begin{equation}
    \bm{F}^\mathrm{b}_{\mu} = \bm{F}^\mathrm{b}_{\mu, \mu} + \bm{F}^\mathrm{b}_{\mu, \mu-1} + \bm{F}^\mathrm{b}_{\mu, \mu+1}
\end{equation}
since a bead feels effective forces from both its `own' included angle, as well as that of adjacent beads (see \fref{fig:angles schematic}).
Further defining $\bm{u}_\mu$ as the unit vector from bead $\mu+1$ to bead $\mu$ with length $Q_\mu$, the overall force $\bm{F}^\mathrm{b}_{\mu}$ can be shown to be:
\begin{widetext}
\begin{multline}
    \bm{F}^\mathrm{b}_\mu = \frac{\partial \phi_{\mathrm{b}, \mu}}{\partial \theta_{\mu}} \frac{1}{\sin{\theta_\mu}} \left[\frac{1}{Q_\mu} \left( \bm{u}_\mu \cos{ \theta_\mu}  - \bm{u}_{\mu-1} \right) + \frac{1}{Q_{\mu-1}} \left( - \bm{u}_{\mu-1} \cos{ \theta_\mu}  + \bm{u}_{\mu} \right) \right]\\
    + \frac{\partial \phi_{\mathrm{b}, \mu-1}}{\partial \theta_{\mu-1}} \frac{1}{\sin{\theta_{\mu-1}}} \left[\frac{1}{Q_{\mu-1}} \left( - \bm{u}_{\mu-1} \cos{\theta_{\mu-1}}  + \bm{u}_{\mu-2} \right) \right] + \frac{\partial \phi_{\mathrm{b}, \mu+1}}{\partial \theta_{\mu+1}} \frac{1}{\sin{\theta_{\mu+1}}} \left[\frac{1}{Q_{\mu}} \left(\bm{u}_{\mu} \cos{\theta_{\mu+1}}  - \bm{u}_{\mu+1} \right) \right] \\
\end{multline}
for an arbitrary bending potential $\phi_{\mathrm{b}, \mu}$, where $\theta_{\mu}$ is the included angle at bead $\mu$ (note that this means $\mu$ begins at 2).
For our specific form of the bending potential given in \eref{eq:bending potential}, we have that:
\begin{equation}
    -\frac{\partial \phi_{\mathrm{b},\mu}}{\partial \theta_{\mu}} = -k_\mathrm{B} T C \sin{\theta_\mu}
\end{equation}
and so therefore, we can see that:
\begin{multline}
\label{Forces for 1-cos potential}
    \frac{\bm{F}^\mathrm{b}_\mu}{k_\mathrm{B} T C} = \left[\frac{1}{Q_\mu} \left(  \bm{u}_\mu \cos{\theta_\mu}  - \bm{u}_{\mu-1} \right) + \frac{1}{Q_{\mu-1}} \left( - \bm{u}_{\mu-1} \cos{\theta_\mu}   + \bm{u}_{\mu} \right) \right]\\
    + \left[\frac{1}{Q_{\mu-1}} \left( - \bm{u}_{\mu-1} \cos{\theta_{\mu-1}}  + \bm{u}_{\mu-2} \right) \right]  + \left[\frac{1}{Q_{\mu}} \left( \bm{u}_{\mu} \cos{\theta_{\mu+1}}  - \bm{u}_{\mu+1} \right) \right] \\
\end{multline}
\end{widetext}
Note that this expression is slightly different from that given by Saadat and Khomami \cite{Saadat2016}, which apparently contains a minor typo.
We have verified that our expression gives the correct equilibrium distribution of angles for a 3-bead trumbbell, which can be determined analytically from the bending potential.
The analytical expression has the form:
\begin{equation}
    \psi_{eq}(\theta) = \frac{ \sin{\theta}  e^{-\phi_\mathrm{b}/k_\mathrm{B} T}}{\int_\theta \sin{\theta} e^{-\phi_\mathrm{b}/k_\mathrm{B} T} d\theta} = \left[\frac{C}{1-e^{-2 C}}\right] \sin{\theta} e^{-C\left(1-\cos{\theta}\right)}
\end{equation}
which we have compared with our simulated values in \fref{fig: C curves no HI 3 beads}.

\section{Methods for calculating tumbling times}
\label{Appendix: Tumbling}

When a polymer chain undergoes shear flow, the rotational component of the velocity field causes end-on-end tumbling of the chain.
Since this is an integral feature of shear flow, we wish to have a method to quantify this tumbling frequency (or its inverse, tumbling time $\tau_\mathrm{tumble}$).
There are three general methods for doing so in the literature.
The first, which we will not describe in detail, is to use the peak in the power spectral density (PSD), which was often calculated in early BD studies \cite{schroeder2005dynamics}.
The other two methods are what we will call direct end-on-end calculation, and the cross-correlation of the gyration tensor.

\begin{figure}
    \centering
    \includegraphics[width=8.5cm,height=!]{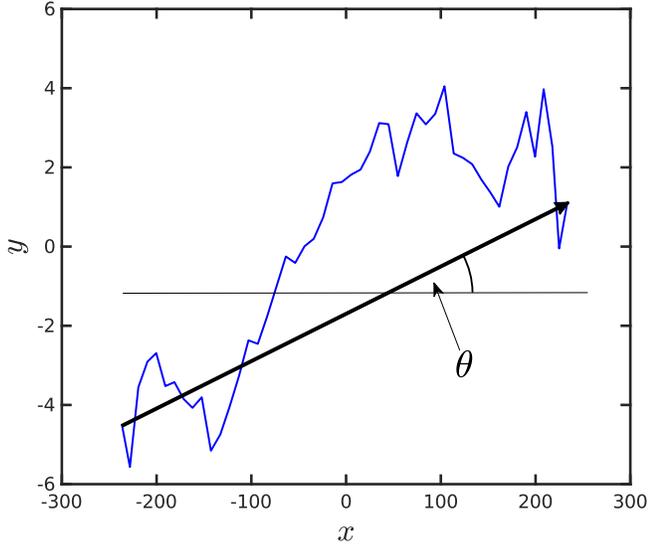}
    \caption[Comparison of TTCF results with shear flow calculations]{Schematic of polymer chain in shear flow, showing the angle the end-to-end vector makes with the flow direction for calculation of the tumbling period. This angle is plotted over time for a single trajectory in \fref{fig:tumbling theta}}
    \label{fig:End tumbling schematic}
\end{figure}

The direct calculation method is straightforward - one simply finds the end-to-end vector of the total polymer chain, and calculates its total average rotational velocity.
This method was employed by Dalal et al. in a BD simulation study \cite{Dalal2012tumbling}, and also by Huber et al. in an experimental study directly imaging actin molecules \cite{Huber2014}.
To demonstrate the procedure, we display a schematic of a polymer chain stretched in shear flow in \fref{fig:End tumbling schematic}.
The angle $\theta$ is of the end-to-end vector with respect to the flow direction at a particular timestep.

\begin{figure}
    \centering
    \includegraphics[width=8.5cm,height=!]{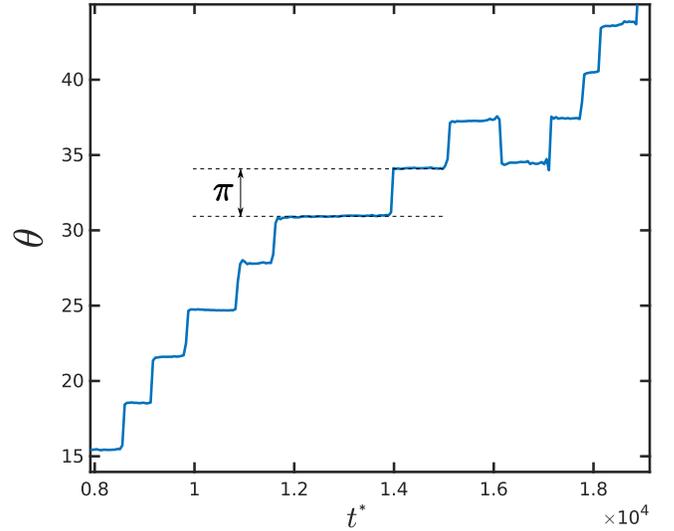}
    \caption[Comparison of TTCF results with shear flow calculations]{Cumulative change in $\theta$ over time for a single example trajectory in shear flow, as per the definition of $\theta$ in \fref{fig:End tumbling schematic}. A single `tumble' is identified by a change of $\theta$ by $\pi$ radians, shown on the figure for a tumble at $t^* \approx 1.4 \times 10^4$. An example trajectory during a tumbling step, when the end-on-end vector rotates by $\pi$, is given in \fref{fig: chain contour tumbling}.}
    \label{fig:tumbling theta}
\end{figure}

\begin{figure}[!ht]
    \centering
    \includegraphics[width=8cm,height=!]{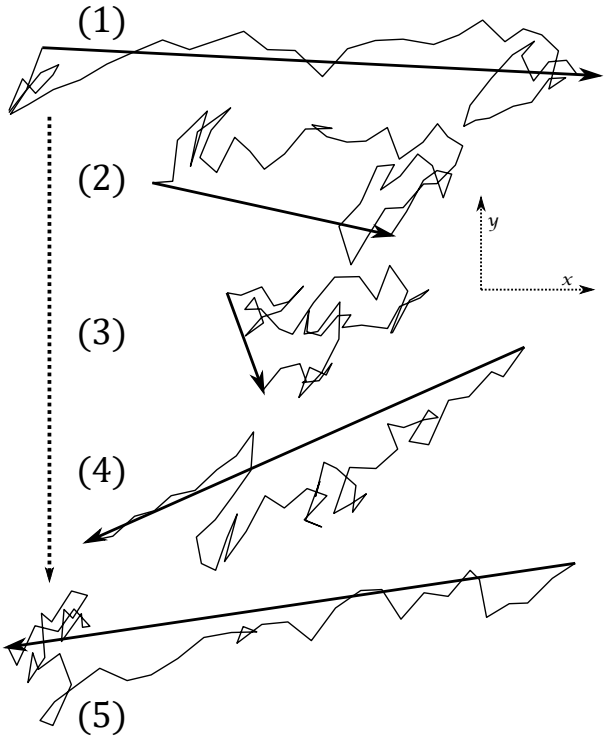}
    \caption{Example of chain contour during an end-on-end tumbling event. Numbers represent successive times, where (1) represents the state prior to a `step' in \fref{fig:tumbling theta}, while (5) represents the state after the `step'. The $x$ and $y$ axes correspond to flow and gradient directions respectively. End-to-end vector is displayed via an arrow from bead $\mu=1$ to bead $\mu=N$.}
    \label{fig: chain contour tumbling}
\end{figure}

To find the tumbling time, we first plot the cumulative angle the end to end vector has swept out in some time $t$.
This is displayed in \fref{fig:tumbling theta}, which shows $\theta$ as a function of dimensionless time.
Note the clear `steps' in $\theta$, which are of $\pi$ radians, corresponding to half a revolution of the chain.
An example revolution is shown in \fref{fig: chain contour tumbling}, where one can clearly identify the half-revolution of the end-to-end vector.
It is this half-revolution that we call a `tumble', and the tumbling period is simply given by:
\begin{equation}
    \tau_\mathrm{tumble} = \frac{\Delta \theta}{\Delta t} \frac{1}{\pi}
\end{equation}
where $\Delta \theta$ is the cumulative change in the rotation angle $\theta$, and $\Delta t$ is the total time over which sums the change in $\theta$.
For example, in \fref{fig:tumbling theta} the total change $\Delta \theta \approx 30$, while the change in time $\Delta t \approx 1 \times 10^4$, so we have $\tau_\mathrm{tumble} \approx 3 \times 10^3$.
This is then averaged over all trajectories to obtain a mean tumbling time.

\begin{figure}
    \centering
    \includegraphics[width=8.5cm,height=!]{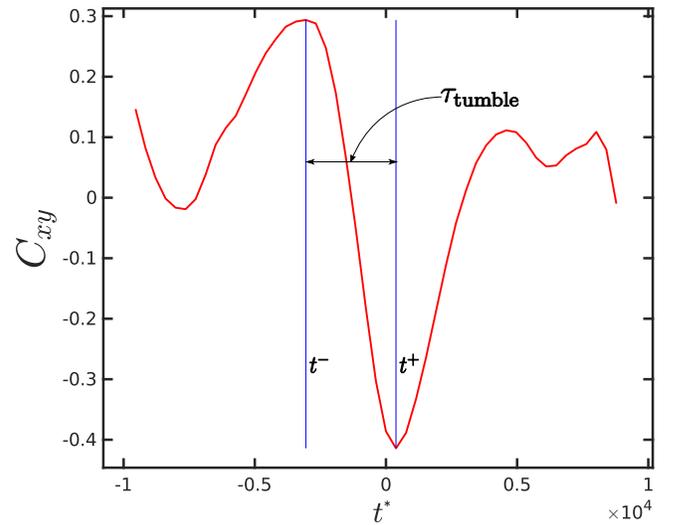}
    \caption[Comparison of TTCF results with shear flow calculations]{Plot of the cross-correlation function $C_{x,y}$ as described in \eref{tumbling cross-correlation}, for an ensemble of trajectories in shear flow. The locations of the peak and trough around $t = 0$, labelled as $t^-$ and $t^+$, are given as vertical blue lines on the figure. The difference in $t^-$ and $t^+$ is identified as the tumbling period.}
    \label{fig:tumbling cross correlation}
\end{figure}

The second method is to use the cross-correlation of the flow and gradient components of the gyration tensor \cite{Chen2013, teixeira2005shear, Huang2011}.
We first define a function $C_{x,y}(t)$, given by the following formula:
\begin{equation}
\label{tumbling cross-correlation}
    C_{x,y}(t) = \frac{\langle \delta G_{xx}(t_0) \delta G_{yy}(t_0+t) \rangle}{\sqrt{\langle \delta G_{xx}^2(t_0) \rangle \langle \delta G_{yy}^2(t_0) \rangle }}
\end{equation}
where $G_{\alpha \alpha}$ for $\alpha = \{x,y,z\}$ is the given component of the gyration tensor, and $\delta G_{\alpha \alpha} = G_{\alpha \alpha} - \langle G_{\alpha \alpha} \rangle$.
We can imagine that as a polymer chain tumbles, it begins in an extended state in the flow direction, and then coils up and expands slightly in the gradient direction as it flips end on end.
This can be seen in \fref{fig: chain contour tumbling}, where the stretched chain conformations before and after the tumble at (1) and (5) have a greater $x$-extent and slightly smaller $y$-extent, while the conformations during the tumble, particularly (3), are far more compact.
Therefore, the time lag in the peaks of this correlation function should give us some sense of the tumbling time.
This can be seen in \fref{fig:tumbling cross correlation}, which gives an example $C_{x,y}(t)$ in shear flow.
The locations of two peaks around $t = 0$ have been labelled as $t^+$ and $t^-$, and it is the difference between these two values which gives us our tumbling time.

In general, these times have similar qualitative behaviour (particularly in terms of scaling with shear rate), although they are not necessarily exactly identical.
Both of these methods have been employed in the study of ring polymers \cite{Chen2013}, in which the cross-correlation defines tumbling motion, while picking a point on the ring and observing the cumulative angle it sweeps out could also be associated with tank-treading motion, rather than pure tumbling.

\section{Calculations of zero-shear viscosity and relaxation time}
\label{Appendix TTCF}

There are several possible methods to obtain the zero-shear viscosity and/or relaxation times from BD simulations.
Relaxation times can be found from:
\begin{itemize}
    \item The exponential decay of the chain size (either end-to-end distance, radius of gyration, or stretch \cite{Sunthar2005parameterfree}) after imposing some external force and then letting the system return to equilibrium.
    \item The autocorrelation of the chain size at equilibrium, which should also exhibit an exponential decay.
    \item The value of some other dynamic property at or near equilibrium, such as the zero-shear viscosity, or hydrodynamic radius \cite{bird1987dynamics}.
\end{itemize}
In general, these may be sums of exponentials, and we are usually interested in the longest relaxation time.
The zero-shear viscosity can be determined in several ways \cite{bird1987dynamics}:
\begin{itemize}
    \item Simulations at low but finite shear rate, potentially including extrapolation to zero shear.
    \item Green-Kubo relations, which give the zero-shear viscosity in terms of the stress autocorrelation at equilibrium.
    \item Finite-shear extensions of Green-Kubo, namely the Transient Time Correlation Functions (TTCF).
    \item The stress decay after a step strain.
\end{itemize}
Finally, although we will not describe them in detail here, we note that there are similar relations for the diffusivity.
Diffusivity can be calculated directly using the mean-squared displacement, as well as Kirkwood's static expression with Fixman's correction for fluctuating hydrodynamic interactions \cite{Sunthar2006}.

We will describe how we have applied the methods listed above to calculate relaxation times and zero-shear viscosity with two examples.
The first is that of a FENE spring with Gaussian EV, hydrodynamic interactions and a bending potential, including nearly all of the physical effects mentioned in the body of this paper.
The second is a stiff FENE-Fraenkel spring of $\sigma^* = 9$ and $\delta Q^* = 1$ with $h^* = \sqrt{3} \chi$, for which it is considerably harder to obtain accurate predictions.

While we have already given explicit expressions for the viscosity at finite shear rate in \eref{eq: material functions eta, psi}, the zero-shear viscosity can be calculated using integrals over the stress autocorrelation at equilibrium $C(t)$, or the relaxation modulus $G(t)$.
These are defined as \cite{bird1987dynamics, Pan2014zeroshear, Fixman1981}:
\begin{equation}
    C(t) = \left\langle \tau_{xy}(0) \tau_{xy}(t) \right \rangle 
\end{equation}
where $\tau_{xy}$ is given by \eref{eq: stress tensor expression}.
At equilibrium, the configuration is isotropic and so the average is also taken over $\tau_{xz}$ and $\tau_{yz}$.
Additionally,
\begin{equation}
    G(t) = \lim_{\gamma_0 \rightarrow 0} \frac{ \left\langle \tau_{xy}(t) \right \rangle}{\gamma_0}
\end{equation}
where an instantaneous strain of $\gamma_0$ is applied at $t = 0$, such that $\dot{\gamma} = \gamma_0 \delta(t)$, and $\delta(t)$ is the dirac delta function.
We then find that:
\begin{equation}
    \eta_0 = \int_0^\infty \left\{ G(t), C(t) \right\} dt
\end{equation}
where by $\{ G(t), C(t) \}$ we mean either $G(t)$ or $C(t)$.

In our simulations, we can measure both $G(t)$ and $C(t)$ at the same time through a variance reduction procedure \cite{Ottinger1996, wagner1997accurate, Kumar2003}.
An equilibrium configuration is simulated along two separate trajectories using the same set of random numbers.
In one trajectory, there is a very rapid step-strain applied at $t = 0$, while the other is kept at equilibrium.
The equilibrium trajectory can be used to calculate $C(t)$, while the decay of the stress after the step-strain gives us $G(t)$.
Since the average stress at equilibrium is zero, we can subtract the stress for the equilibrium trajectory from the stress for the strained trajectory at each timestep to obtain the same $G(t)$ curve with considerably reduced error.
We cannot impose a truly instantaneous step strain, so we instead apply an extremely rapid shear at some large shear rate $\dot{\gamma} \gg 1$, and check that results are independent of $\dot{\gamma}$.

\begin{figure}
    \centering
    \includegraphics[width=8.5cm,height=!]{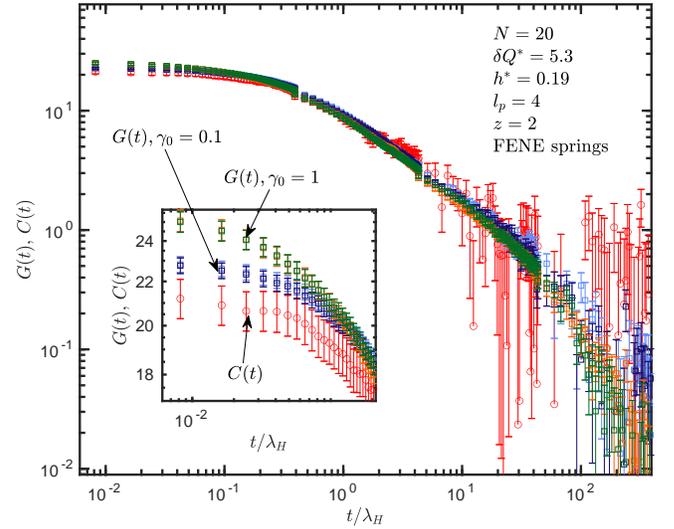}
    \caption[Comparison of TTCF results with shear flow calculations]{Plot of $G(t)$ and $C(t)$ for two different step strain magnitudes, at two different shear rates for the step strain. Light and dark blue curves are $G(t)$ at $\gamma_0 = 0.1$ with $\dot{\gamma}^* = 10^3$ and $\dot{\gamma}^* = 10^5$ respectively. Green and yellow curves are $G(t)$ at $\gamma_0 = 0.1$ with $\dot{\gamma}^* = 10^3$ and $\dot{\gamma}^* = 10^5$ respectively. Red curve is $C(t)$ calculated using the equilibrium trajectories. Inset is curves at small time, showing the variation with $\gamma_0$.}
    \label{fig: Gt Ct FENE HI EV bend}
\end{figure}

This is shown in \fref{fig: Gt Ct FENE HI EV bend} for a FENE spring with HI, EV and a bending potential.
Results are independent of $\dot{\gamma}$, but have not converged in $G(t)$ for $\gamma_0 = 0.1$ and $\gamma_0 = 1$.
It is immediately obvious that error bars are considerably larger for $C(t)$ due to the lack of variance reduction.
The integrals are computed using simple trapezoidal integration:
\begin{equation}
    \eta_0 = \frac{1}{2} \sum\nolimits_{i = 1}^{N_\mathrm{samples}-1} \left(t_{i+1} - t_i\right) \left[ G(t_{i+1}) + G(t_{i}) \right]
\end{equation}
where the error is:
\begin{equation}
    \Delta \eta_0 =\frac{1}{2} \sqrt{ \sum\nolimits_{i = 1}^{N_\mathrm{samples}-1} \left(t_{i+1} - t_i\right)^2 \left[ \Delta G(t_{i+1})^2 + \Delta G(t_{i})^2 \right] }
\end{equation}
with $\Delta G(t)$ as the standard error in $G(t)$ over the ensemble of trajectories.
Since errors accumulate monotonically when integrating an equilibrated trajectory, we often truncate this sum at some $t_\mathrm{max}$ which is less than the total simulated period to obtain reasonable precision.

\begin{figure}
    \centering
    \includegraphics[width=8.5cm,height=!]{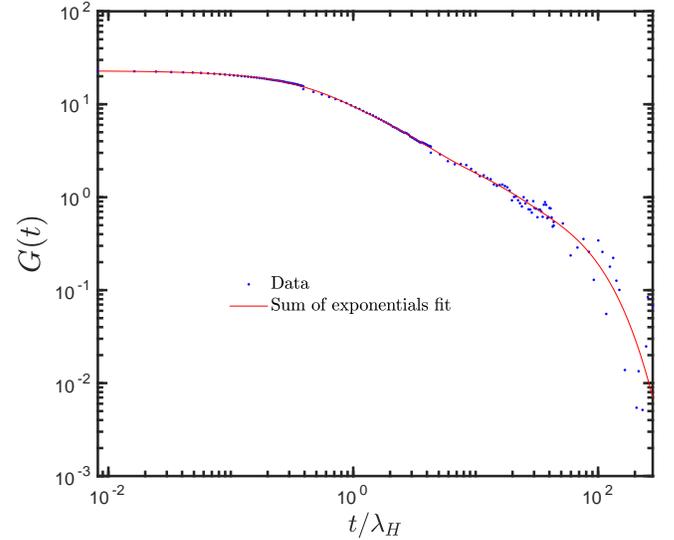}
    \caption[Comparison of TTCF results with shear flow calculations]{$\gamma_0 = 0.1$, $G(t)$ curve in \fref{fig: Gt Ct FENE HI EV bend} fit with a sum of 5 exponentials.}
    \label{fig: Gt fit sum exp}
\end{figure}

It is also possible to fit some function to $G(t)$ or $C(t)$ and then analytically integrate the resulting function to infinity, as done by Pan et al. \cite{Pan2014zeroshear}.
We have used the peeling method to fit a sum of exponentials.
This involves first fitting a single exponential to the tail of the data, subtracting this fit away from the data, and then fitting a new exponential to the tail of the modified data.
This process is continued for however many exponentials is needed for a reasonable fit, generally found to be 3 to 6.
An example fit is shown in \fref{fig: Gt fit sum exp} for the $G(t)$ data in \fref{fig: Gt Ct FENE HI EV bend} with a sum of 5 exponential functions.
Despite the apparent accuracy of this fit, we generally find that direct trapezoidal integration is sufficient, and simply use the exponential fit as a check against the direct result.

An extension of the Green-Kubo relations, the so-called Transient Time Correlation Functions (TTCF), can be used to obtain the viscosity at finite shear rates without the associated increase in error at very low shear rates.
These functions have been used in non-equilibrium molecular dynamics (NEMD) simulations for years, but have not (as far as we are aware) been applied to BD simulations to date.
A excellent review of NEMD simulations, as well as the statistical mechanical foundations of TTCFs, can be found in the textbook by Daivis and Todd \cite{todd_daivis_2017}.
Essentially, the technique involves integrating the correlation between some phase space variable at equilibrium and after the inception of an external field.
For the specific case of shear flow, the time-dependent behaviour of some phase variable $B$ is given by:
\begin{equation}
    \langle B(t) \rangle = \langle B(0) \rangle - \frac{1}{k_\mathrm{B} T} \dot{\gamma} V \int^t_0 \left\langle B(s) \tau_{xy}(0) \right \rangle \mathrm{d}s
\end{equation}
where $V$ is the system volume, $\dot{\gamma}$ is the flowrate, and $\tau_{x,y}$ is the component of the stress tensor in the flow and gradient directions.
For the specific case of viscosity, we can find the average over $\tau_{xy}$ and apply Newton's law of viscosity \cite{todd_daivis_2017}:
\begin{equation}
\label{TTCF viscosity shear flow}
    \eta(t;\dot{\gamma}) = -\frac{V}{k_\mathrm{B}T} \int^t_0 \left\langle \tau_{xy}(s;\dot{\gamma}) \tau_{xy}(0;\dot{\gamma} = 0) \right \rangle \mathrm{d}s
\end{equation}
where $\tau_{xy}(t,\dot{\gamma})$ is the $xy$ component of the stress tensor at time $t$ and shear rate $\dot{\gamma}$.
Notably, we do not have to explicitly divide by shear rate in this expression.
For the direct calculation where $\tau_{xy}$ becomes smaller and smaller at steady state, divided by a smaller and smaller shear rate, leading to very large relative errors as the absolute error in $\tau_{xy}$ stays constant.

\begin{figure}
    \centering
    \includegraphics[width=8.5cm,height=!]{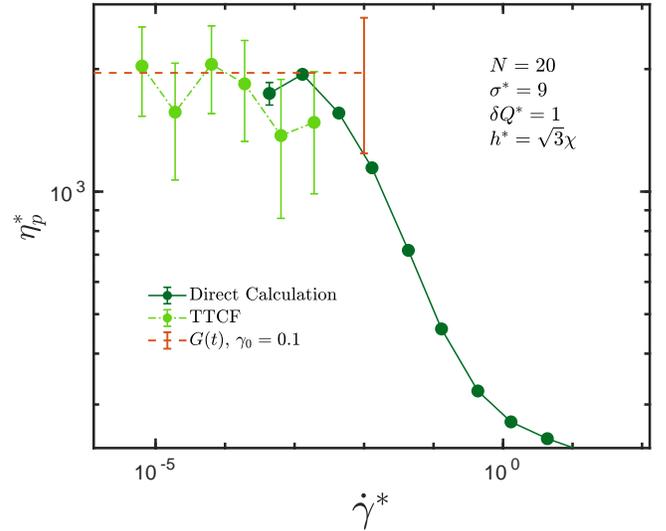}
    \caption{Comparison of listed methods to obtain zero-shear viscosity. The $C(t)$ result is similar to that for $G(t)$, but with considerably larger error bars.}
    \label{fig:TTCF sigma 9 N20}
\end{figure}

Comparison of these results is given in \fref{fig:TTCF sigma 9 N20}.
It is clear that for a stiff spring with $\sigma^* = 9$, although the direct calculation is difficult to fully extrapolate to zero shear, it has considerably smaller error bars than the alternate methods.
Therefore, generally estimates of the zero-shear viscosity for the relaxation time are made using $C(t)$ or $G(t)$, and then exact results plotted using the viscosity at low $Wi$.

\providecommand{\noopsort}[1]{}\providecommand{\singleletter}[1]{#1}%

\end{document}